\documentclass[reprint,aps, prl,superscriptaddress,amsmath,amssymb]{revtex4-2}
\usepackage{graphicx}
\usepackage{dcolumn}
\usepackage{bm}
\usepackage[mathlines]{lineno}
\usepackage{etoolbox}
\usepackage[usenames,dvipsnames]{color}
\usepackage{gensymb}
\usepackage{graphicx}
\usepackage{subfigure}
\usepackage{chemformula}
\usepackage{amsmath}
\usepackage{amssymb}
\usepackage{dsfont}
\usepackage{float}
\usepackage{hyperref}
\usepackage{multirow}
\usepackage{rotating}
\usepackage{epstopdf}
\usepackage{textcomp}
\usepackage{gensymb}
\usepackage{color}
\usepackage{wrapfig}
\usepackage{fancyhdr}
\usepackage{lipsum}
\usepackage{tabularx}
\usepackage[normalem]{ulem}
\usepackage{soul}
\usepackage[capitalize]{cleveref}

\newcommand{\IACS}{\affiliation{School of Physical Sciences, Indian Association for the Cultivation of Science, Jadavpur, Kolkata 700032, India}}

\newcommand{\SNBose}{\affiliation{Department of Condensed Matter and Materials Physics, S. N. Bose National Centre for Basic Sciences, Kolkata 700106, India}}

\usepackage[T1]{fontenc}
\DeclareUnicodeCharacter{2212}{-}
\DeclareUnicodeCharacter{03B4}{$\delta$}

\begin{document}	
\preprint{APS}	
\title{Exceptionally Slow, Long Range, and Non-Gaussian Critical Fluctuations Dominate the Charge Density Wave Transition}

\author{Sk Kalimuddin}
\IACS
\author{Sudipta Chatterjee}
\SNBose
\author{Arnab Bera}
\IACS
\author{Hasan Afzal}
\IACS
\author{Satyabrata Bera}
\IACS
\author{Deep Singha Roy}
\IACS
\author{Soham Das}
\IACS
\author{Tuhin Debnath}
\IACS
\author{Bhavtosh Bansal}
\email[Author to whom correspondence should be addressed:]{bhavtosh@iiserkol.ac.in}
\affiliation{Indian Institute of Science Education and Research Kolkata, Mohanpur, Nadia, West Bengal 741246, India}
\author{Mintu Mondal}
\email[Author to whom correspondence should be addressed:]{sspmm4@iacs.res.in}
\IACS
\date{\today}

\begin{abstract}
\ch{(TaSe4)2I} is a well-studied quasi-one-dimensional compound long-known to have a charge-density wave (CDW) transition around 263 K. 
We argue that the critical fluctuations of the pinned CDW order parameter near the transition can be inferred from the resistance noise on account of their coupling to the dissipative normal carriers. Remarkably, the critical fluctuations of the CDW order parameter are slow enough to survive the thermodynamic limit and dominate the low-frequency resistance noise.
The noise variance and relaxation time show rapid growth (critical opalescence and critical slowing down)  within a temperature window of $ \varepsilon \approx \pm 0.1$, where $\varepsilon$ is the reduced temperature. This is very wide but consistent with the Ginzburg criterion.  We further show that this resistance noise can be quantitatively used to extract the associated critical exponents. Below $|\varepsilon | \lesssim 0.02$, we observe a crossover from mean-field to a fluctuation-dominated regime with the critical exponents taking anomalously low values. The distribution of fluctuations in the critical transition region is skewed and strongly non-Gaussian. This non-Gaussianity is interpreted as the breakdown of the validity of the central limit theorem as the diverging coherence volume becomes comparable to the macroscopic sample size. The large magnitude critical fluctuations observed over an extended temperature range, as well as the crossover from the mean-field to the fluctuation-dominated regime highlight the role of the quasi-one dimensional character in controlling the phase transition.
 \end{abstract}
\maketitle
Charge density waves (CDWs) are fascinating cooperative phenomena characterized by periodic modulation of the conduction electrons accompanied by lattice distortion \cite{peierls1955quantum,GRUNERRevModPhys.60.1129,monceau2012electronic,gruner2018density}. Within the framework of the Ginzburg-Landau theory, the CDW state is represented by a superconductor-like two-component order parameter, $\Psi(\Vec{r},t)~=~\Delta (\Vec{r},t)~e^{i \phi (\Vec{r},t)}$ \cite{GRUNERRevModPhys.60.1129, monceau2012electronic,supply}, where $\Delta (\Vec{r},t)$ is the amplitude of the spatio-temporal modulated charge density and $\phi (\Vec{r},t)$ its phase.

CDWs have been extensively investigated for their crucial role in Fermi surface instability \cite{FermisurfaceInstability,brouet2004fermi,tournier2013electronic}, metal-insulator transitions \cite{TaS2PhysRevLett.105.187401,hellmann2012time}, and superconductivity \cite{Tan2021cdwsuperconductivity,chen2022chargesuperconductivity}.  
Over the years, the dynamics of collective CDW modes have been studied \cite{GRUNERRevModPhys.60.1129, gruner2018density,monceau2012electronic} using techniques as varied as neutron scattering \cite{Moncton1975neutron,Weber2011neutron}, Raman spectroscopy  \cite{sugai1985TSI2Raman,sugai2006phasonRaman,song2023phononslowing}, ultra-fast time-domain THz spectroscopy \cite{kim2023observation,warawa2023combined}, optical pump-probe experiments  \cite{demsar1999single, Schaefer2014Collective,zong2019dynamical}, ultra-fast x-ray scattering\cite{Nguyen_Ultrafast_xRay2023} and time-resolved photoemission spectroscopy  \cite{schmitt2008transient, liu2022observation}. There has also been a recent upsurge of interest due to recent claims of an axial anomaly of collective CDW modes \cite{Gooth2019,Zhang2020firstprinciple,Shi2021}.

The CDW phase is often observed in low-dimensional materials where we expect no long-range order at finite temperature \cite{Hohenberg1965}. This apparent violation of the Hohenberg-Mermin-Wagner theorem is prevented by the fact that the chain-like CDW materials such as TTF-TCNQ, NbTe$_4$,  TaTe$_4$, and \ch{(TaSe4)2I} are not strictly one-dimensional but merely anisotropic, with non-negligible inter-chain interactions. Nevertheless, the vestiges of the quasi-one-dimensional nature remain in the strong role that fluctuations play in significantly lowering the transition temperature $T_{CDW}^{3D}$ much below what is predicted by the mean-field theory, $T_{CDW}^{MF}=2\Delta / (3.53~k_B)$\cite{gruner2018density}. Fluctuations are also known to modify the density of states and other physical properties \cite{McKenzie_PRB1995,monien2001exact,gruner2018density}. Yet, direct measurements of the fluctuations themselves have been elusive.

In this Letter, we investigate the temperature-dependent resistance noise spectra in  \ch{(TaSe4)2I} and carefully map out the critical scaling of the fluctuations over two decades of the reduced temperature on both sides of the CDW transition, which is found to be 263.03 K. \ch{(TaSe4)2I} belongs to the quasi-1D series of transition-metal-tetra-chalcogenides, (MX$_4$)$_n$I (M = Nb, Ta; X = S, Se;  $n$ = 2, 3, 10/3) which has recently garnered attention for intriguing electronic properties \cite{1DbandinsulatorPhysRevLett.84.1272,DielectricPhysRevLett.96.046402,Gooth2019,Shi2021,cheng2024chirality}, including unusual phase coexistence of superconductivity and ferromagnetism \cite{MMondalnTSI}.

We establish that the \ch{(TaSe4)2I} undergoes  a fluctuation-dominated transition at $T_{CDW}=263.03~K$ \cite{gressier1984characterization,Gooth2019}. The critical fluctuations associated with this CDW condensation of the electrons survive in the thermodynamic limit and manifest in measurements involving time scales of seconds and length scales comparable to the sample size, over an extremely wide temperature window ($\simeq 52 $ K). Remarkably,  we observe the crossover from the mean-field behavior to a fluctuation-dominated regime \cite{McKenzie_PRB1995} with a different set of critical exponents. The anomalously low values of the critical exponents in the fluctuation-dominated regime may even suggest a crossover to a hysteresis-free weak first-order transition. In the process, we establish that resistance noise spectroscopy which has been a very popular tool to study phase transitions \cite{mullerPhysRevLett.114.216403,hemantaPhysRevLett.119.226802,satyakiPhysRevLett.124.095703,machado2022quantum,ghosh2004set,supply}, can be used for quantitative study of the critical exponents \cite{machado2022quantum}.

\begin{figure}
\centering
\includegraphics[width=1.0\columnwidth]{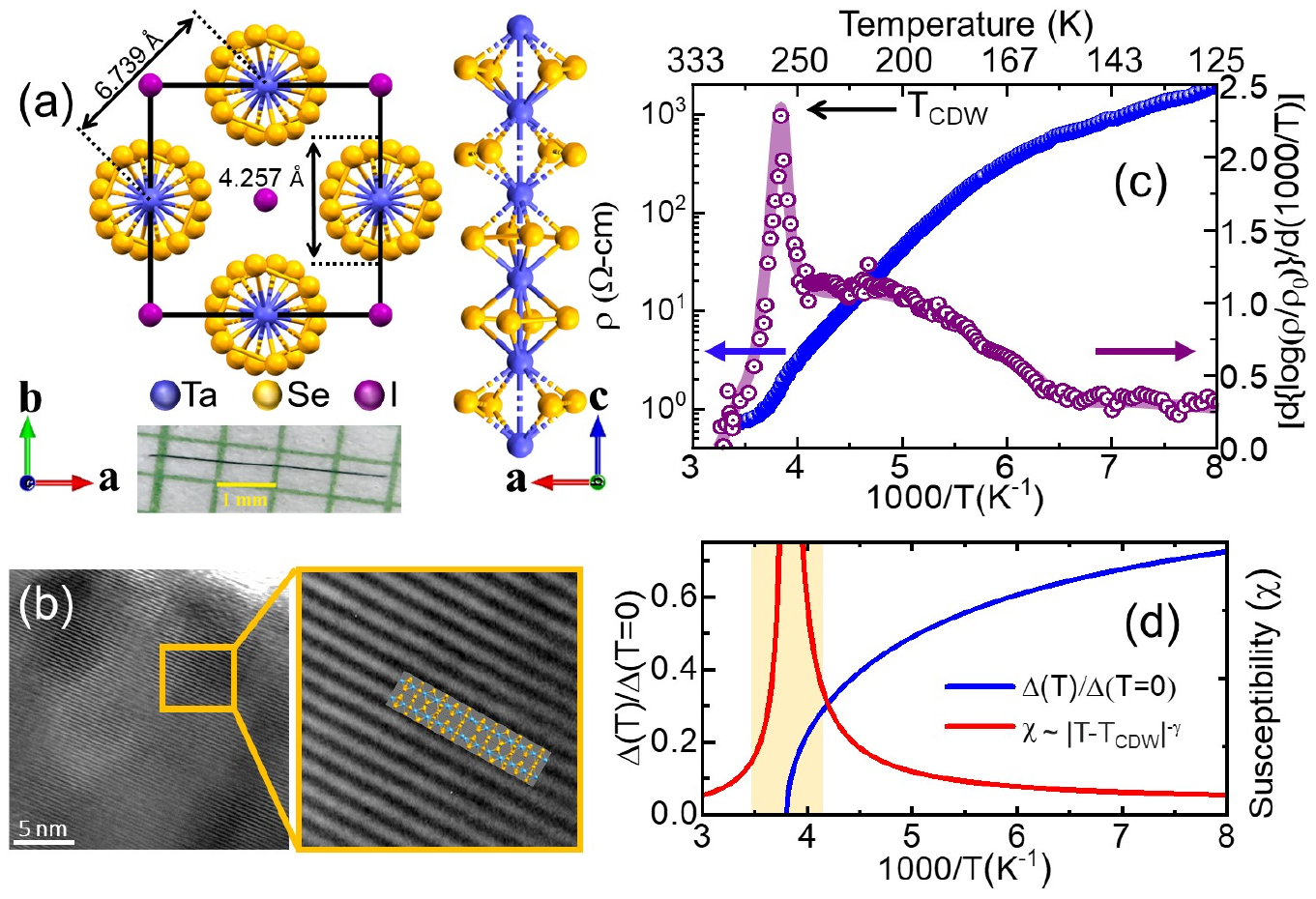}  
\caption{Characterization of the (TaSe$_4$)$_2$I crystal. (a) (top) Schematic of the crystal structure from single crystal X-ray diffraction experiment at $T\sim 300~K$. Projection of the structure of the tetragonal unit cell onto the $ab$- and $ac$-planes respectively to highlight the quasi-1D structure. (bottom) Image of the as-grown single crystal. (b) HRTEM image depicting the crystallinity (c) Resistivity plotted on a logarithmic scale as a function of inverse temperature reveals the activated behavior at low temperatures. The CDW transition ($T_{CDW}\sim 263~K$), barely discernible in the resistivity plot, is clearly seen in the rapid growth of the logarithmic derivative of the resistivity. (d) Schematic of the temperature variation of the order parameter (amplitude) and susceptibility. The yellow shade highlights the fluctuation-dominated region.}
\label{Fig:Characterization}	
\end{figure}

\begin{figure}
\centering
\includegraphics[width=1.0\columnwidth]{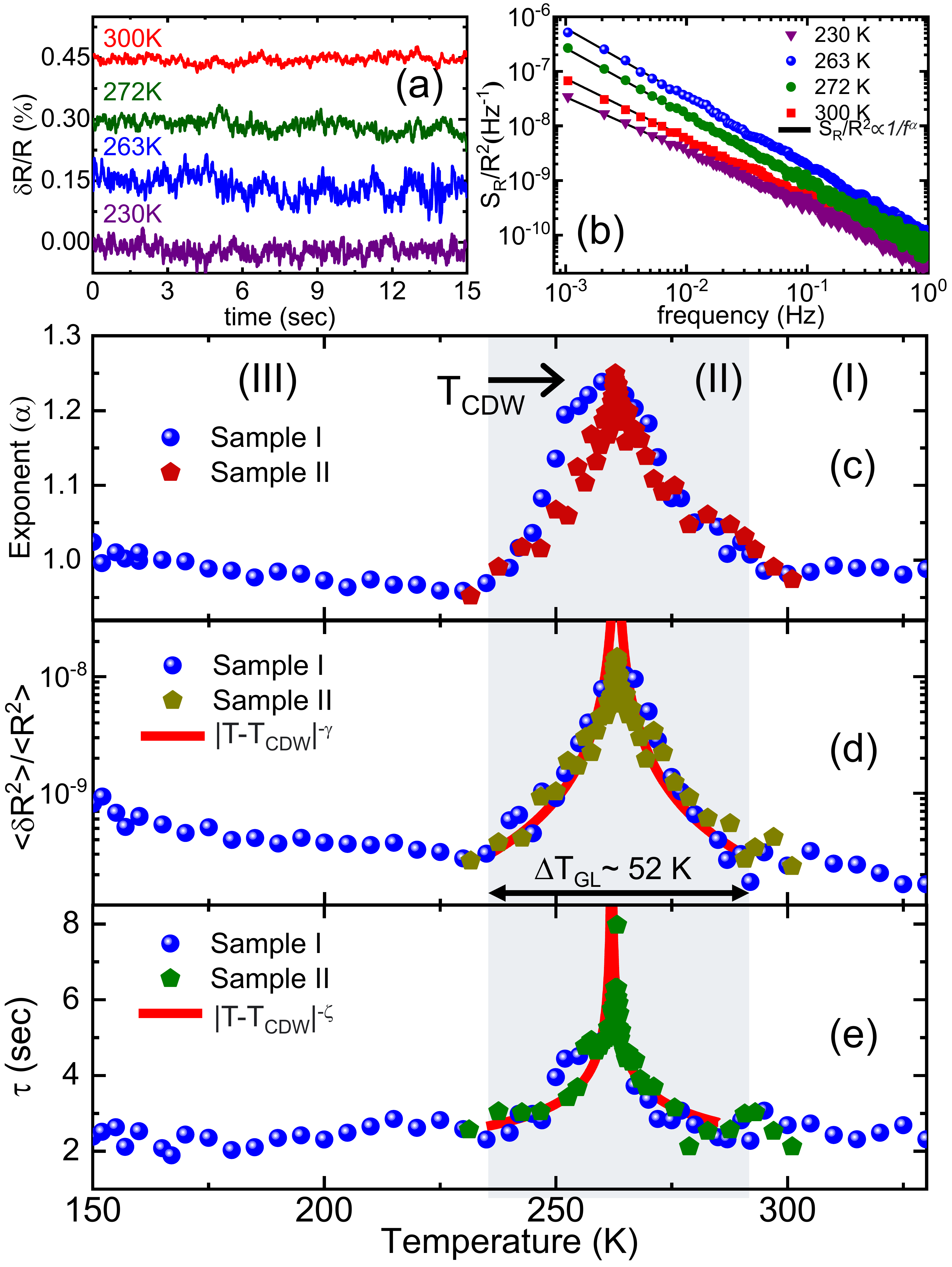} 
\caption{Departure from $1/f$ behavior and critical slowing-down of fluctuations. (a) Representative time-series of resistance fluctuations around $T_{CDW}$ with a vertical shift of 0.15\% for clarity. (b) Normalized PSD (S$_R$(f)/$R^2$) of corresponding resistance fluctuations. (c) Temperature dependence of PSD exponent $\alpha$, where S$_R$(f)/$R^2 \propto 1/ f^{\alpha}$. (d) Temperature variation of the relaxation time ($\tau$) which is estimated from the autocorrelation of resistance fluctuations. (e) The relative variance of resistance fluctuations for two samples. The solid lines (red) in (d)-(e) show the diverging trend at T=T$_{CDW}$.}
\label{Fig:Noise}	
\end{figure}

Single crystals of \ch{(TaSe4)2I}  were grown using chemical vapor transport technique \cite{supply}. 
Schematic of crystal structure, inferred from single-crystal X-ray diffraction, along with a picture of \ch{(TaSe4)2I}  crystal are shown in Fig.~\ref{Fig:Characterization}(a).   High-resolution transmission electron microscopy (HRTEM) image [Fig.~\ref{Fig:Characterization}(b)] clearly indicates the crystallinity and quasi-1D nature. The temperature-dependent resistivity, shown in Fig.~\ref{Fig:Characterization}(c), is non-hysteretic and shows four orders of magnitude increase as the temperature is lowered from 300 K to 125 K. Although barely discernible in this resistivity plot, the logarithmic derivative (right axis) exhibits a prominent peak at $T_{CDW}\approx$ 263 K, consistent with prior investigations \cite{gressier1984characterization}. Below $T_{CDW}$, resistivity exhibits approximately activated behavior as $\rho (T) \approx \rho_0$~e$^{(E_g/k_BT)}$ which is indicative of the growth in n$_{CDW}$ forming the insulating CDW state with decreasing temperature. The conduction process can be approximately thought of as being due to the normal carriers which are created via single-particle excitation across the CDW gap $E_g$, like in a simple semiconductor.  The estimated activation energy over the linear region of the plot is $E_g\simeq$ 216$\pm$2 meV, in very good agreement with previous reports \cite{tournier2013electronic,Gooth2019}. The experimental d[log $\rho_{dc}(T)$/d(1000/T)] exhibits a maximum at T$_{CDW}$ instead of the cusp singularity predicted by the mean-field theory \cite{gruner2018density}. This already highlights the importance of order parameter fluctuations. The fluctuation region around T$_{CDW}$ can be estimated by the Ginzburg criterion ($\Delta$T$_{GL}$). The estimated $\Delta$T$_{GL}$ = $\frac{k_B^2 T_{CDW}}{32\pi^2 (\Delta C V_\xi)^2}\approx$~52 K, where $\Delta$C=0.83~JK$^{-1}$mol$^{-1}$ is specific heat anomaly \cite{starevsinic2002specific,saint2019survey} and V$_\xi$ = 2111~nm$^3$ is the coherence volume of \ch{(TaSe4)2I}.

Here we investigate the critical nature of the order parameter fluctuations across the CDW transition by measuring the low-frequency resistance noise \cite{mullerPhysRevLett.114.216403,hemantaPhysRevLett.119.226802,satyakiPhysRevLett.124.095703,machado2022quantum, scofield1987ac,ghosh2004set,supply}. The time series of the four-probe resistance fluctuations were measured using a lock-in-based phase-sensitive detection under a very small bias \cite{scofield1987ac,ghosh2004set,supply} over a period of 80 minutes with the sample temperature kept fixed to within $\pm 2$ mK \cite{supply}. Fig.  \ref{Fig:Noise}(a) represents the mean-subtracted time series of the resistance fluctuations, $\delta R(t) = R(t)-\langle R(t)\rangle$ at given temperatures.

The normalized power spectral density (PSD) of these representative resistance fluctuations between $f_{min} = 10^{-3}$ Hz to $f_{max} = 1$ Hz, is shown in Fig. \ref{Fig:Noise}(b). PSD exhibits the typical $1/f^{\alpha}$ behavior \cite{ghosh2004set}. The noise exponent ($\alpha$), relative variance $\bigl(\frac{\langle\delta{R^{2}} \rangle}{\langle R^{2}\rangle}= \int_{f_{min}}^{f_{max}} \frac{S_{R}(f)}{R^{2}}df\bigr)$ and relaxation time ($\tau$),  plotted w.r.t. common temperature axis in Fig. \ref{Fig:Noise}(c)-(e). 
The fluctuations reveal three distinct regimes: (I) uncorrelated Gaussian fluctuations in the high-temperature normal state, (II) a strongly correlated fluctuation regime (critical regime) around the CDW transition, and (III) suppressed resistance fluctuations in the long-range ordered CDW state at lower temperatures. The noise observed in regions I (normal state) and III (CDW ordered state)  exhibits a moderate magnitude and nearly Gaussian $1/f^{1.00 \pm 0.05}$ type spectrum as shown in Fig. \ref{Fig:Noise} (c). This is clearly the usual $1/f$ flicker noise \cite{kogan_noise,Weissman1988Noise} and defines the {\em noise floor}.

We now focus on region II where the critical thermodynamic fluctuations emerge above the aforementioned noise floor and dominate the resistance noise $\delta R (t)$. The critical regime extends over a broad temperature window \cite{monien2001exact},  $\varepsilon = (T-T_{CDW})/T_{CDW} \approx [-0.1, +0.1]$ \textcolor{red}, which agrees rather well with a naive estimate based on Ginzburg criterion, $\Delta T_{GL} \approx 52$ K. The noise exponent $\alpha$ increases to $\alpha$~$\sim$~1.25 indicating a substantial shift in the spectral weight of fluctuations to lower frequencies in the vicinity of \textit{T$_{CDW}$}.  The relative variance also grows by an order of magnitude around $T_{CDW}$ [Fig. \ref{Fig:Noise}(d)]. The rapid growth in the variance (critical opalescence) around $T_{CDW}$ clearly indicates that these are critical fluctuations \cite{mullerPhysRevLett.114.216403}. The corresponding critical slowing down is directly observed in Fig. \ref{Fig:Noise}(d) where the relaxation time $\tau$ for the decay of the autocorrelation, C(t)=$\langle \delta$R(t$'$)$\cdot$$\delta$R(t$'$+t)$\rangle_{t'}$ is plotted as the function of temperature  \cite{satyakiPhysRevLett.124.095703, supply}. 

Critical opalescence and critical slowing around phase transitions continue to be of interest across a range of systems spanning different length scales —from quark-gluon matter \cite{Hydrodynamic.PhysRevLett.127.072301,QCDPhysRevLett.102.032301, QCDPhysRevLett.97.032002}, to jamming transitions \cite{PhysRevLett.124.118001, PhysRevLett.131.178201}, dynamical systems \cite{gomez2017critical, PhysRevLett.125.134102}, ecology and biology \cite{scheffer2009critical,scheffer2012anticipating}, and, of course, a diverse array of condensed matter systems \cite{zong2019dynamical,mullerPhysRevLett.114.216403,NiermannPRL2015} 
 
\begin{figure}
\centering
\includegraphics[width=1.0\columnwidth]{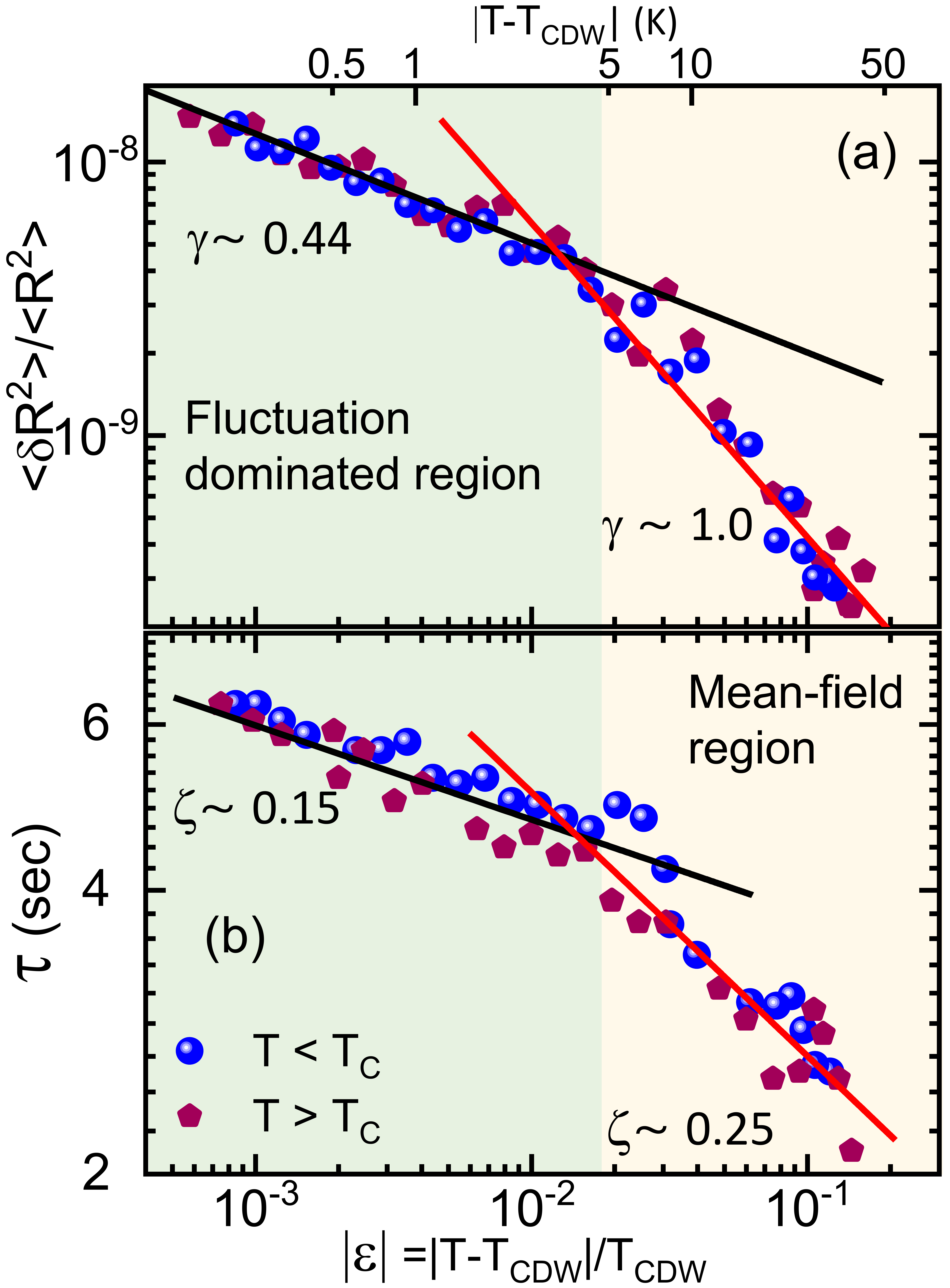}
\caption{{\em Power-law 
 scaling analysis: Crossover from the mean field to the critical fluctuation-dominated regime}. The fluctuation data from region II shown in Fig. 2 (d--e) are replotted as a function of the reduced temperature ($\varepsilon$) in logarithmic scale. (a) The normalized variance $\frac{<\delta R^2>}{<R^2>}$, being proportional to the isothermal susceptibility has the scaling $\chi{_{_T}}\sim |\varepsilon|^{-\gamma}$, and (b) the relaxation time $\tau \sim |\varepsilon|^{-\zeta}$, where $\zeta$ is the mode-softening exponent\cite{gruner2018density}. Solid lines (black and red) are guides to the eye. Note that the mean-field theory predicts $\gamma=1$ and $\zeta=1/4$\cite{gruner2018density}. There is a clear crossover to different exponent values as one enters the critical fluctuation-dominated regime.} 
\label{Fig:Exponent}	
\end{figure}

The low-frequency resistance fluctuations can arise from fluctuations in either the carrier mobility $\delta \mu$ or in the density of electrons $\delta n$, viz., $\delta R(t)= \frac{\partial R}{\partial n}  \delta n(t) +\frac{\partial R}{\partial \mu}  \delta \mu (t)$ \cite{Hooge19941byfnoise}. Within the two-fluid model of CDW ground state \cite{two_fluidPhysRevLett.62.2032,supply,gruner2018density}, the total electron density is the conserved sum of normal $n_{normal}$ and the condensed fraction, $n_{CDW}$ i.e. $  n = n_{normal} +n_{CDW}$ . At low bias, the $n_{CDW}$ is pinned and therefore insulating. The temporal changes in mobility are also negligible because they are non-critical and a result of the scattering events occurring on the microscopic time scales of $10^{-12}-10^{-9}$ seconds. They would thus average out on our time scale of interest. 

Thus $\delta R(t) \simeq \frac{\partial R}{\partial n_{normal}}  \delta n_{normal}(t)$. Since the total density $n$ is conserved \cite{tinkham2004introduction, gruner2018density,two_fluidPhysRevLett.62.2032, supply},  $\delta n_{CDW}(t) \simeq - \delta n_{normal}(t)$.  Thus $\delta R(t)  \simeq - \frac{\partial R}{\partial n_{normal}}  \delta n_{CDW} (t)$ \cite{supply}. Furthermore, as $\delta n_{CDW} = \delta (Re[\Psi])$ \cite{supply}, the critical fluctuations in the order parameter directly translate into the experimentally measured resistance fluctuations.
\begin{figure}
\centering
\includegraphics[width=1.0\columnwidth]{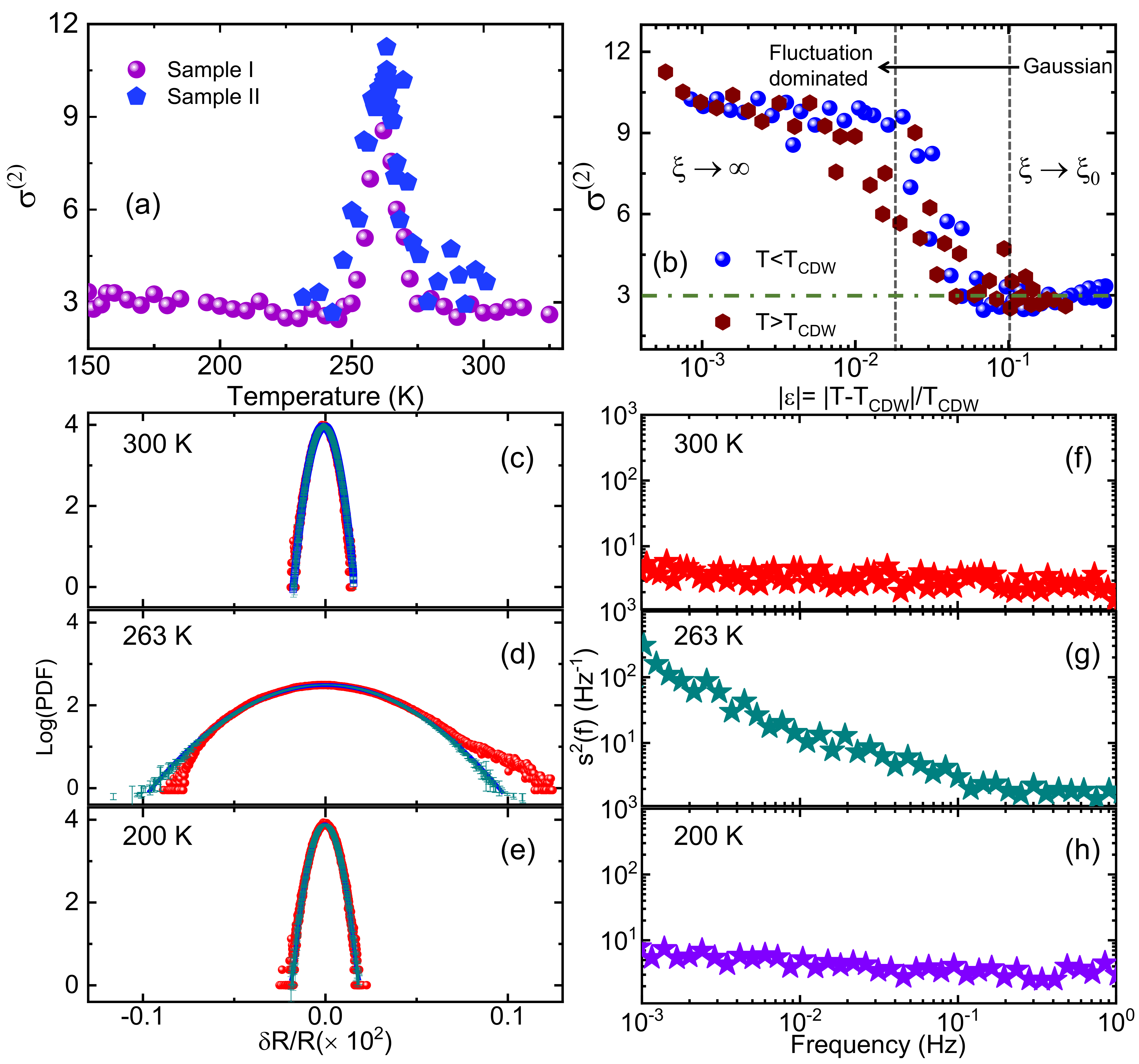}  
\caption{{\em Correlated fluctuations and non-Gaussianity at the critical point}.---(a) The normalized ``second-variance'' $\sigma^{(2)}$ as a function of temperature. $\sigma^{(2)}$$\simeq$ 3 is expected for the uncorrelated Gaussian fluctuations. (b) $\sigma^{(2)}$, now plotted as the function of $|\varepsilon |$. A build-up of non-Gaussianity is observed as the correlation length grows in the vicinity of the critical point. Compare with the observations in Fig. 3. (c--e) The PDF of the normalized resistance fluctuations plotted along with the Gaussian distribution with zero mean, and (f--h) frequency-resolved second spectra $S^2(f)$ above, below, and at the critical temperature. Note that the qualitative difference at 263 K where the fluctuations have a much larger variance and the PDF is clearly non-Gaussian.}
\label{Fig:Sigma2}	
\end{figure}

We can thus understand our experimental results within the thermodynamics of critical transitions. The resistance fluctuations  [Fig. \ref{Fig:Noise}(d)] essentially reflect the order parameter fluctuations, viz.,  $\frac{\langle\delta R^{2} \rangle}{\langle R^{2} \rangle} \propto \langle \delta n_{CDW}^{2} \rangle\simeq \langle \delta \Delta^{2} \rangle \simeq \int d^3 r\, G(r)=k_BT\chi_T$.  The fluctuation-dissipation theorem has been used in the last term and $G(r)$ is the equal-time two-point correlation function of the order parameter  \cite{goldenfeld2018lectures}.  Since the isothermal susceptibility $\chi_{_{T}}$ should also diverge as $\chi_{_{T}}\propto |T-T_{CDW}|^{-\gamma}$, we expect $\frac{\langle\delta R^{2} \rangle}{\langle R^{2} \rangle} \propto |T-T_{CDW}|^{-\gamma}$ with a mean-field value of $\gamma = 1$. This rapid growth of the variance is the manifestation of critical opalescence and is indeed seen in Fig. \ref{Fig:Noise}(d) and \ref{Fig:Exponent}(a). 

Similarly, the auto-correlations of the resistance fluctuations capture the autocorrelation of the CDW order parameter fluctuations. Growth of their relaxation time $\tau$ [Fig. \ref{Fig:Noise}(e)] is the signature of critical slowing down.  While such critical slowing down is usually understood within the dissipative time-dependent Landau-Ginzburg dynamics \cite{chaikin1995principles,satyakiPhysRevLett.124.095703,Schaefer2014Collective,goldenfeld2018lectures}, for a CDW system both the amplitude and phase modes are inertial; the free energy has a kinetic energy term that supports wave-like solutions \cite{gruner2018density}. In such systems, criticality is instead manifested via mode softening \cite{gilmore1993catastrophe} i.e. $\omega_{\Delta}$~$\propto (T-T_{C})^{-\zeta}$\cite{gruner2018density}, where $\zeta = \frac{1}{4}$ from mean-field. Here, $\zeta$ represents the exponent of the characteristic time scale associated with the propagating amplitude mode. Consequently, the characteristic time scale ($\tau \sim 2 \pi /\omega_{\Delta}$) for the propagation of the fluctuations diverges as the rigidity of the amplitude mode collapses.

For such critical exponent analysis of power law divergence, Fig.  \ref{Fig:Noise} (c-e) also shows measurements on a second sample with closely spaced (linearly spaced on a logarithmic scale) data points within the $\Delta T_{GL}$. The values of the critical exponents $\gamma$ and $\zeta$, inferred by an error minimization algorithm [see supplementary for details \cite{supply}]  are shown in Fig.~\ref{Fig:Exponent}. This analysis \cite{supply} also yields a more precise $T_{CDW}=$ ($263.03 \pm 0.05$)~K. The analysis also shows that the transition is non-hysteretic to 50 millikelvins and thus a second-order or a weak first-order transition, unlike most CDW compounds  \cite{Wilson1975,Lee2019_TaS2}.

While we had inferred the value of the Ginzburg temperature to be around 52 K, in Fig. \ref{Fig:Noise}, Fig. \ref{Fig:Exponent} shows that the fluctuations play a more subtle role in such low dimensional systems \cite{McKenzie_PRB1995,gruner2018density}. Relatively far from the critical point ($0.02\lesssim |\varepsilon|\lesssim 0.12$ ), the critical exponents start with a mean-field value ($\gamma = 1.06 \pm 0.08$ for $T^+_{CDW}$; $\gamma = 1.05 \pm 0.09$ for $T^-_{CDW}$) and for the autocorrelation time $\tau$, $\zeta = 0.25 \pm 0.02$ for $T^+_{CDW}$; $\zeta = 0.26 \pm 0.03$ for $T^-_{CDW}$) consistent with the mean-filed predictions of $1$ and $1/4$, respectively. This is observed on both sides of the transition as expected from conventional phase transition theory. But close to the critical point ($|\varepsilon|\lesssim 0.02$) we observe a crossover to a regime where the exponents are distinctly different; the system exhibits non-mean field behavior with exponents $\gamma = 0.44 \pm 0.09$ and $\zeta = 0.15 \pm 0.03$. Note that the value of $\gamma$ is anomalously low compared to $\gamma \approx 1.3$, the value expected for the three-dimensional XY model \cite{pelissetto2002critical}. 

Note that this could also indicate a very weak first-order transition, which may be experimentally indistinguishable from a genuine critical transition. While such a change in the order of the transition can arise on account of the changes in the lattice structure at the transition, previous studies have concluded that the symmetry arguments do not necessitate it \cite{van2001structure}. Also, the observation of large fluctuations, scaling of the physical quantities, and the absence of hysteresis is striking. Alternatively,  the thermal fluctuations themselves \cite{Halperin1974, fluctuationinduced2013}, given the similarity of the CDW order parameter to the superconductor and liquid crystal order parameter, could yield a very weak first-order transition via the Halperin-Lubensky-Ma mechanism  \cite{Halperin1974}. We tentatively note that the low energy band structure of \ch{(TaSe4)2I} is predicted to have Weyl fermions \cite{Gooth2019,Zhang2020firstprinciple, Shi2021}. In these systems, fluctuating pseudo-gauge fields (e.g., on the account of strain) \cite{yu2021pseudo} could give such a pathway.

Fig. \ref{Fig:Sigma2} summarizes the statistics of the fluctuations at different temperatures. Fig. \ref{Fig:Sigma2} (a) shows the Fourier transform of the four-point correlation function, the second spectrum \cite{seidler1996non} S$^{(2)}$(f) = $\int_{0}^{\infty} \langle \delta R^{2}(t) \delta R^{2}(t+\varsigma)\rangle  \cos (2\pi f\varsigma) d\varsigma$  \cite{supply,AKRPhysRevLett.91.216603,chadniPhysRevLett.102.025701, NbNPhysRevLett.111.197001}. The departures of the normalized ``second-variance'' $ \sigma^{(2)} = \int_{0}^{f_{H} - f_{L}} S^{(2)}(f)df/\big|\int_{f_{L}}^{f_{H}} S_{R}(f)df\big|^2$ from 3, is a convenient measure of the non-Gaussianity of the fluctuations \cite{chadniPhysRevLett.102.025701, mullerPhysRevLett.114.216403,sudiptaPhysRevB.104.155101} within the frequency band (f$_L$, f$_H$).  The temperature dependence of $\sigma^{(2)}$, computed with f$_L$ = 20~mHz and f$_H$ = 0.2~Hz, is shown in Fig. \ref{Fig:Sigma2}(a) \cite{supply}.  Note the rapid increase in $\sigma^{(2)}$ from the background value of  $\sim$ 3 near $T_{CDW}$.
The non-Gaussian nature of fluctuations signifies the breakdown of the central limit theorem as the coherence volumes approach the sample size on the account of the growth of two
correlation lengths $\xi_{||}$ and $\xi_\perp$ defined along-the-chain and out-of-the-chain directions respectively, as they both diverge $\propto |T-T_{CDW}|^{-\nu}$\cite{goldenfeld2018lectures}. In Fig \ref{Fig:Sigma2} (b), we now plot $\sigma^{(2)}$ as a function of $\varepsilon$ on both sides of the transition. We find that the Gaussian to the non-Gaussian crossover (departure of $\sigma^{(2)}$ from 3) which begins at the boundary of region II (of Fig \ref{Fig:Noise}) is complete in the fluctuation-dominated regime. The value of the temperature where $\sigma^{(2)}$ saturates is the same as where the critical exponents changed value in Fig. \ref{Fig:Exponent}.

The representative plots of the probability density function (PDF) and $S^2(f)$ are shown in Fig. \ref{Fig:Sigma2} (c-h). Both PDF and $S^2(f)$ reveal the non-Gaussian fluctuations in the critical region II.  Finally, we note that the PDF of the noise around the transition region is asymmetric [Fig. \ref{Fig:Sigma2}(c-e)]. Such skewed non-Gaussian behavior at criticality has been experimentally seen in disparate contexts and is perhaps indicative of extreme value statistics \cite{Antal2001PRLExtreme,GarnierPhysRevLett.100.180601}.  The non-Gaussian fluctuations and critical opalescence have been posited as possible experimental signatures of the elusive quantum chromodynamics critical point \cite{Hydrodynamic.PhysRevLett.127.072301, QCDPhysRevLett.102.032301, QCDPhysRevLett.97.032002}.

\noindent
\underline{{\em Conclusions.}}---Noise measurements have in the past been a powerful tool to study phase transitions \cite{mullerPhysRevLett.114.216403, bogdanovich2002onset,jaroszynski2002universal, satyakiPhysRevLett.124.095703, sudiptaPhysRevB.104.155101}. However, phase transitions often accompanied additional features not simply reducible to the order parameter fluctuations. For example, both the correlation-driven (Mott) as well as the disorder-driven (Anderson), metal-insulator transitions  \cite{mullerPhysRevLett.114.216403, bogdanovich2002onset,jaroszynski2002universal, kogan_noise} are usually accompanied with kinetic arrest, glassy dynamics, two-level behavior, structural (martensitic) transformation, and/or percolative transitions \cite{chadniPhysRevLett.102.025701}.

In contrast, our findings on the continuous and hysteresis-free CDW transition in \ch{(TaSe4)2I} suggest a purely electronic origin  (apart from an accompanying Peierls instability). It is thus remarkable that the manifestations of criticality persist over the scale of the sample size ($\sim$ millimeters). Critical opalescence is evident in the large variance of the fluctuations even in the macroscopic resistivity.  Similarly, the critical slowing down of the correlation times stretches from the picoseconds \cite{zong2019dynamical,song2023phononslowing} to the scale of seconds [Fig. \ref{Fig:Sigma2}(d)]. Moreover, these resistance fluctuations could be directly mapped onto the CDW order parameter fluctuations to quantitatively infer the isothermal susceptibility and the dynamic mode softening critical exponents [Fig. \ref{Fig:Exponent}]. The beginning of the critical region gave exponents in remarkable agreement with the mean-field values. The growing correlation lengths were indirectly inferred from the growth of the non-Gaussian character of the fluctuations [Fig. \ref{Fig:Sigma2}]. While one expects a dimensional crossover from the regime of weakly coupled quasi-one dimensional chains to the system effectively becoming three-dimensional in the critical regime on the account of the growing transverse correlation lengths \cite{gruner2018density}, the quasi-one-dimensional character of the system was nevertheless manifest in the dominant role played by fluctuations. We observed a crossover to a non-mean field regime on a closer approach to the critical point ($\varepsilon\approx [-0.02, +0.02]$). While theoretically expected \cite{McKenzie_PRB1995}, such a crossover has been hard to experimentally observe. The anomalously small values of the exponents in this regime do not rule out a very weak first-order transition. This may be caused by an accompanying structural transition or may itself be fluctuation-induced \cite{Halperin1974}.    

\section{Acknowledgments}
This work was partly supported by the 'Department of Science and Technology, Government of India (Grant No. CRG/2023/001100 and CRG/2022/008662) and CSIR-Human Resource Development Group (HRDG) (03/1511/23/EMR-II).  S.K. thanks IACS for PhD fellowship. M.M. and A.B. thank Subhadeep Datta for the technical help. We thank Krishnendu Sengupta, Arnab Das, Nandini Trivedi, Mohit Randeria, and Sumilan Banerjee for the fruitful discussions. 


%


\newpage

\vspace{2cm}

\begin{center}

\onecolumngrid

\newpage
\textbf{\underline{\large{Supplemental Material}}} \\
\vspace{0.5cm}
\textbf{Exceptionally Slow, Long Range, and Non-Gaussian Critical Fluctuations Dominate the Charge Density Wave Transition} \\ 
\vspace{0.5cm}

\end{center}

\renewcommand{\thesection}{S\arabic{section}}
\renewcommand{\thefigure}{S\arabic{figure}}
\renewcommand{\thetable}{S\arabic{table}}
\setcounter{section}{0}
\setcounter{figure}{0}
\setcounter{table}{0}

\tableofcontents

\vspace{2cm}

\twocolumngrid
\clearpage
{\section{Crystal Growth and Characterization}}
{\subsection{Single crystal Growth}}

\tableofcontents

{\section{Crystal Growth and Characterization}}
{\subsection{Single crystal Growth}}

\begin{figure}
	\centering\includegraphics[width=1\columnwidth]{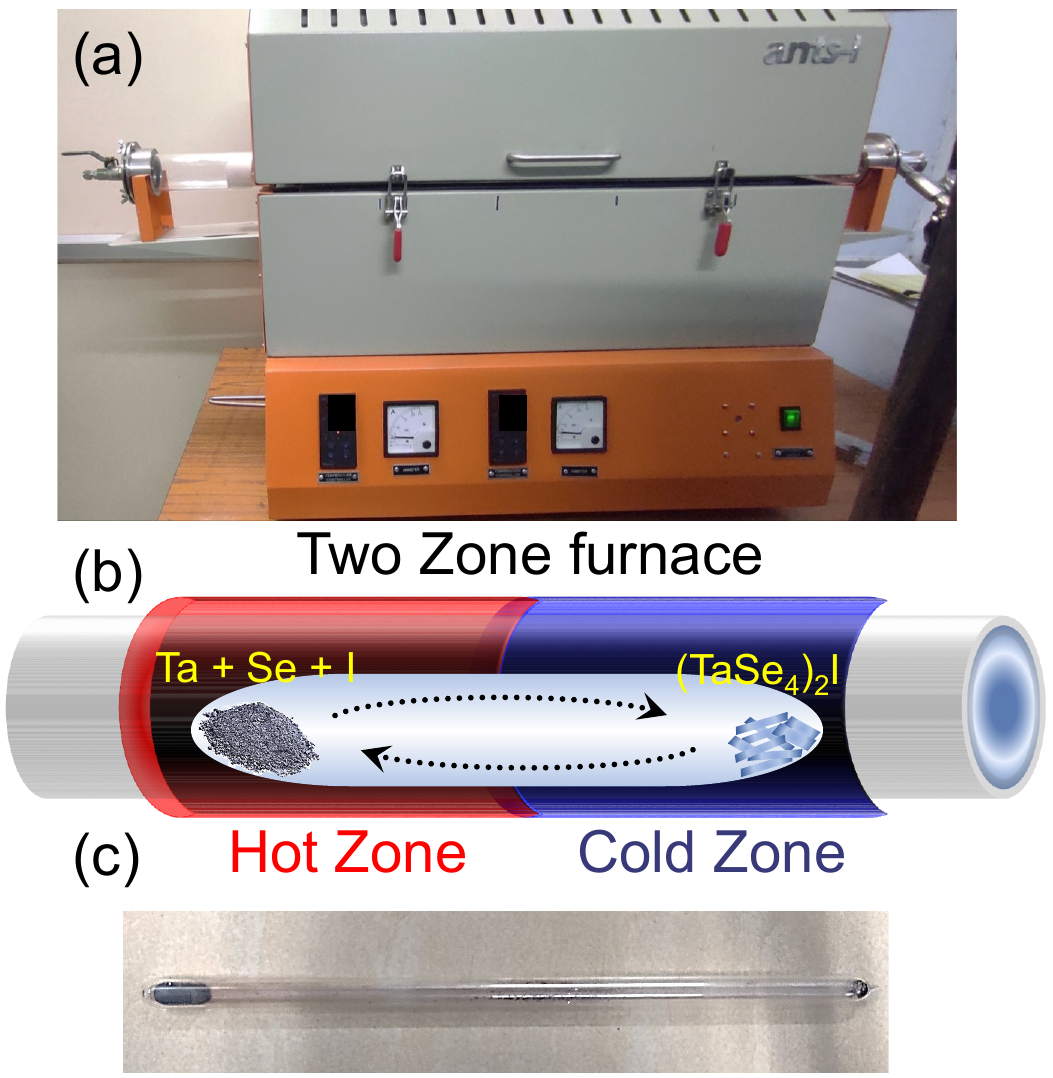} \caption{\textbf{Single crystal growth.} (a) Two-zone furnace used for the single crystal growth. (b) Schematic of the source and sink zone. (c) Quartz tube after crystal formation.}
	\label{Fig:Growth}	
\end{figure}

For the present study, high-quality single crystals of \ch{(TaSe4)2I} have been grown by the chemical vapor transport (CVT) method using I$_2$ as a transport agent.  The growth procedure closely follows the method described in previously reported work \cite{gressier1984characterization.S}. Stoichiometric mixtures of [Ta (99.99$\%$ pure), Se (99.99$\%$ pure)] powders and I$_2$ granules in a molar ratio Ta:Se:I = 2:8:1 has been grounded in a mortar pestle. The grounded powder along with the transport agent I$_2$ of quantity 2 mg/cm$^3$, is vacuum-sealed in a quartz tube of length 20 cm having a diameter 12/15 mm. The sealed quartz tube was then placed inside a two-zone furnace, where the source and growth zone temperatures were maintained at 500 and 400 $\degree$C, respectively. The schematic of the growth process is shown in Fig [\ref{Fig:Growth}(a)-(c)]. The temperatures of the source and growth zones have been kept constant with minimal temperature fluctuations ($\pm 1^{\circ} C$), maintaining a constant temperature gradient for seven days to complete the growth process and finally cooled naturally. Thereafter, the quartz tube is broken under ambient conditions. Good quality needle-like single crystal has been obtained in the cold zone with lengths up to 3$-$4 mm.\\

\subsection{Structural Characterization}

The crystal structure of the material is determined from single-crystal x-ray diffraction (XRD) using a Bruker D8 VENTURE Microfocus diffractometer which is equipped with a PHOTON II Detector. The diffraction is performed on a Parabar oil-coated sample mounted on the instrument. Mo K$_\alpha$ radiation with a wavelength of 0.71073 \AA\ was used, and data acquisition and control are managed by the APEX III (v201730) software package. The structure determination involved direct methods followed by full-matrix least-squares refinement on F2 using the SHELXL\cite{Sheldrick2008.S} program suite and the OLEX2\cite{Dolomanov2009.S} interface.

The single crystal XRD studies that have been conducted at room temperature reveal that the material crystallizes in a body-centered tetragonal structure (space group  \textit{I422}, no 97) having an empirical formula of \ch{(TaSe4)2I}. The unit cell of \ch{(TaSe4)2I} consists of parallel chains of \ch{TaSe4} units with a Tantalum(Ta) metal chain along the c-axis where\ch{Se4} of \ch{TaSe4} forms two dimers aligned in a plane forming a rectangle. Each Ta atom is sandwiched at the center between two adjacent \ch{Se4} planes. Consecutive rectangular planes of \ch{Se4} are skew rotated by $\sim$ 45$\degree $.  These well-separated \ch{TaSe4} chains provide a quasi-1D structure to the system. Each unit cell contains two units of \ch{MX4} chains. Here I$^-$ plays the role of a spacer. The  \ch{(TaSe4)2I} compound possesses a tetragonal crystal structure with an \textit{I}422 space group that is chiral and lacks an inversion center. Infinite parallel skew rotated \ch{TaSe4} chains produce quasi-1D structure and substantial anisotropy in the electronic structure. 

\subsection{Morphological Characterization}

\begin{figure}
	\centering
	\includegraphics[width=1\columnwidth]{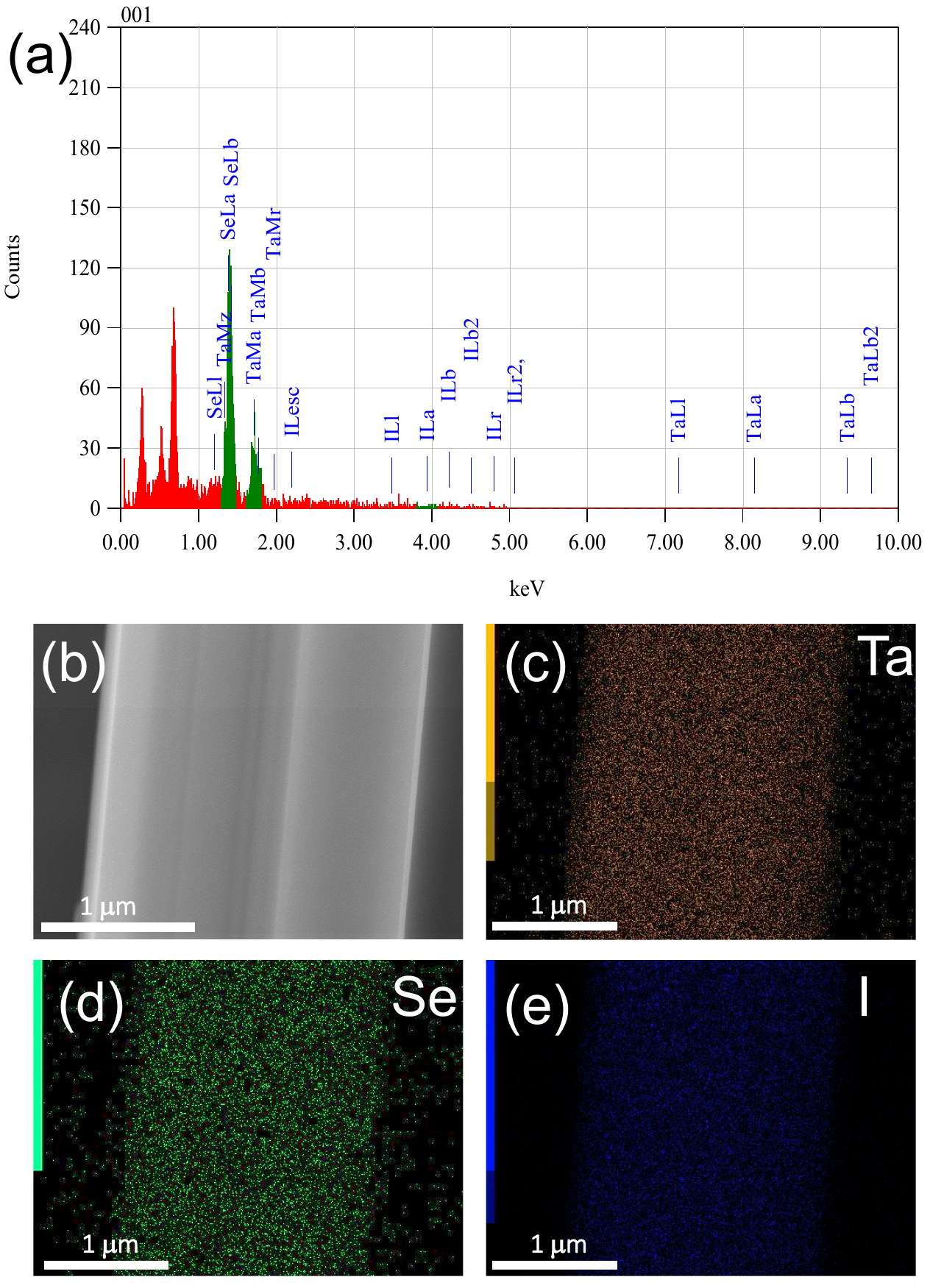}   \caption{\textbf{Morphology and composition.} (a) EDS spectra of a \ch{(TaSe4)2I} single crystal. (b) Image of the selected region for elemental mapping. (c-e) The elemental mapping of the corresponding elements (Tantalum)Ta, (Selenium)Se, and (Iodine)I respectively.}
	\label{Fig:SEM}	
\end{figure}

The surface morphology of the \ch{(TaSe4)2I} single crystal is studied using a Field Emission Scanning Electron Microscope(FESEM; JEOL; MODEL JSM1T300HR) which is equipped with energy-dispersive X-Ray spectroscopy (EDS). In the EDS spectra as shown in Fig. [\ref{Fig:SEM}(a)] we observe elements like Tantalum (Ta) and Selenium(Se) in the higher count as Iodine (I) is observed in small quantities as the elemental percentage (9\%) is quite small.  A magnified image of \ch{(TaSe4)2I} wire-like single crystal image with a scan area of 2.5 × 1.5 $\mu m$ is shown in Fig. [\ref{Fig:SEM} (b)] and it is conspicuous that the surface of the \ch{(TaSe4)2I} single crystal is quite uniform. The corresponding elemental mapping of the selected surface area is shown in Fig. [\ref{Fig:SEM}(c)-(e)]. It is to be noted that dominant counts in the elemental mapping can be attributed to the dominance of the elemental percentage in the crystal.

\begin{table*}
\caption{\textbf{EDS characterization of the CVT grown \ch{(TaSe4)2I} single crystals}}
\begin{tabular}{||p{4.0cm}|p{2.5cm}|p{2.5cm}|p{2.5cm}|p{5.0cm}||}
 \hline
 \hline
\textbf{Sample} & \multicolumn{4}{|c|}{\textbf{\ch{(TaSe4)2I}}}|\\
\hline
\hline
\multirow{ 3}{*}{\textbf{\ch{(TaSe4)2I}}} & & &  & \\
 & \textbf{Ta \%}& \textbf{Se \%} & \textbf{I \%} & \textbf{Formula} (Normalized to Ta)\\
  & & &  & \\
 \hline
& & &  & \\
\textbf{Theoretical}  & \textbf{18.2}& \textbf{72.7} & \textbf{9.1} & \textbf{Ta$_2$Se$_8$I}\\
  & & &  & \\
\hline

\multirow{ 3}{*}{} & & &  & \\

 \textbf{Experimental \hspace{1.5cm}(Averaged of 10 scans)} & \textbf{19.7}& \textbf{72.4} & \textbf{7.9} & \textbf{Ta$_2$Se$_{7.4}$I$_{0.8}$}\\
  & & &  & \\
\hline
\end{tabular}
\label{Table:EDS}
\end{table*}
\bigskip

{\subsection{TEM Characterization}}
The crystalline quality of the \ch{(TaSe4)2I} is observed under Ultra High-Resolution Field Emission Gun Transmission Electron Microscope (UHR-FEG-TEM)(HRTEM model: Tecnai G2 F30 STWIN) equipped with field emission gun and with an electron accelerating voltage of 300 kV). 
\begin{figure}
\centering
\includegraphics[width=1\columnwidth]{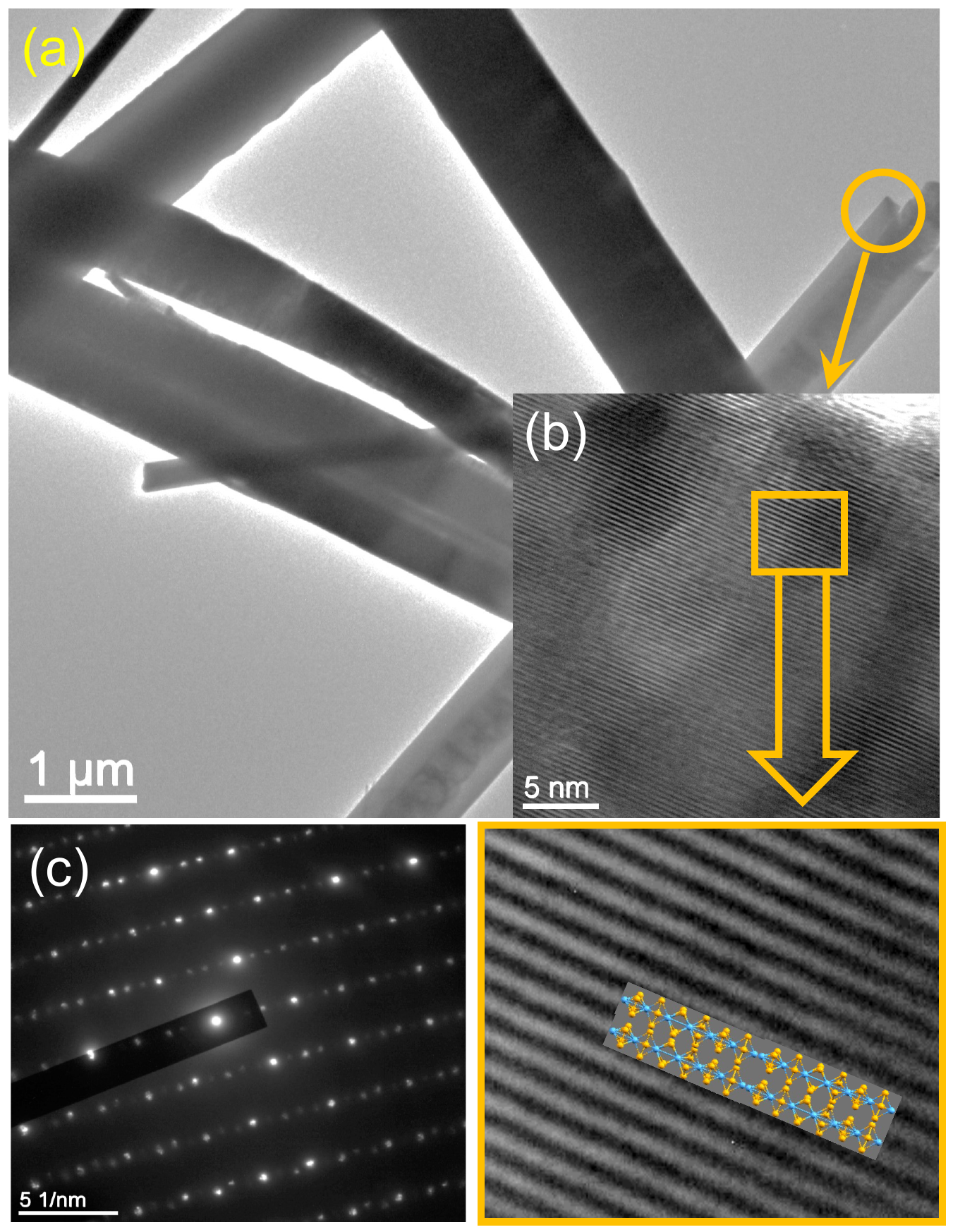}   
\caption{\textbf{TEM micrographs.} (a) Magnified TEM image of \ch{(TaSe4)2I} single crystals. (b) A magnified HRTEM image shows a parallel chain structure of  (TaSe$_4$)$_2$I single crystals. (c) Corresponding Selected Area Electron Diffraction (SAED) of  (TaSe$_4$)$_2$I crystal.  }
\label{Fig:TEM}	
\end{figure}
Wire-like morphology of the \ch{(TaSe4)2I} single crystals is observed under TEM microscopy which is presented in Fig. [\ref{Fig:TEM}(a)]. HRTEM image of  (TaSe$_4$)$_2$I crystals indicates the presence of infinite parallel chains of \ch{(TaSe4)} as shown in Fig. [\ref{Fig:TEM}(b)]. The corresponding Selected Area Electron Diffraction (SAED) pattern is shown in Fig. [\ref{Fig:TEM}(c)]. The SAED pattern indicates a single phase and good crystallinity of \ch{(TaSe4)2I.}

{\section{Resistance noise spectroscopy}}
To understand the underlying fluctuation dynamics of CDW condensation, lock-in-based phase-sensitive detection has been implemented for the low-frequency $1/f$ resistance-noise spectroscopy (LFRNS). The noise spectroscopy allows us to measure both background and sample noise contribution to the total noise \cite{ghosh2004set.S}. This technique has already been employed to study various types of phase transition\cite{kogan_noise.S,AKRPhysRevLett.91.216603.S,Koushik_R.2013,chadniPhysRevLett.102.025701.S,satyakiPhysRevLett.124.095703.S,sudiptaPhysRevB.104.155101.S}.   

\begin{figure}[h]
	\centering
	\includegraphics[width=1\columnwidth]{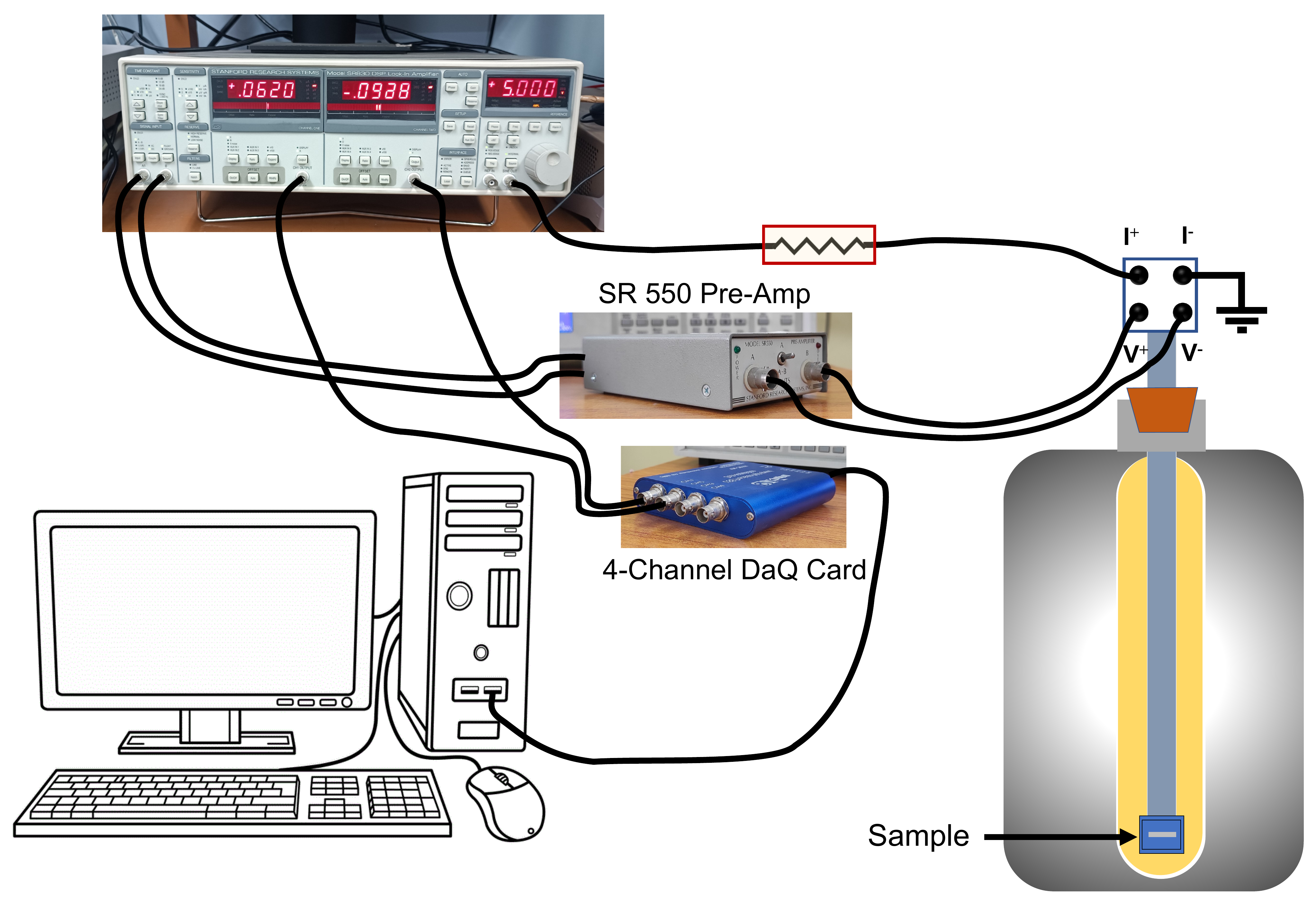}   	
	\caption{\textbf{Low-frequency $1/f$ noise spectroscopy setup.} Schematic representation of lock-in-based LFRNS setup. }
	\label{Fig:Schematic}	
\end{figure}

{\subsection{Low frequency $1/f$ noise: Measurement protocol}}
We have used lock-in-based detection to carry out the noise characteristics for this study. Resistance noise measurement is carried out in standard four-probe geometry under a constant ac current. In order to bias the sample with constant ac current a lock$-$in amplifier(LIA model SRS$-$830) is used with a voltage-to-current converter resistance of 1 M$\Omega$.  The voltage drop across inner probes of the sample is amplified using a low-noise ac coupled pre$-$amplifier(model SR550) in differential mode and subsequently, outputs of the pre$-$amplifier are further detected by LIA. The real-time resistance fluctuations(from measured voltage fluctuations) at each temperature are digitized for 80 minutes at a sampling rate of 1024 Hz (data points per second) using a 16-bit DaQ card (Analogue to digital converter). Each time series data acquisition is made at a constant temperature.\\
To rule out the temperature fluctuations on the real$\-$time resistance fluctuations (TSRF), temperature stability was enhanced to less than $\pm$ 2 mK.
The power spectrum ($S_R$) of the TSRF data is computed using Eq (\ref{Eq:SRsupp}) over a bandwidth of 1 mHz -16 Hz from the filtered time series using the Welch periodogram method. Estimated normalized power spectrum($S_R/R^2$) from the TSRF after background elimination shows typical $\sim$ $1/f^{\alpha}$ nature where $\alpha\sim$ 1 away from transitions.    

\begin{equation}
S_R(f)=\lim_{T \to \infty}\Big (\frac{1}{T}\Big )\Bigg |\int_{-T/2}^{T/2} R(t)e^{-i2\pi ft}dt\Bigg |^2
\label{Eq:SRsupp}
\end{equation}

To avoid mechanical noise in the experiments the noise measurement is carried out in a liquid nitrogen-cooled bath cryostat having a temperature range of 80 - 350 K. Copper wire contacts in four-probe geometry by using the silver epoxy is used for the measurement.  
The schematic of the noise measurement setup is shown in Fig [\ref{Fig:Schematic}]. The measurement is carried out with a constant ac-current of 5~$\mu$A.
\begin{figure}
	\centering	\includegraphics[width=0.55\columnwidth]{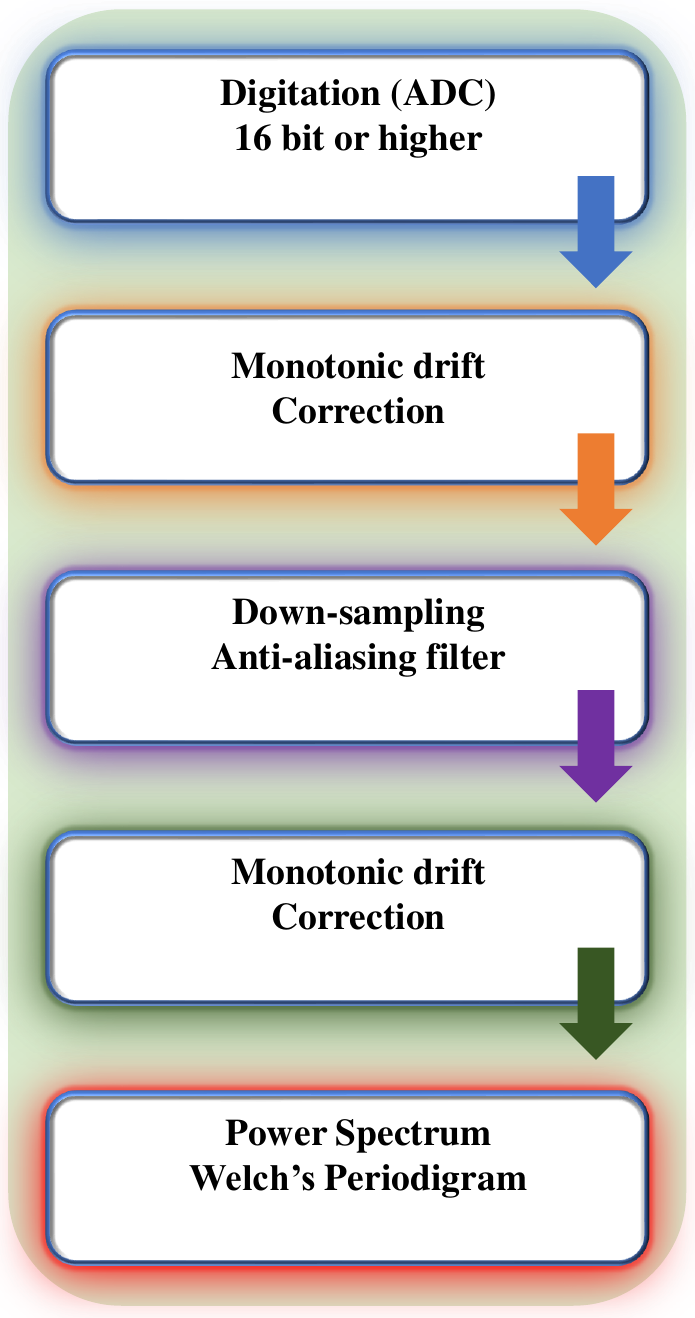}   \caption{\textbf{Noise measurement flowchart. } A systematic flow chart depicting the steps involved in the digital signal processing till the final power spectral Power spectrum $\frac{S_R}{R^2}$.}
	\label{Fig:Flow-chart}	
\end{figure}

The details of the measurement protocol and data acquisitions and corresponding involvement of the digital signal processing (DSP) techniques are summarized as a flow chart as shown in Fig. [\ref{Fig:Flow-chart}]. Each time series data acquisition was made at a constant temperature. 

{\section{Time vs Temperature}}
The \ch{(TaSe4)2I} undergoes from a semi-metallic to an insulating state below the Peierls' transition. A small drift of the temperature from the desired set point can dilute the electronic state and may corrupt the signal. Therefore proper temperature stability is quite essential throughout the duration of data acquisition. To rule out the effect of the temperature fluctuations we set the temperature to desired set point and make sure that the temperature stability of the heat bath is less than $\pm$~2mK (especially for the acquisition over the insulating region). In addition, a sample that is immediately ramped from a different temperature however the temperature may have reached to stability regime has a high probability of unstable electronic superstructure. Therefore an additional delay of 15 minutes is added at the desired set point, before the data acquisition.\\

\begin{figure}
	\centering
\includegraphics[width=1.0\columnwidth]{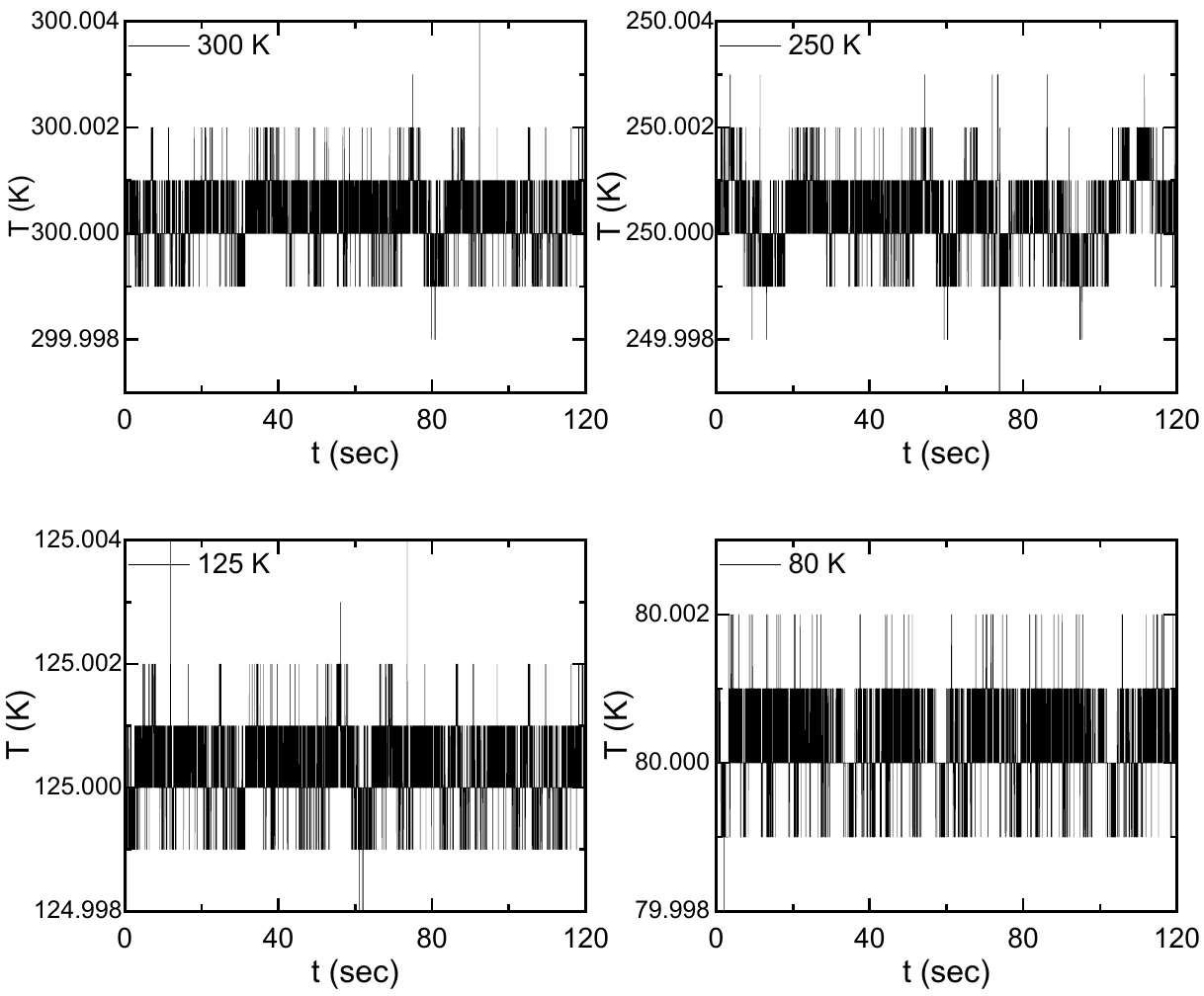}
\caption{\textbf{Temperature stability data. } Temperature stability with time during the measurement to ensure minimal contribution from temperature fluctuations.}
	\label{Fig:Temp}	
\end{figure}

{\section{Autocorrelation: Slow dynamics}}
For a time series of resistance fluctuations $\delta$R(t), the autocorrelation is estimated using,

\begin{equation}
C(\varsigma)=\lim_{T \to \infty}\frac{1}{T} \int_{-T/2}^{T/2} \delta R(t) \delta R(t+\varsigma)dt
\label{Eq:Autocorrelationsupp}
\end{equation}

Here, $\varsigma$ is the lag between two points. By analyzing the autocorrelation function, we can extract the characteristic decay time of autocorrelation, often referred to as the relaxation time ($\tau$). This relaxation time serves as a quantitative measure of dynamics in the system. The autocorrelation of the resistance fluctuations is decaying in nature. The critical slowing down is captured in a system via a diverging increase in relaxation time ($\tau$).\\

\begin{table*}
\begin{center}
\caption{Dynamics and phase transitions in CDW systems: Relaxation time scales and order of CDW transitions.}
\label{table:comparisonCDW}
\begin{tabular}{|p{2cm}|p{2.5cm}|p{2.5cm}|p{2.5cm}|p{6cm}|}
\hline
\hspace{0.4cm}System&Experimental technique & Time scale of relaxation (Transient dynamics) & Order of the transition& \hspace{1.5cm}References \\
\hline
\hline
Rb$_{0.3}$\ch{MoO3} & trARPES & 60 fs & \hspace{0.5cm} --- &  Phys. Rev. Lett. \textbf{125}, 266402 \cite{RbMoO3.PhysRevLett.125.266402.S} \\
\hline
K$_{0.3}$\ch{MoO3} & trARPES & 300 fs & Second-order &  Phys. Rev. Lett. \textbf{102}, 066404 \cite{KMoO3.PhysRevLett.102.066404.S} \\
\hline
1\textit{T}-\ch{TaS2} & trARPES & 20 fs & First-order &  Phys. Rev. Lett. \textbf{97}, 067402 \cite{TaS2.PhysRevLett.97.067402.S} \\
\hline 
\ch{LaTe3} & transient optical spectroscopy & --- & \hspace{0.5cm} --- &  Phys. Rev. Lett. \textbf{123}, 097601 \cite{N.Gedik.PhysRevLett.123.097601} \\
\hline
\ch{TbTe3} & trARPES & 20 fs & First-order &  Science, \textbf{321}(5896),1649-1652 \cite{schmitt.science.1160778} \\
\hline
2\textit{H}-\ch{NbSe2} & optical pump-probe & 100 ps & Second-order & Phys. Rev. B \textbf{102}, 205139 \cite{NbSe2.PhysRevB.102.205139.S} \\
\hline
\end{tabular}
\end{center}
\end{table*}

Several studies in pre-existing literature delved into the dynamical responses exhibited by collective CDW modes within highly non-equilibrium settings, particularly through pump-probe experiments in various systems. A concise compilation of these studies on CDW systems is presented in Table S \ref{table:comparisonCDW}. Notably, the signatures of criticality have been identified in some of these works, with the typical time scale of the dynamics ranging from femtoseconds to picoseconds. Our work stands out from the aforementioned reports by unveiling the critical singularity through the direct measurement of order parameter fluctuations in equilibrium.   Most importantly, the time scale of fluctuations at the critical singularity persists to even seconds dominating the low-frequency noise, consistent with the theoretical anticipation of a diverging nature.

\begin{figure}
	\centering
    \includegraphics[width=1\columnwidth]{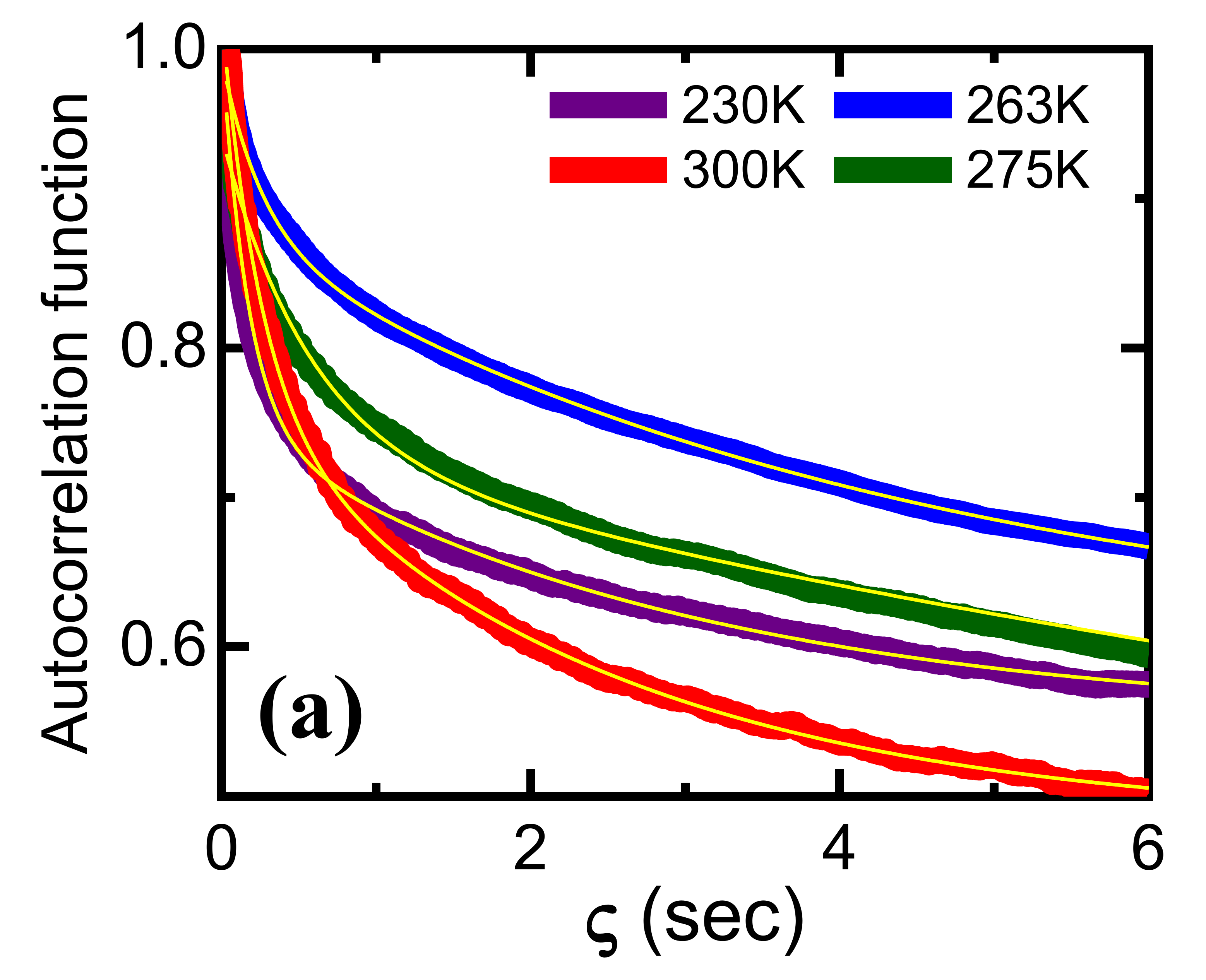}  
    \caption{\textbf{Autocorrelation: Slow dynamics.} autocorrelation functions vs correlation time~($\varsigma$) with fitting(yellow) at a few representable temperatures including the T$_{CDW}$ temperature.}
	\label{Fig:AutocorrelationFit}	
\end{figure}

\bigskip

\section{Resistance noise: Order parameter fluctuations}
Resistance fluctuations in a system commonly arise from two distinct sources, namely,  temporal fluctuations in (a) the carrier density $\langle\delta n\rangle$ and (b) the mobility $\langle\delta \mu\rangle$ \cite{333808,Hooge1981,Weissman1988Noise.S}.
\begin{equation}
\delta R= \frac{\partial R}{\partial n}  \delta n +\frac{\partial R}{\partial \mu}  \delta \mu 
\end{equation}
Therefore, the mean square resistance fluctuations $\frac{\langle \delta R^{2}\rangle}{\langle R ^2\rangle}$ can be written as  
\begin{align}
\frac{\langle \delta R^{2}\rangle}{\langle R ^2\rangle}
  & = \frac{\langle \delta n^{2}\rangle}{\langle n^2\rangle} + \frac{\langle \delta \mu  ^2\rangle}{\langle \mu^2 \rangle }+2\frac{\langle \delta n\rangle \langle \delta \mu\rangle}{\langle n\rangle \langle \mu\rangle}\\
  & \simeq \frac{\langle \delta n^{2}\rangle}{\langle n^2 \rangle} + \frac{\langle \delta \mu^{2}\rangle}{\langle \mu^2 \rangle} \\
\end{align}
Here we have used the fact that $\langle\delta n\rangle$ and  $\langle\delta \mu\rangle$ are uncorrelated to drop the cross term. Furthermore, the temporal fluctuations in mobility are not relevant on account of the very different time scales. We note that the mobility fluctuations are not critical fluctuations; the mobility is controlled by the collisions and is therefore a result of the physics on the relaxation time scale (picoseconds to nanoseconds). Fluctuations on this nanosecond scale would average out to be unimportant for the millisecond to seconds time scale noise spectroscopy. Therefore on account of the criticality and different time-scales, $\frac{\langle \delta R^{2}\rangle}{\langle R ^2\rangle}$ can be written as (\ref{Eq:nfluctuation1}) 

\begin{equation}
 \frac{\langle \delta R ^2\rangle}{\langle R^2 \rangle} \simeq \frac{\langle \delta n^{2}\rangle}{\langle n^2 \rangle } 
 \label{Eq:nfluctuation1}
\end{equation}
Usually, thermodynamic phase transitions are characterized by strong enhancement in the 1/f noise level of such fluctuations. However, it is quite important to look for the origin of such excess noise close to the transition. Moreover, it is essential to relate the mean square fluctuations to thermodynamic variables via the fluctuation-dissipation theorem.

For CDW ground state, the dynamics of the fluctuations can be described using a two-fluid model of the conserved sum of n$_{normal}$ and n$_{CDW}$ as \cite{gruner2018density.S},

\begin{equation}
n = n_{normal} +n_{CDW} 
\end{equation}

Any fluctuation in n$_{CDW}$ is expected to be locally equilibrated with a well-defined local momentum and keep the total carrier density (conserved). At a low bias field, the condensate fraction $n_{CDW}$ is pinned and therefore insulating. Additionally, the fluctuation in mobility is also negligible. Thus $\delta R(t) \simeq \frac{\partial R}{\partial n_{normal}}  \delta n_{normal}(t)$. The conservation in density ($n$) yields,  $\delta n_{CDW}(t) \simeq - \delta n_{normal}(t)$.  Thus $\delta R(t)  \simeq - \frac{\partial R}{\partial n_{normal}}  \delta n_{CDW} (t)$. Since $\delta n_{CDW} = \delta (Re[\Psi])$. Therefore, critical fluctuations in the order parameter directly translate into the experimentally measured resistance fluctuations.\\

The variance of resistance fluctuations, 
\begin{align}
\frac{\langle \delta R^{2}\rangle}{\langle R ^2\rangle}
   & \simeq \frac{\langle \delta n^{2}\rangle}{\langle n^2\rangle} + \frac{\langle \delta \mu  ^2\rangle}{\langle \mu^2 \rangle }\\
   & \propto \langle \delta n^{2}\rangle\\
   & \propto \langle \delta n_{CDW} ^2\rangle\\
   & \propto \langle (\delta Re[\Psi (\Vec{r},t)])^2 \rangle \\
   & \propto \langle \delta \Delta(\Vec{r},t)^{2} \rangle \simeq \int d^3 r\, G(r) \\
   & =k_BT\chi_T.
\end{align}

Here, $\Psi(\Vec{r},t)~=~\Delta (\Vec{r},t)~e^{i \phi (\Vec{r},t)}$ \cite{gruner2018density.S} is the two-component order parameter of the CDW state, where $\Delta (\Vec{r},t)$ is the amplitude of the spatio-temporal modulated charge density and $\phi (\Vec{r},t)$ is the phase of the same. G(r) is the two-point correlation function of the order parameter.  Therefore, the normalized variance of the resistance fluctuations can be easily reduced to the order parameter fluctuations and the ($\frac{S_R}{\langle R^2\rangle} $) mimics the mean square order parameter fluctuations.

\begin{figure}
	\centering
	\includegraphics[width=1\columnwidth]{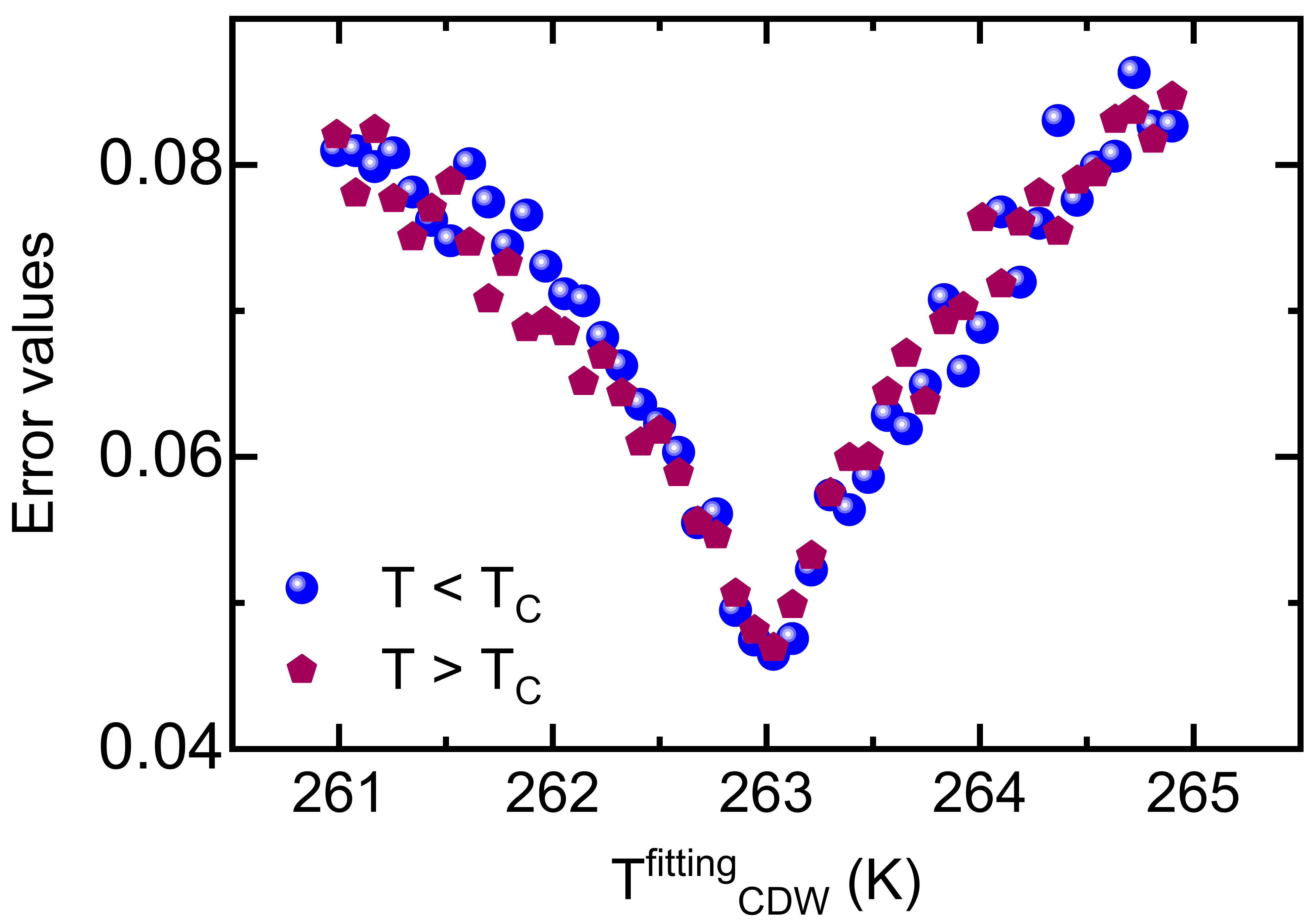}   	
	\caption{\textbf{Determination of critical point:} Error value corresponds to the errors of the linear fit of log($\frac{<\delta R^2>}{<R^2>}$) vs log ($\epsilon$) data by taking the variation in critical temperature. Each T$_{CDW}$ has two sets of data on either side of it. Corresponding errors of data sets for T < T$_{CDW}$ (blue) and T> T$_{CDW}$ (wine). The critical point of \ch{(TaSe4)2I} is identified to be T$_{CDW} \sim 263.033 K$ from the minimum error.}
	\label{Fig:Error}	
\end{figure}

{\section{Critical point determination}}
The determination of critical exponents using power-law formalism requires an exact identification of the critical point. Very careful measurements conducted continuously over a period exceeding two and half months, with closely spaced data points, ensure precise characterization of the power-law behavior in the vicinity of the critical point. The exact critical point is estimated through error minimization of the fitting on both sides of T$_{CDW}$ based on a linear fit on the log-log scale for power law divergence. Figure \ref{Fig:Error} illustrates the error as a function of $T_{CDW}^{fitting}$. The critical point of \ch{(TaSe4)2I} is identified to be T$_{CDW} \sim 263.03 K$ from the minimum error as seen in Fig. \ref{Fig:Error}.

\vspace{0.4cm}

{\section{Mean-field theory and beyond}}

\begin{figure}
	\centering
	\includegraphics[width=0.85\columnwidth]{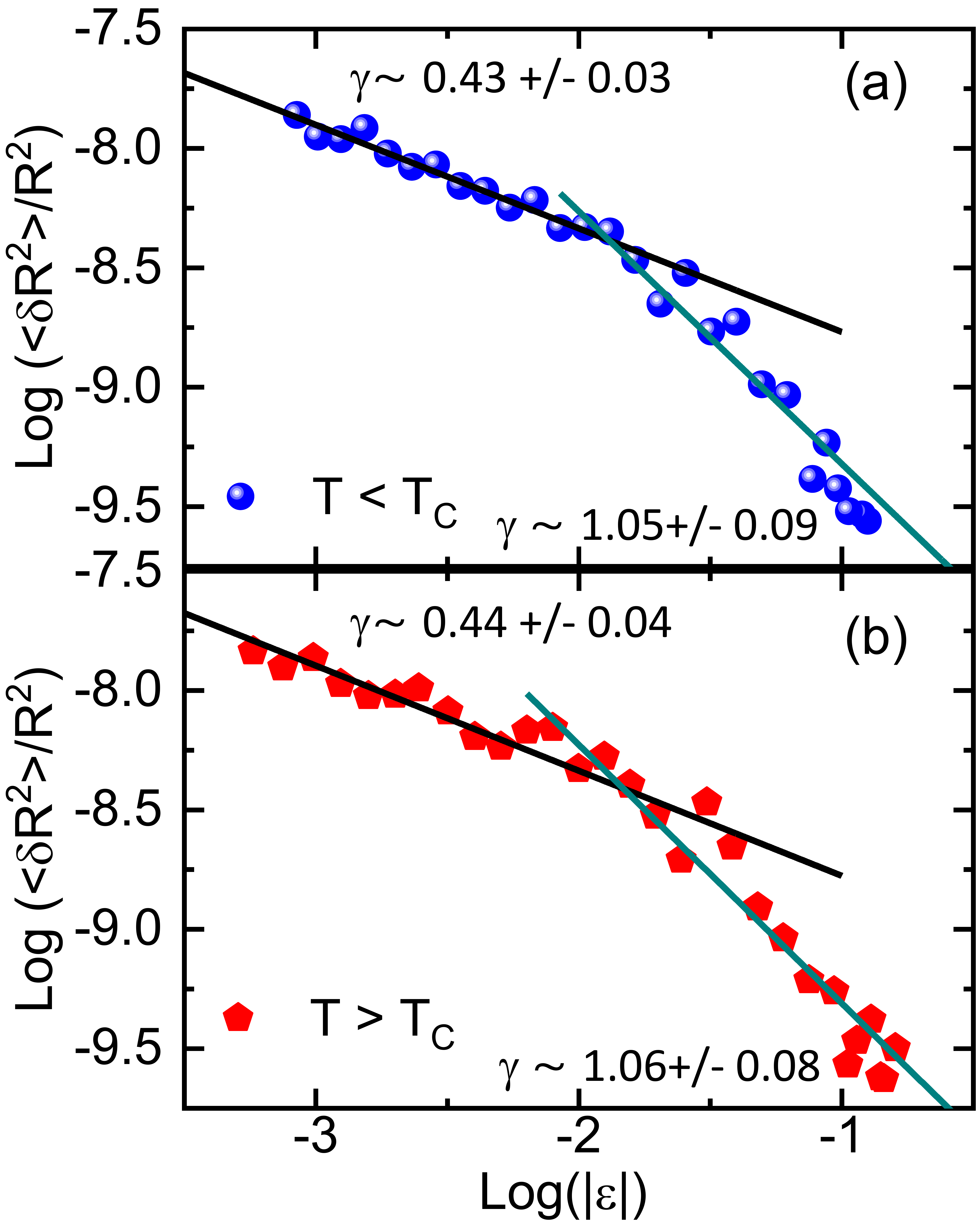}   	
	\caption{\textbf{Power-law scaling analysis of variance}. The variance of thermodynamic fluctuation data is plotted against the reduced temperature ($\epsilon$) on a logarithmic scale. The normalized variance $\frac{<\delta R^2>}{<R^2>}$ is proportional to the isothermal susceptibility which exhibits $\chi{_{_T}} \sim \epsilon^{-\gamma}$. Solid lines represent the linear fits to the data for extracting the exponent. It's noteworthy that mean-field theory predicts $\gamma=1$ \cite{gruner2018density.S}. A distinct crossover to different exponent values is observed upon entering the critical fluctuation-dominated regime. All the logarithmic scales are on base 10.}
	\label{Fig:Variance}	
\end{figure}

The plot of the variance vs reduced temperature ($\epsilon$) on a log-log scale is shown in Fig\ref{Fig:Variance} ( Fig. 3 (a)  in the main text). Interestingly, we observe two distinct regions within the critical fluctuation window defined by the Ginzburg criterion ($\Delta T_{GL} \sim 52K$). Two regions within $\Delta T_{GL}$ are identified with different coloured shades one is the mean-field region and the second is the fluctuation-dominated region. The region slightly away from the $T_{CDW}$ on either side exhibits a clear mean-filed like behaviour with the experimentally determined slopes of $\gamma \sim 1.05 \pm 0.09$ (for T < T$_C$) and $\gamma \sim 1.08 \pm 0.08$ (for T > T$_C$). We identify these two regions as mean-filed regions on either side of T$_C$ and mean-filed theory accurately describes the ground state.

\begin{figure}
	\centering
	\includegraphics[width=0.85\columnwidth]{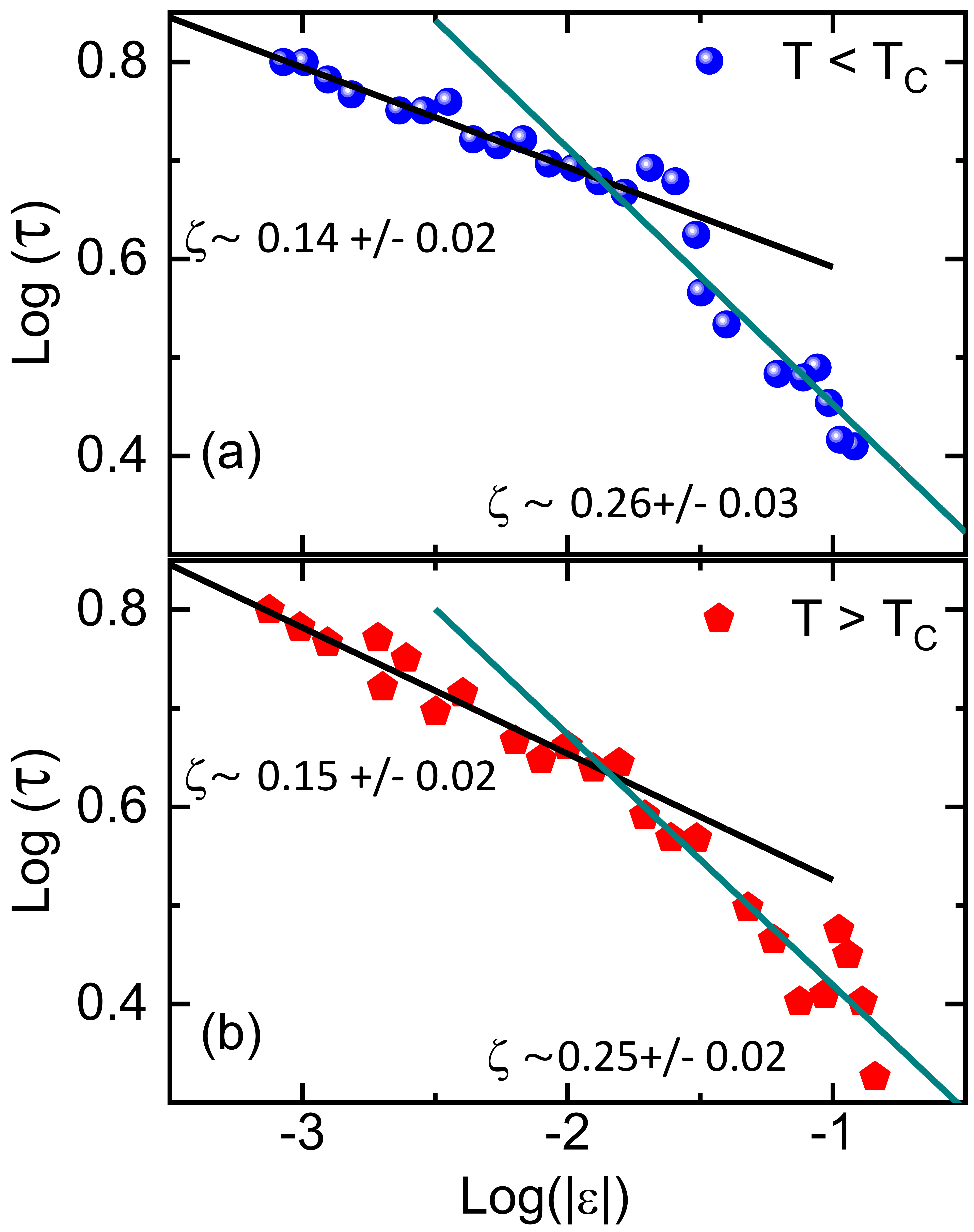}\caption{\textbf{Power-law scaling analysis of relaxation time.} The relaxation times, $\tau$, are plotted against the reduced temperature ($\epsilon$) on a logarithmic scale. $\tau$ can be related to the characteristic time scale of propagation of the amplitude mode as $\tau \sim 2 \pi / \omega_{\Delta}$, which exhibits $\omega_{\Delta} \sim \epsilon^{-\gamma}$ \cite{gruner2018density.S}. Solid lines represent linear fits to the data for extracting the exponent. A distinct crossover to different exponent values is observed upon entering the critical fluctuation-dominated regime. All the logarithmic scales are on base 10.}
	\label{Fig:Tau}	
\end{figure}

A similar analysis for $\tau$ is carried out followed by accurate determination of the $T_{CDW}$. Interestingly, critical points identified from both variance and $\tau$ coincide. The plot of the $\tau$ vs reduced temperature ($\epsilon$) on a log-log scale is shown in Fig. \ref{Fig:Tau} (Fig. 3 (b). in the main text). In terms of the power-law divergence, $\tau$ also follows a similar trend of two-slope behaviour within $\Delta T_{GL}$. The exponents in the mean-field region are $\zeta \sim 0.26 \pm 0.02$ (for T < T$_C$) and $\zeta \sim 0.24 \pm 0.02$ (for T > T$_C$). Notably, the mean-field to non-mean-field crossover in variance as well as in $\tau$ are identified to lie in the same temperature region. The breakdown of the mean-field theory near the transition is also reflected in $\tau$ over a temperature window of $\pm$ 5~K same as the variance. The narrow temperature window near T$_{CDW}$ where the mean-field breaks down is captured via carefully measured noise spectra. 

\vspace{0.6cm}

{\section{Non-Gaussian Fluctuations}}
{\subsection{Second spectrum: Measure of non-Gaussian fluctuations}
Fluctuations of most observables typically follow a Gaussian distribution. However, certain physical systems display interesting characteristics with correlated fluctuators, resulting in non-Gaussian fluctuations.  Consequently, resistance and/or voltage noise in these systems exhibit non-Gaussian behavior\cite{AKRPhysRevLett.91.216603.S,chadniPhysRevLett.102.025701.S}. Understanding the extent of non-Gaussianity is vital for correlating noise output to the dynamics of the system.

One of the measures for non-Gaussian components (NGC) in a signal is to compute its higher-order moments like skewness, kurtosis, etc \cite{Koushik_R.2013}. But computing these quantities directly from time series can be tricky as the time series we measure has contributions from both the sample and background fluctuations as well as components of unwanted bands beyond one's range of interest. Hence, we have used the technique of second spectrum \cite{SeidlerPhysRevB.53.9753} to calculate the NGC present in our noise measurements. The second spectrum is fourth order moment of resistance fluctuations ($\delta$R(t)) given by,

\begin{equation}
    S^{(2)}_R (f_2)= \int_{0}^{\infty} \langle \delta R^{2}(t) \delta R^{2}(t+\varsigma)\rangle  \cos (2\pi f_2\varsigma)d\varsigma
\label{Eq:S2R}
\end{equation}
\begin{figure}
	\centering
    \includegraphics[width=1\columnwidth]{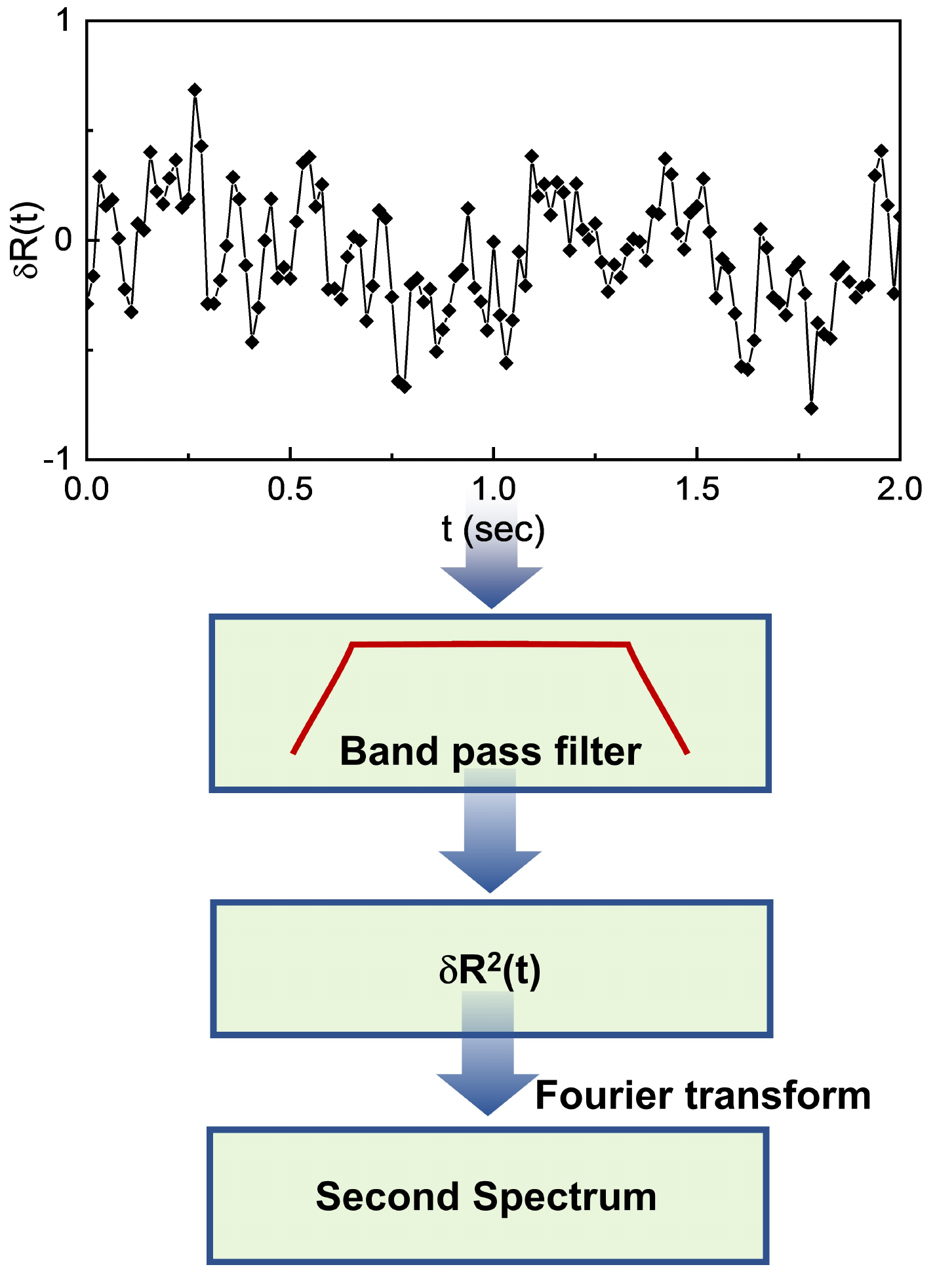}   	
	\caption{\textbf{Second spectra.} Scheme based on Ref \cite{SeidlerPhysRevB.53.9753} for calculating the second spectrum. }
	\label{Fig:S2Schematic}	
\end{figure}

{\subsection{Second spectrum}}
For the calculation of the second spectra, we have chosen the method proposed by Seidler et. al \cite{SeidlerPhysRevB.53.9753}. The scheme of which is presented in Fig.[\ref{Fig:S2Schematic}]. The decimated data is passed through a digital band pass filter over a chosen octave. The bandwidth is determined by the sensitivity of the frequency regimes over which the second spectrum is sensitive to non-Gaussian fluctuations. More often the non-Gaussianity of a signal is reflected in the low frequencies. The filtered signal is then squared point-by-point and the PSD of the resultant sequence gives the second spectrum. In effect, the second spectrum calculated using the method stated in [\ref{Fig:S2Schematic}] represents fluctuations in the instantaneous noise power.

In Eq. [\ref{Eq:S2R}], S$^{(2)}_R$(f$_2$) represents the ‘‘spectral wandering’’ or fluctuations in the noise power in a chosen frequency band (f$_L$, f$_H$), and (f$_2$) represents the spectral frequencies corresponding to the second spectrum. A common way to represent the second spectrum is to calculate the normalized second spectrum (s$^{(2)}_R$(f$_2$)) and its variance $\sigma^{(2)}$ which are defined as
\begin{equation}
    s^{(2)}_R (f_2) = \frac{S^{(2)}(f_2)}{\big|\int_{f_{L}}^{f_{H}} S_{R}(f)df\big|^2}
    \label{Eq:s2normalized}
\end{equation}
\begin{equation}
    \sigma^{(2)}= \int_{0}^{f_{H} - f_{L}} s^{(2)}_R (f_2)df_2
    \label{Eq:sigma2}
\end{equation}
where S$_R$(f) represents the PSD of resistance fluctuations and f$_L$; f$_H$ represents the lower and upper frequencies in a chosen octave respectively. It is important to choose an octave where the excess noise from the sample is considerably larger than the background to minimize any corruption due to the background which is generally Gaussian in nature. The definition of $\sigma^{(2)}$ is analogous to that of kurtosis in other words it represents the normalized variance of the second spectrum and thus represents NGC contribution in noise. In this work, we have used the method proposed by Seidler et. al \cite{SeidlerPhysRevB.53.9753} for the estimation of the second spectrum.\\

\subsection{Probability distribution function (PDF)}
An alternative technique to check for non-Gaussian noise is to plot its probability distribution function. This P(|$\delta$R|) is the probability of occurrence of a resistance fluctuation of magnitude |$\delta$R| due to the fluctuating resistance. For a Gaussian noise with zero mean the PDF (P(x)) is given by,
\begin{equation}
    P(x)=\frac{1}{\sqrt{2\pi \sigma}}e^{-x^2/2\sigma^2}
\label{Eq:PDF}
\end{equation}

where $\sigma^2$ is the variance of the distribution. Theoretically, log(PDF) vs x$^2$ is a purely straight line.  In the experimental case, any deviation from the linearity of log(P(x)) in x$^2$ on a semi-log plot is taken as evidence for non-Gaussian noise. This technique works well when the sample noise is larger than the background fluctuations in the measurement bandwidth. In our case, the sample resistance is significantly higher than the background noise and therefore reduces many technical challenges.\\

\begin{figure}
	\centering
	\includegraphics[width=1\columnwidth]{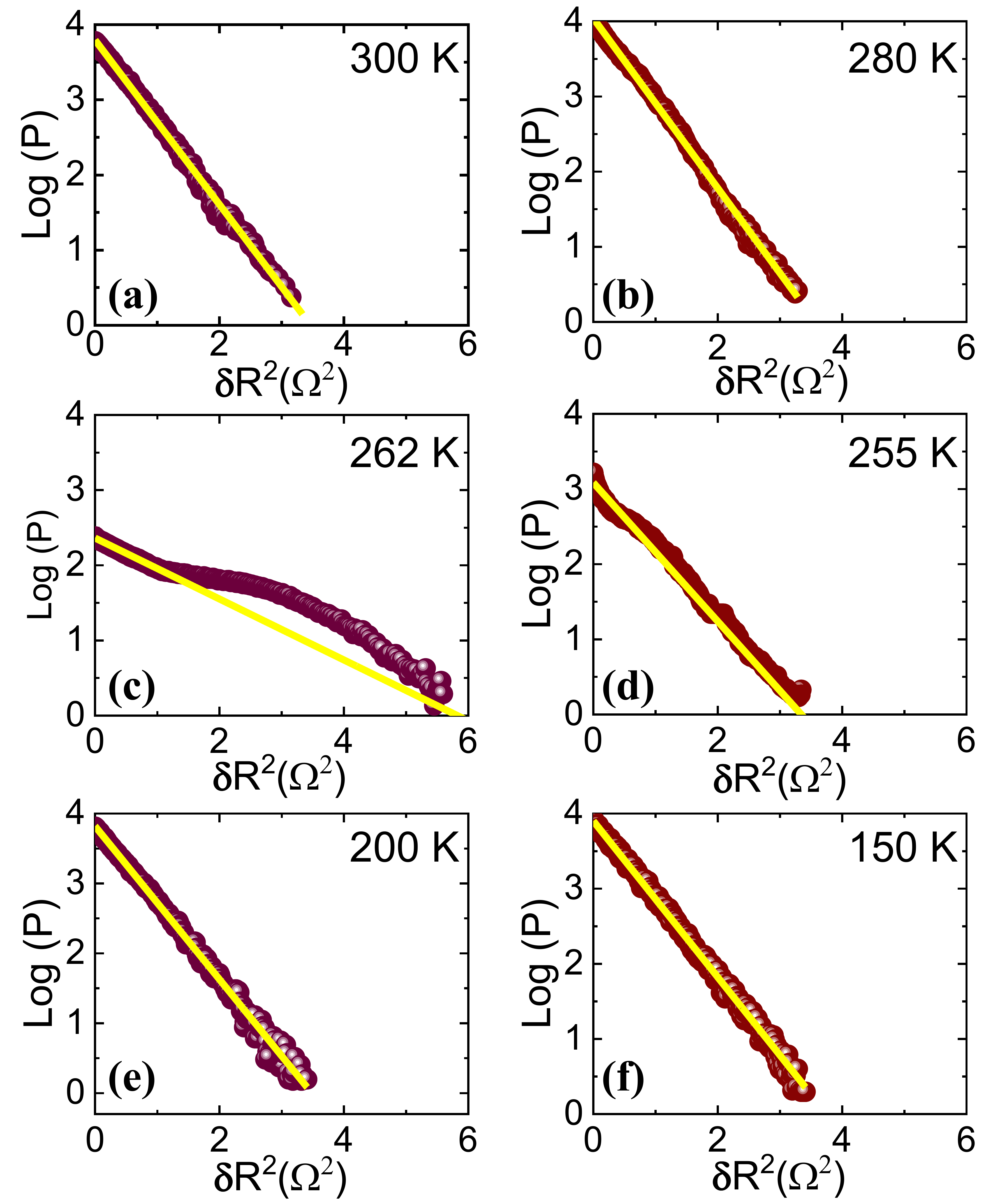}   	
	\caption{\textbf{Non-Gaussian fluctuations.} The log(PDF) vs $\delta R^2$ having strong non-Gaussian nature around $T_{CDW}$.}
	\label{Fig:PDF}	
\end{figure}
To visualize the non-Gaussian nature of the resistance fluctuations we plotted the log(PDF) vs the squared bins of the resistance fluctuations. For purely Gaussian fluctuations of the resistance Log(PDF) vs $\delta$R$^2$ is a straight line whereas the presence of non-Gaussian fluctuations makes it deviate from the linear behavior. Semi-log plot of PDF vs $\delta$R$^2$ is presented at a few temperatures including the transition temperature where the strong non-Gaussian nature of the PDF is observed. Away from the transition temperature the log(PDF) varies purely linear fashion with the squared bins as shown in Fig. [\ref{Fig:PDF}]. Therefore both $\sigma^{(2)}$ as well as PDF signify the correlated nature of the fluctuations in the vicinity of $T_{CDW}$.

\begin{thebibliography}{85}%
\makeatletter
\providecommand \@ifxundefined [1]{%
 \@ifx{#1\undefined}
}%
\providecommand \@ifnum [1]{%
 \ifnum #1\expandafter \@firstoftwo
 \else \expandafter \@secondoftwo
 \fi
}%
\providecommand \@ifx [1]{%
 \ifx #1\expandafter \@firstoftwo
 \else \expandafter \@secondoftwo
 \fi
}%
\providecommand \natexlab [1]{#1}%
\providecommand \enquote  [1]{``#1''}%
\providecommand \bibnamefont  [1]{#1}%
\providecommand \bibfnamefont [1]{#1}%
\providecommand \citenamefont [1]{#1}%
\providecommand \href@noop [0]{\@secondoftwo}%
\providecommand \href [0]{\begingroup \@sanitize@url \@href}%
\providecommand \@href[1]{\@@startlink{#1}\@@href}%
\providecommand \@@href[1]{\endgroup#1\@@endlink}%
\providecommand \@sanitize@url [0]{\catcode `\\12\catcode `\$12\catcode `\&12\catcode `\#12\catcode `\^12\catcode `\_12\catcode `\%12\relax}%
\providecommand \@@startlink[1]{}%
\providecommand \@@endlink[0]{}%
\providecommand \url  [0]{\begingroup\@sanitize@url \@url }%
\providecommand \@url [1]{\endgroup\@href {#1}{\urlprefix }}%
\providecommand \urlprefix  [0]{URL }%
\providecommand \Eprint [0]{\href }%
\providecommand \doibase [0]{https://doi.org/}%
\providecommand \selectlanguage [0]{\@gobble}%
\providecommand \bibinfo  [0]{\@secondoftwo}%
\providecommand \bibfield  [0]{\@secondoftwo}%
\providecommand \translation [1]{[#1]}%
\providecommand \BibitemOpen [0]{}%
\providecommand \bibitemStop [0]{}%
\providecommand \bibitemNoStop [0]{.\EOS\space}%
\providecommand \EOS [0]{\spacefactor3000\relax}%
\providecommand \BibitemShut  [1]{\csname bibitem#1\endcsname}%
\let\auto@bib@innerbib\@empty
\bibitem [{\citenamefont {Peierls}\ and\ \citenamefont {Peierls}(1955)}]{peierls1955quantum}%
  \BibitemOpen
  \bibfield  {author} {\bibinfo {author} {\bibfnamefont {R.}~\bibnamefont {Peierls}}\ and\ \bibinfo {author} {\bibfnamefont {R.~E.}\ \bibnamefont {Peierls}},\ }\href@noop {} {\emph {\bibinfo {title} {Quantum theory of solids}}}\ (\bibinfo  {publisher} {Oxford University Press},\ \bibinfo {year} {1955})\BibitemShut {NoStop}%
\bibitem [{\citenamefont {Gr\"uner}(1988)}]{GRUNERRevModPhys.60.1129}%
  \BibitemOpen
  \bibfield  {author} {\bibinfo {author} {\bibfnamefont {G.}~\bibnamefont {Gr\"uner}},\ }\bibfield  {title} {\bibinfo {title} {The dynamics of charge-density waves},\ }\href {https://doi.org/10.1103/RevModPhys.60.1129} {\bibfield  {journal} {\bibinfo  {journal} {Rev. Mod. Phys.}\ }\textbf {\bibinfo {volume} {60}},\ \bibinfo {pages} {1129} (\bibinfo {year} {1988})}\BibitemShut {NoStop}%
\bibitem [{\citenamefont {Monceau}(2012)}]{monceau2012electronic}%
  \BibitemOpen
  \bibfield  {author} {\bibinfo {author} {\bibfnamefont {P.}~\bibnamefont {Monceau}},\ }\bibfield  {title} {\bibinfo {title} {Electronic crystals: an experimental overview},\ }\href {https://doi.org/https://doi.org/10.1080/00018732.2012.719674} {\bibfield  {journal} {\bibinfo  {journal} {Advances in Physics}\ }\textbf {\bibinfo {volume} {61}},\ \bibinfo {pages} {325} (\bibinfo {year} {2012})}\BibitemShut {NoStop}%
\bibitem [{\citenamefont {Gruner}(2018)}]{gruner2018density}%
  \BibitemOpen
  \bibfield  {author} {\bibinfo {author} {\bibfnamefont {G.}~\bibnamefont {Gruner}},\ }\href@noop {} {\emph {\bibinfo {title} {Density waves in solids}}}\ (\bibinfo  {publisher} {CRC press},\ \bibinfo {year} {2018})\BibitemShut {NoStop}%
\bibitem [{sup()}]{supply}%
  \BibitemOpen
  \href@noop {} {\bibinfo {title} {See the supplemental material............for details of the sample growth and characterizations as well as for the details of low-frequency resistance noise spectroscopy measurements, which includes {R}efs. \cite{Sheldrick2008,Dolomanov2009,RbMoO3.PhysRevLett.125.266402,KMoO3.PhysRevLett.102.066404,TaS2.PhysRevLett.97.067402,NbSe2.PhysRevB.102.205139,hooge1981experimental}}}\BibitemShut {NoStop}%
\bibitem [{\citenamefont {Yeom}\ \emph {et~al.}(1999)\citenamefont {Yeom}, \citenamefont {Takeda}, \citenamefont {Rotenberg}, \citenamefont {Matsuda}, \citenamefont {Horikoshi}, \citenamefont {Schaefer}, \citenamefont {Lee}, \citenamefont {Kevan}, \citenamefont {Ohta}, \citenamefont {Nagao},\ and\ \citenamefont {Hasegawa}}]{FermisurfaceInstability}%
  \BibitemOpen
  \bibfield  {author} {\bibinfo {author} {\bibfnamefont {H.~W.}\ \bibnamefont {Yeom}}, \bibinfo {author} {\bibfnamefont {S.}~\bibnamefont {Takeda}}, \bibinfo {author} {\bibfnamefont {E.}~\bibnamefont {Rotenberg}}, \bibinfo {author} {\bibfnamefont {I.}~\bibnamefont {Matsuda}}, \bibinfo {author} {\bibfnamefont {K.}~\bibnamefont {Horikoshi}}, \bibinfo {author} {\bibfnamefont {J.}~\bibnamefont {Schaefer}}, \bibinfo {author} {\bibfnamefont {C.~M.}\ \bibnamefont {Lee}}, \bibinfo {author} {\bibfnamefont {S.~D.}\ \bibnamefont {Kevan}}, \bibinfo {author} {\bibfnamefont {T.}~\bibnamefont {Ohta}}, \bibinfo {author} {\bibfnamefont {T.}~\bibnamefont {Nagao}},\ and\ \bibinfo {author} {\bibfnamefont {S.}~\bibnamefont {Hasegawa}},\ }\bibfield  {title} {\bibinfo {title} {Instability and charge density wave of metallic quantum chains on a silicon surface},\ }\href {https://doi.org/10.1103/PhysRevLett.82.4898} {\bibfield  {journal} {\bibinfo  {journal} {Phys. Rev. Lett.}\ }\textbf {\bibinfo {volume} {82}},\ \bibinfo {pages}
  {4898} (\bibinfo {year} {1999})}\BibitemShut {NoStop}%
\bibitem [{\citenamefont {Brouet}\ \emph {et~al.}(2004)\citenamefont {Brouet}, \citenamefont {Yang}, \citenamefont {Zhou}, \citenamefont {Hussain}, \citenamefont {Ru}, \citenamefont {Shin}, \citenamefont {Fisher},\ and\ \citenamefont {Shen}}]{brouet2004fermi}%
  \BibitemOpen
  \bibfield  {author} {\bibinfo {author} {\bibfnamefont {V.}~\bibnamefont {Brouet}}, \bibinfo {author} {\bibfnamefont {W.~L.}\ \bibnamefont {Yang}}, \bibinfo {author} {\bibfnamefont {X.~J.}\ \bibnamefont {Zhou}}, \bibinfo {author} {\bibfnamefont {Z.}~\bibnamefont {Hussain}}, \bibinfo {author} {\bibfnamefont {N.}~\bibnamefont {Ru}}, \bibinfo {author} {\bibfnamefont {K.~Y.}\ \bibnamefont {Shin}}, \bibinfo {author} {\bibfnamefont {I.~R.}\ \bibnamefont {Fisher}},\ and\ \bibinfo {author} {\bibfnamefont {Z.~X.}\ \bibnamefont {Shen}},\ }\bibfield  {title} {\bibinfo {title} {Fermi surface reconstruction in the cdw state of \ch{CeTe3} observed by photoemission},\ }\href {https://doi.org/https://link.aps.org/doi/10.1103/PhysRevLett.93.126405} {\bibfield  {journal} {\bibinfo  {journal} {Physical review letters}\ }\textbf {\bibinfo {volume} {93}},\ \bibinfo {pages} {126405} (\bibinfo {year} {2004})}\BibitemShut {NoStop}%
\bibitem [{\citenamefont {Tournier-Colletta}\ \emph {et~al.}(2013)\citenamefont {Tournier-Colletta}, \citenamefont {Moreschini}, \citenamefont {Aut\`es}, \citenamefont {Moser}, \citenamefont {Crepaldi}, \citenamefont {Berger}, \citenamefont {Walter}, \citenamefont {Kim}, \citenamefont {Bostwick}, \citenamefont {Monceau}, \citenamefont {Rotenberg}, \citenamefont {Yazyev},\ and\ \citenamefont {Grioni}}]{tournier2013electronic}%
  \BibitemOpen
  \bibfield  {author} {\bibinfo {author} {\bibfnamefont {C.}~\bibnamefont {Tournier-Colletta}}, \bibinfo {author} {\bibfnamefont {L.}~\bibnamefont {Moreschini}}, \bibinfo {author} {\bibfnamefont {G.}~\bibnamefont {Aut\`es}}, \bibinfo {author} {\bibfnamefont {S.}~\bibnamefont {Moser}}, \bibinfo {author} {\bibfnamefont {A.}~\bibnamefont {Crepaldi}}, \bibinfo {author} {\bibfnamefont {H.}~\bibnamefont {Berger}}, \bibinfo {author} {\bibfnamefont {A.~L.}\ \bibnamefont {Walter}}, \bibinfo {author} {\bibfnamefont {K.~S.}\ \bibnamefont {Kim}}, \bibinfo {author} {\bibfnamefont {A.}~\bibnamefont {Bostwick}}, \bibinfo {author} {\bibfnamefont {P.}~\bibnamefont {Monceau}}, \bibinfo {author} {\bibfnamefont {E.}~\bibnamefont {Rotenberg}}, \bibinfo {author} {\bibfnamefont {O.~V.}\ \bibnamefont {Yazyev}},\ and\ \bibinfo {author} {\bibfnamefont {M.}~\bibnamefont {Grioni}},\ }\bibfield  {title} {\bibinfo {title} {Electronic instability in a zero-gap semiconductor: The charge-density wave in \ch{(TaSe4)2I}},\ }\href
  {https://doi.org/10.1103/PhysRevLett.110.236401} {\bibfield  {journal} {\bibinfo  {journal} {Physical review letters}\ }\textbf {\bibinfo {volume} {110}},\ \bibinfo {pages} {236401} (\bibinfo {year} {2013})}\BibitemShut {NoStop}%
\bibitem [{\citenamefont {Hellmann}\ \emph {et~al.}(2010)\citenamefont {Hellmann}, \citenamefont {Beye}, \citenamefont {Sohrt}, \citenamefont {Rohwer}, \citenamefont {Sorgenfrei}, \citenamefont {Redlin}, \citenamefont {Kall\"ane}, \citenamefont {Marczynski-B\"uhlow}, \citenamefont {Hennies}, \citenamefont {Bauer}, \citenamefont {F\"ohlisch}, \citenamefont {Kipp}, \citenamefont {Wurth},\ and\ \citenamefont {Rossnagel}}]{TaS2PhysRevLett.105.187401}%
  \BibitemOpen
  \bibfield  {author} {\bibinfo {author} {\bibfnamefont {S.}~\bibnamefont {Hellmann}}, \bibinfo {author} {\bibfnamefont {M.}~\bibnamefont {Beye}}, \bibinfo {author} {\bibfnamefont {C.}~\bibnamefont {Sohrt}}, \bibinfo {author} {\bibfnamefont {T.}~\bibnamefont {Rohwer}}, \bibinfo {author} {\bibfnamefont {F.}~\bibnamefont {Sorgenfrei}}, \bibinfo {author} {\bibfnamefont {H.}~\bibnamefont {Redlin}}, \bibinfo {author} {\bibfnamefont {M.}~\bibnamefont {Kall\"ane}}, \bibinfo {author} {\bibfnamefont {M.}~\bibnamefont {Marczynski-B\"uhlow}}, \bibinfo {author} {\bibfnamefont {F.}~\bibnamefont {Hennies}}, \bibinfo {author} {\bibfnamefont {M.}~\bibnamefont {Bauer}}, \bibinfo {author} {\bibfnamefont {A.}~\bibnamefont {F\"ohlisch}}, \bibinfo {author} {\bibfnamefont {L.}~\bibnamefont {Kipp}}, \bibinfo {author} {\bibfnamefont {W.}~\bibnamefont {Wurth}},\ and\ \bibinfo {author} {\bibfnamefont {K.}~\bibnamefont {Rossnagel}},\ }\bibfield  {title} {\bibinfo {title} {Ultrafast melting of a charge-density wave in the mott insulator
  1\ch{T-TaS2}},\ }\href {https://doi.org/10.1103/PhysRevLett.105.187401} {\bibfield  {journal} {\bibinfo  {journal} {Phys. Rev. Lett.}\ }\textbf {\bibinfo {volume} {105}},\ \bibinfo {pages} {187401} (\bibinfo {year} {2010})}\BibitemShut {NoStop}%
\bibitem [{\citenamefont {Hellmann}\ \emph {et~al.}(2012)\citenamefont {Hellmann}, \citenamefont {Rohwer}, \citenamefont {Kall{\"a}ne}, \citenamefont {Hanff}, \citenamefont {Sohrt}, \citenamefont {Stange}, \citenamefont {Carr}, \citenamefont {Murnane}, \citenamefont {Kapteyn}, \citenamefont {Kipp} \emph {et~al.}}]{hellmann2012time}%
  \BibitemOpen
  \bibfield  {author} {\bibinfo {author} {\bibfnamefont {S.}~\bibnamefont {Hellmann}}, \bibinfo {author} {\bibfnamefont {T.}~\bibnamefont {Rohwer}}, \bibinfo {author} {\bibfnamefont {M.}~\bibnamefont {Kall{\"a}ne}}, \bibinfo {author} {\bibfnamefont {K.}~\bibnamefont {Hanff}}, \bibinfo {author} {\bibfnamefont {C.}~\bibnamefont {Sohrt}}, \bibinfo {author} {\bibfnamefont {A.}~\bibnamefont {Stange}}, \bibinfo {author} {\bibfnamefont {A.}~\bibnamefont {Carr}}, \bibinfo {author} {\bibfnamefont {M.}~\bibnamefont {Murnane}}, \bibinfo {author} {\bibfnamefont {H.}~\bibnamefont {Kapteyn}}, \bibinfo {author} {\bibfnamefont {L.}~\bibnamefont {Kipp}}, \emph {et~al.},\ }\bibfield  {title} {\bibinfo {title} {Time-domain classification of charge-density-wave insulators},\ }\href {https://doi.org/10.1038/ncomms2078} {\bibfield  {journal} {\bibinfo  {journal} {Nature communications}\ }\textbf {\bibinfo {volume} {3}},\ \bibinfo {pages} {1069} (\bibinfo {year} {2012})}\BibitemShut {NoStop}%
\bibitem [{\citenamefont {Tan}\ \emph {et~al.}(2021)\citenamefont {Tan}, \citenamefont {Liu}, \citenamefont {Wang},\ and\ \citenamefont {Yan}}]{Tan2021cdwsuperconductivity}%
  \BibitemOpen
  \bibfield  {author} {\bibinfo {author} {\bibfnamefont {H.}~\bibnamefont {Tan}}, \bibinfo {author} {\bibfnamefont {Y.}~\bibnamefont {Liu}}, \bibinfo {author} {\bibfnamefont {Z.}~\bibnamefont {Wang}},\ and\ \bibinfo {author} {\bibfnamefont {B.}~\bibnamefont {Yan}},\ }\bibfield  {title} {\bibinfo {title} {Charge density waves and electronic properties of superconducting kagome metals},\ }\href {https://doi.org/10.1103/PhysRevLett.127.046401} {\bibfield  {journal} {\bibinfo  {journal} {Phys. Rev. Lett.}\ }\textbf {\bibinfo {volume} {127}},\ \bibinfo {pages} {046401} (\bibinfo {year} {2021})}\BibitemShut {NoStop}%
\bibitem [{\citenamefont {Chen}\ \emph {et~al.}(2022)\citenamefont {Chen}, \citenamefont {Chen}, \citenamefont {Schnelle}, \citenamefont {Felser},\ and\ \citenamefont {Gaulin}}]{chen2022chargesuperconductivity}%
  \BibitemOpen
  \bibfield  {author} {\bibinfo {author} {\bibfnamefont {Q.}~\bibnamefont {Chen}}, \bibinfo {author} {\bibfnamefont {D.}~\bibnamefont {Chen}}, \bibinfo {author} {\bibfnamefont {W.}~\bibnamefont {Schnelle}}, \bibinfo {author} {\bibfnamefont {C.}~\bibnamefont {Felser}},\ and\ \bibinfo {author} {\bibfnamefont {B.~D.}\ \bibnamefont {Gaulin}},\ }\bibfield  {title} {\bibinfo {title} {Charge density wave order and fluctuations above t cdw and below superconducting \ch{T}$_c$ in the kagome metal csv 3 sb 5},\ }\href {https://journals.aps.org/prl/abstract/10.1103/PhysRevLett.129.056401} {\bibfield  {journal} {\bibinfo  {journal} {Physical Review Letters}\ }\textbf {\bibinfo {volume} {129}},\ \bibinfo {pages} {056401} (\bibinfo {year} {2022})}\BibitemShut {NoStop}%
\bibitem [{\citenamefont {Moncton}\ \emph {et~al.}(1975)\citenamefont {Moncton}, \citenamefont {Axe},\ and\ \citenamefont {DiSalvo}}]{Moncton1975neutron}%
  \BibitemOpen
  \bibfield  {author} {\bibinfo {author} {\bibfnamefont {D.~E.}\ \bibnamefont {Moncton}}, \bibinfo {author} {\bibfnamefont {J.~D.}\ \bibnamefont {Axe}},\ and\ \bibinfo {author} {\bibfnamefont {F.~J.}\ \bibnamefont {DiSalvo}},\ }\bibfield  {title} {\bibinfo {title} {Study of superlattice formation in 2\ch{H}-\ch{NbSe2} and 2\ch{H}-\ch{TaSe2} by neutron scattering},\ }\href {https://doi.org/10.1103/PhysRevLett.34.734} {\bibfield  {journal} {\bibinfo  {journal} {Phys. Rev. Lett.}\ }\textbf {\bibinfo {volume} {34}},\ \bibinfo {pages} {734} (\bibinfo {year} {1975})}\BibitemShut {NoStop}%
\bibitem [{\citenamefont {Weber}\ \emph {et~al.}(2011)\citenamefont {Weber}, \citenamefont {Rosenkranz}, \citenamefont {Castellan}, \citenamefont {Osborn}, \citenamefont {Hott}, \citenamefont {Heid}, \citenamefont {Bohnen}, \citenamefont {Egami}, \citenamefont {Said},\ and\ \citenamefont {Reznik}}]{Weber2011neutron}%
  \BibitemOpen
  \bibfield  {author} {\bibinfo {author} {\bibfnamefont {F.}~\bibnamefont {Weber}}, \bibinfo {author} {\bibfnamefont {S.}~\bibnamefont {Rosenkranz}}, \bibinfo {author} {\bibfnamefont {J.-P.}\ \bibnamefont {Castellan}}, \bibinfo {author} {\bibfnamefont {R.}~\bibnamefont {Osborn}}, \bibinfo {author} {\bibfnamefont {R.}~\bibnamefont {Hott}}, \bibinfo {author} {\bibfnamefont {R.}~\bibnamefont {Heid}}, \bibinfo {author} {\bibfnamefont {K.-P.}\ \bibnamefont {Bohnen}}, \bibinfo {author} {\bibfnamefont {T.}~\bibnamefont {Egami}}, \bibinfo {author} {\bibfnamefont {A.~H.}\ \bibnamefont {Said}},\ and\ \bibinfo {author} {\bibfnamefont {D.}~\bibnamefont {Reznik}},\ }\bibfield  {title} {\bibinfo {title} {Extended phonon collapse and the origin of the charge-density wave in 2\ch{H}-\ch{NbSe2}},\ }\href {https://doi.org/10.1103/PhysRevLett.107.107403} {\bibfield  {journal} {\bibinfo  {journal} {Phys. Rev. Lett.}\ }\textbf {\bibinfo {volume} {107}},\ \bibinfo {pages} {107403} (\bibinfo {year} {2011})}\BibitemShut {NoStop}%
\bibitem [{\citenamefont {Sugai}\ \emph {et~al.}(1985)\citenamefont {Sugai}, \citenamefont {Sato},\ and\ \citenamefont {Kurihara}}]{sugai1985TSI2Raman}%
  \BibitemOpen
  \bibfield  {author} {\bibinfo {author} {\bibfnamefont {S.}~\bibnamefont {Sugai}}, \bibinfo {author} {\bibfnamefont {M.}~\bibnamefont {Sato}},\ and\ \bibinfo {author} {\bibfnamefont {S.}~\bibnamefont {Kurihara}},\ }\bibfield  {title} {\bibinfo {title} {Interphonon interactions at the charge-density-wave phase transitions in \ch{(TaSe4)2I} and \ch{(NbSe4)2I}},\ }\href {https://journals.aps.org/prb/abstract/10.1103/PhysRevB.32.6809} {\bibfield  {journal} {\bibinfo  {journal} {Physical Review B}\ }\textbf {\bibinfo {volume} {32}},\ \bibinfo {pages} {6809} (\bibinfo {year} {1985})}\BibitemShut {NoStop}%
\bibitem [{\citenamefont {Sugai}\ \emph {et~al.}(2006)\citenamefont {Sugai}, \citenamefont {Takayanagi},\ and\ \citenamefont {Hayamizu}}]{sugai2006phasonRaman}%
  \BibitemOpen
  \bibfield  {author} {\bibinfo {author} {\bibfnamefont {S.}~\bibnamefont {Sugai}}, \bibinfo {author} {\bibfnamefont {Y.}~\bibnamefont {Takayanagi}},\ and\ \bibinfo {author} {\bibfnamefont {N.}~\bibnamefont {Hayamizu}},\ }\bibfield  {title} {\bibinfo {title} {Phason and amplitudon in the charge-density-wave phase of one-dimensional charge stripes in \ch{La}$_{2-x}$\ch{Sr}$_x$\ch{CuO4}},\ }\href {https://journals.aps.org/prl/abstract/10.1103/PhysRevLett.96.137003} {\bibfield  {journal} {\bibinfo  {journal} {Physical review letters}\ }\textbf {\bibinfo {volume} {96}},\ \bibinfo {pages} {137003} (\bibinfo {year} {2006})}\BibitemShut {NoStop}%
\bibitem [{\citenamefont {Song}\ \emph {et~al.}(2023)\citenamefont {Song}, \citenamefont {Wu}, \citenamefont {Chen}, \citenamefont {He}, \citenamefont {Uchiyama}, \citenamefont {Li}, \citenamefont {Cao}, \citenamefont {Guo}, \citenamefont {Cao},\ and\ \citenamefont {Birgeneau}}]{song2023phononslowing}%
  \BibitemOpen
  \bibfield  {author} {\bibinfo {author} {\bibfnamefont {Y.}~\bibnamefont {Song}}, \bibinfo {author} {\bibfnamefont {S.}~\bibnamefont {Wu}}, \bibinfo {author} {\bibfnamefont {X.}~\bibnamefont {Chen}}, \bibinfo {author} {\bibfnamefont {Y.}~\bibnamefont {He}}, \bibinfo {author} {\bibfnamefont {H.}~\bibnamefont {Uchiyama}}, \bibinfo {author} {\bibfnamefont {B.}~\bibnamefont {Li}}, \bibinfo {author} {\bibfnamefont {S.}~\bibnamefont {Cao}}, \bibinfo {author} {\bibfnamefont {J.}~\bibnamefont {Guo}}, \bibinfo {author} {\bibfnamefont {G.}~\bibnamefont {Cao}},\ and\ \bibinfo {author} {\bibfnamefont {R.}~\bibnamefont {Birgeneau}},\ }\bibfield  {title} {\bibinfo {title} {Phonon softening and slowing-down of charge density wave fluctuations in \ch{BaNi2As2}},\ }\href {https://journals.aps.org/prb/abstract/10.1103/PhysRevB.107.L041113} {\bibfield  {journal} {\bibinfo  {journal} {Physical Review B}\ }\textbf {\bibinfo {volume} {107}},\ \bibinfo {pages} {L041113} (\bibinfo {year} {2023})}\BibitemShut {NoStop}%
\bibitem [{\citenamefont {Kim}\ \emph {et~al.}(2023)\citenamefont {Kim}, \citenamefont {Lv}, \citenamefont {Sun}, \citenamefont {Zhao}, \citenamefont {Bielinski}, \citenamefont {Murzabekova}, \citenamefont {Qu}, \citenamefont {Duncan}, \citenamefont {Nguyen}, \citenamefont {Trigo} \emph {et~al.}}]{kim2023observation}%
  \BibitemOpen
  \bibfield  {author} {\bibinfo {author} {\bibfnamefont {S.}~\bibnamefont {Kim}}, \bibinfo {author} {\bibfnamefont {Y.}~\bibnamefont {Lv}}, \bibinfo {author} {\bibfnamefont {X.-Q.}\ \bibnamefont {Sun}}, \bibinfo {author} {\bibfnamefont {C.}~\bibnamefont {Zhao}}, \bibinfo {author} {\bibfnamefont {N.}~\bibnamefont {Bielinski}}, \bibinfo {author} {\bibfnamefont {A.}~\bibnamefont {Murzabekova}}, \bibinfo {author} {\bibfnamefont {K.}~\bibnamefont {Qu}}, \bibinfo {author} {\bibfnamefont {R.~A.}\ \bibnamefont {Duncan}}, \bibinfo {author} {\bibfnamefont {Q.~L.}\ \bibnamefont {Nguyen}}, \bibinfo {author} {\bibfnamefont {M.}~\bibnamefont {Trigo}}, \emph {et~al.},\ }\bibfield  {title} {\bibinfo {title} {Observation of a massive phason in a charge-density-wave insulator},\ }\href {https://www.nature.com/articles/s41563-023-01504-5} {\bibfield  {journal} {\bibinfo  {journal} {Nature Materials}\ ,\ \bibinfo {pages} {1}} (\bibinfo {year} {2023})}\BibitemShut {NoStop}%
\bibitem [{\citenamefont {Warawa}\ \emph {et~al.}(2023)\citenamefont {Warawa}, \citenamefont {Christophel}, \citenamefont {Sobolev}, \citenamefont {Demsar}, \citenamefont {Roskos},\ and\ \citenamefont {Thomson}}]{warawa2023combined}%
  \BibitemOpen
  \bibfield  {author} {\bibinfo {author} {\bibfnamefont {K.}~\bibnamefont {Warawa}}, \bibinfo {author} {\bibfnamefont {N.}~\bibnamefont {Christophel}}, \bibinfo {author} {\bibfnamefont {S.}~\bibnamefont {Sobolev}}, \bibinfo {author} {\bibfnamefont {J.}~\bibnamefont {Demsar}}, \bibinfo {author} {\bibfnamefont {H.~G.}\ \bibnamefont {Roskos}},\ and\ \bibinfo {author} {\bibfnamefont {M.~D.}\ \bibnamefont {Thomson}},\ }\bibfield  {title} {\bibinfo {title} {Combined investigation of collective amplitude and phase modes in a quasi-one-dimensional charge density wave system over a wide spectral range},\ }\href {https://doi.org/10.1103/PhysRevB.108.045147} {\bibfield  {journal} {\bibinfo  {journal} {Phys. Rev. B}\ }\textbf {\bibinfo {volume} {108}},\ \bibinfo {pages} {045147} (\bibinfo {year} {2023})}\BibitemShut {NoStop}%
\bibitem [{\citenamefont {Demsar}\ \emph {et~al.}(1999)\citenamefont {Demsar}, \citenamefont {Biljakovi{\'c}},\ and\ \citenamefont {Mihailovic}}]{demsar1999single}%
  \BibitemOpen
  \bibfield  {author} {\bibinfo {author} {\bibfnamefont {J.}~\bibnamefont {Demsar}}, \bibinfo {author} {\bibfnamefont {K.}~\bibnamefont {Biljakovi{\'c}}},\ and\ \bibinfo {author} {\bibfnamefont {D.}~\bibnamefont {Mihailovic}},\ }\bibfield  {title} {\bibinfo {title} {Single particle and collective excitations in the one-dimensional charge density wave solid \ch{K}$_{0.3}$\ch{MoO3} probed in real time by femtosecond spectroscopy},\ }\href {https://journals.aps.org/prl/abstract/10.1103/PhysRevLett.83.800} {\bibfield  {journal} {\bibinfo  {journal} {Physical review letters}\ }\textbf {\bibinfo {volume} {83}},\ \bibinfo {pages} {800} (\bibinfo {year} {1999})}\BibitemShut {NoStop}%
\bibitem [{\citenamefont {Schaefer}\ \emph {et~al.}(2014)\citenamefont {Schaefer}, \citenamefont {Kabanov},\ and\ \citenamefont {Demsar}}]{Schaefer2014Collective}%
  \BibitemOpen
  \bibfield  {author} {\bibinfo {author} {\bibfnamefont {H.}~\bibnamefont {Schaefer}}, \bibinfo {author} {\bibfnamefont {V.~V.}\ \bibnamefont {Kabanov}},\ and\ \bibinfo {author} {\bibfnamefont {J.}~\bibnamefont {Demsar}},\ }\bibfield  {title} {\bibinfo {title} {Collective modes in quasi-one-dimensional charge-density wave systems probed by femtosecond time-resolved optical studies},\ }\href {https://doi.org/10.1103/PhysRevB.89.045106} {\bibfield  {journal} {\bibinfo  {journal} {Phys. Rev. B}\ }\textbf {\bibinfo {volume} {89}},\ \bibinfo {pages} {045106} (\bibinfo {year} {2014})}\BibitemShut {NoStop}%
\bibitem [{\citenamefont {Zong}\ \emph {et~al.}(2019)\citenamefont {Zong}, \citenamefont {Dolgirev}, \citenamefont {Kogar}, \citenamefont {Erge\ifmmode~\mbox{\c{c}}\else \c{c}\fi{}en}, \citenamefont {Yilmaz}, \citenamefont {Bie}, \citenamefont {Rohwer}, \citenamefont {Tung}, \citenamefont {Straquadine}, \citenamefont {Wang}, \citenamefont {Yang}, \citenamefont {Shen}, \citenamefont {Li}, \citenamefont {Yang}, \citenamefont {Park}, \citenamefont {Hoffmann}, \citenamefont {Ofori-Okai}, \citenamefont {Kozina}, \citenamefont {Wen}, \citenamefont {Wang}, \citenamefont {Fisher}, \citenamefont {Jarillo-Herrero},\ and\ \citenamefont {Gedik}}]{zong2019dynamical}%
  \BibitemOpen
  \bibfield  {author} {\bibinfo {author} {\bibfnamefont {A.}~\bibnamefont {Zong}}, \bibinfo {author} {\bibfnamefont {P.~E.}\ \bibnamefont {Dolgirev}}, \bibinfo {author} {\bibfnamefont {A.}~\bibnamefont {Kogar}}, \bibinfo {author} {\bibfnamefont {E.}~\bibnamefont {Erge\ifmmode~\mbox{\c{c}}\else \c{c}\fi{}en}}, \bibinfo {author} {\bibfnamefont {M.~B.}\ \bibnamefont {Yilmaz}}, \bibinfo {author} {\bibfnamefont {Y.-Q.}\ \bibnamefont {Bie}}, \bibinfo {author} {\bibfnamefont {T.}~\bibnamefont {Rohwer}}, \bibinfo {author} {\bibfnamefont {I.-C.}\ \bibnamefont {Tung}}, \bibinfo {author} {\bibfnamefont {J.}~\bibnamefont {Straquadine}}, \bibinfo {author} {\bibfnamefont {X.}~\bibnamefont {Wang}}, \bibinfo {author} {\bibfnamefont {Y.}~\bibnamefont {Yang}}, \bibinfo {author} {\bibfnamefont {X.}~\bibnamefont {Shen}}, \bibinfo {author} {\bibfnamefont {R.}~\bibnamefont {Li}}, \bibinfo {author} {\bibfnamefont {J.}~\bibnamefont {Yang}}, \bibinfo {author} {\bibfnamefont {S.}~\bibnamefont {Park}}, \bibinfo {author} {\bibfnamefont
  {M.~C.}\ \bibnamefont {Hoffmann}}, \bibinfo {author} {\bibfnamefont {B.~K.}\ \bibnamefont {Ofori-Okai}}, \bibinfo {author} {\bibfnamefont {M.~E.}\ \bibnamefont {Kozina}}, \bibinfo {author} {\bibfnamefont {H.}~\bibnamefont {Wen}}, \bibinfo {author} {\bibfnamefont {X.}~\bibnamefont {Wang}}, \bibinfo {author} {\bibfnamefont {I.~R.}\ \bibnamefont {Fisher}}, \bibinfo {author} {\bibfnamefont {P.}~\bibnamefont {Jarillo-Herrero}},\ and\ \bibinfo {author} {\bibfnamefont {N.}~\bibnamefont {Gedik}},\ }\bibfield  {title} {\bibinfo {title} {Dynamical slowing-down in an ultrafast photoinduced phase transition},\ }\href {https://journals.aps.org/prl/abstract/10.1103/PhysRevLett.123.097601} {\bibfield  {journal} {\bibinfo  {journal} {Physical review letters}\ }\textbf {\bibinfo {volume} {123}},\ \bibinfo {pages} {097601} (\bibinfo {year} {2019})}\BibitemShut {NoStop}%
\bibitem [{\citenamefont {Nguyen}\ \emph {et~al.}(2023)\citenamefont {Nguyen}, \citenamefont {Duncan}, \citenamefont {Orenstein}, \citenamefont {Huang}, \citenamefont {Krapivin}, \citenamefont {de~la Pe\~na}, \citenamefont {Ornelas-Skarin}, \citenamefont {Reis}, \citenamefont {Abbamonte}, \citenamefont {Bettler}, \citenamefont {Chollet}, \citenamefont {Hoffmann}, \citenamefont {Hurley}, \citenamefont {Kim}, \citenamefont {Kirchmann}, \citenamefont {Kubota}, \citenamefont {Mahmood}, \citenamefont {Miller}, \citenamefont {Osaka}, \citenamefont {Qu}, \citenamefont {Sato}, \citenamefont {Shoemaker}, \citenamefont {Sirica}, \citenamefont {Song}, \citenamefont {Stanton}, \citenamefont {Teitelbaum}, \citenamefont {Tilton}, \citenamefont {Togashi}, \citenamefont {Zhu},\ and\ \citenamefont {Trigo}}]{Nguyen_Ultrafast_xRay2023}%
  \BibitemOpen
  \bibfield  {author} {\bibinfo {author} {\bibfnamefont {Q.~L.}\ \bibnamefont {Nguyen}}, \bibinfo {author} {\bibfnamefont {R.~A.}\ \bibnamefont {Duncan}}, \bibinfo {author} {\bibfnamefont {G.}~\bibnamefont {Orenstein}}, \bibinfo {author} {\bibfnamefont {Y.}~\bibnamefont {Huang}}, \bibinfo {author} {\bibfnamefont {V.}~\bibnamefont {Krapivin}}, \bibinfo {author} {\bibfnamefont {G.}~\bibnamefont {de~la Pe\~na}}, \bibinfo {author} {\bibfnamefont {C.}~\bibnamefont {Ornelas-Skarin}}, \bibinfo {author} {\bibfnamefont {D.~A.}\ \bibnamefont {Reis}}, \bibinfo {author} {\bibfnamefont {P.}~\bibnamefont {Abbamonte}}, \bibinfo {author} {\bibfnamefont {S.}~\bibnamefont {Bettler}}, \bibinfo {author} {\bibfnamefont {M.}~\bibnamefont {Chollet}}, \bibinfo {author} {\bibfnamefont {M.~C.}\ \bibnamefont {Hoffmann}}, \bibinfo {author} {\bibfnamefont {M.}~\bibnamefont {Hurley}}, \bibinfo {author} {\bibfnamefont {S.}~\bibnamefont {Kim}}, \bibinfo {author} {\bibfnamefont {P.~S.}\ \bibnamefont {Kirchmann}}, \bibinfo {author}
  {\bibfnamefont {Y.}~\bibnamefont {Kubota}}, \bibinfo {author} {\bibfnamefont {F.}~\bibnamefont {Mahmood}}, \bibinfo {author} {\bibfnamefont {A.}~\bibnamefont {Miller}}, \bibinfo {author} {\bibfnamefont {T.}~\bibnamefont {Osaka}}, \bibinfo {author} {\bibfnamefont {K.}~\bibnamefont {Qu}}, \bibinfo {author} {\bibfnamefont {T.}~\bibnamefont {Sato}}, \bibinfo {author} {\bibfnamefont {D.~P.}\ \bibnamefont {Shoemaker}}, \bibinfo {author} {\bibfnamefont {N.}~\bibnamefont {Sirica}}, \bibinfo {author} {\bibfnamefont {S.}~\bibnamefont {Song}}, \bibinfo {author} {\bibfnamefont {J.}~\bibnamefont {Stanton}}, \bibinfo {author} {\bibfnamefont {S.~W.}\ \bibnamefont {Teitelbaum}}, \bibinfo {author} {\bibfnamefont {S.~E.}\ \bibnamefont {Tilton}}, \bibinfo {author} {\bibfnamefont {T.}~\bibnamefont {Togashi}}, \bibinfo {author} {\bibfnamefont {D.}~\bibnamefont {Zhu}},\ and\ \bibinfo {author} {\bibfnamefont {M.}~\bibnamefont {Trigo}},\ }\bibfield  {title} {\bibinfo {title} {Ultrafast x-ray scattering reveals composite amplitude
  collective mode in the weyl charge density wave material \ch{(TaSe4)2I}},\ }\href {https://doi.org/10.1103/PhysRevLett.131.076901} {\bibfield  {journal} {\bibinfo  {journal} {Phys. Rev. Lett.}\ }\textbf {\bibinfo {volume} {131}},\ \bibinfo {pages} {076901} (\bibinfo {year} {2023})}\BibitemShut {NoStop}%
\bibitem [{\citenamefont {Schmitt}\ \emph {et~al.}(2008)\citenamefont {Schmitt}, \citenamefont {Kirchmann}, \citenamefont {Bovensiepen}, \citenamefont {Moore}, \citenamefont {Rettig}, \citenamefont {Krenz}, \citenamefont {Chu}, \citenamefont {Ru}, \citenamefont {Perfetti}, \citenamefont {Lu} \emph {et~al.}}]{schmitt2008transient}%
  \BibitemOpen
  \bibfield  {author} {\bibinfo {author} {\bibfnamefont {F.}~\bibnamefont {Schmitt}}, \bibinfo {author} {\bibfnamefont {P.~S.}\ \bibnamefont {Kirchmann}}, \bibinfo {author} {\bibfnamefont {U.}~\bibnamefont {Bovensiepen}}, \bibinfo {author} {\bibfnamefont {R.~G.}\ \bibnamefont {Moore}}, \bibinfo {author} {\bibfnamefont {L.}~\bibnamefont {Rettig}}, \bibinfo {author} {\bibfnamefont {M.}~\bibnamefont {Krenz}}, \bibinfo {author} {\bibfnamefont {J.-H.}\ \bibnamefont {Chu}}, \bibinfo {author} {\bibfnamefont {N.}~\bibnamefont {Ru}}, \bibinfo {author} {\bibfnamefont {L.}~\bibnamefont {Perfetti}}, \bibinfo {author} {\bibfnamefont {D.}~\bibnamefont {Lu}}, \emph {et~al.},\ }\bibfield  {title} {\bibinfo {title} {Transient electronic structure and melting of a charge density wave in \ch{TbTe3}},\ }\href {https://www.science.org/doi/10.1126/science.1160778} {\bibfield  {journal} {\bibinfo  {journal} {Science}\ }\textbf {\bibinfo {volume} {321}},\ \bibinfo {pages} {1649} (\bibinfo {year} {2008})}\BibitemShut {NoStop}%
\bibitem [{\citenamefont {Liu}\ \emph {et~al.}(2022)\citenamefont {Liu}, \citenamefont {Ma}, \citenamefont {He}, \citenamefont {Li}, \citenamefont {Tan}, \citenamefont {Liu}, \citenamefont {Xu}, \citenamefont {Tang}, \citenamefont {Watanabe}, \citenamefont {Taniguchi} \emph {et~al.}}]{liu2022observation}%
  \BibitemOpen
  \bibfield  {author} {\bibinfo {author} {\bibfnamefont {G.}~\bibnamefont {Liu}}, \bibinfo {author} {\bibfnamefont {X.}~\bibnamefont {Ma}}, \bibinfo {author} {\bibfnamefont {K.}~\bibnamefont {He}}, \bibinfo {author} {\bibfnamefont {Q.}~\bibnamefont {Li}}, \bibinfo {author} {\bibfnamefont {H.}~\bibnamefont {Tan}}, \bibinfo {author} {\bibfnamefont {Y.}~\bibnamefont {Liu}}, \bibinfo {author} {\bibfnamefont {J.}~\bibnamefont {Xu}}, \bibinfo {author} {\bibfnamefont {W.}~\bibnamefont {Tang}}, \bibinfo {author} {\bibfnamefont {K.}~\bibnamefont {Watanabe}}, \bibinfo {author} {\bibfnamefont {T.}~\bibnamefont {Taniguchi}}, \emph {et~al.},\ }\bibfield  {title} {\bibinfo {title} {Observation of anomalous amplitude modes in the kagome metal \ch{CsV3Sb5}},\ }\href {https://www.nature.com/articles/s41467-022-31162-1} {\bibfield  {journal} {\bibinfo  {journal} {Nature communications}\ }\textbf {\bibinfo {volume} {13}},\ \bibinfo {pages} {3461} (\bibinfo {year} {2022})}\BibitemShut {NoStop}%
\bibitem [{\citenamefont {Gooth}\ \emph {et~al.}(2019)\citenamefont {Gooth}, \citenamefont {Bradlyn}, \citenamefont {Honnali}, \citenamefont {Schindler}, \citenamefont {Kumar}, \citenamefont {Noky}, \citenamefont {Qi}, \citenamefont {Shekhar}, \citenamefont {Sun}, \citenamefont {Wang}, \citenamefont {Bernevig},\ and\ \citenamefont {Felser}}]{Gooth2019}%
  \BibitemOpen
  \bibfield  {author} {\bibinfo {author} {\bibfnamefont {J.}~\bibnamefont {Gooth}}, \bibinfo {author} {\bibfnamefont {B.}~\bibnamefont {Bradlyn}}, \bibinfo {author} {\bibfnamefont {S.}~\bibnamefont {Honnali}}, \bibinfo {author} {\bibfnamefont {C.}~\bibnamefont {Schindler}}, \bibinfo {author} {\bibfnamefont {N.}~\bibnamefont {Kumar}}, \bibinfo {author} {\bibfnamefont {J.}~\bibnamefont {Noky}}, \bibinfo {author} {\bibfnamefont {Y.}~\bibnamefont {Qi}}, \bibinfo {author} {\bibfnamefont {C.}~\bibnamefont {Shekhar}}, \bibinfo {author} {\bibfnamefont {Y.}~\bibnamefont {Sun}}, \bibinfo {author} {\bibfnamefont {Z.}~\bibnamefont {Wang}}, \bibinfo {author} {\bibfnamefont {B.~A.}\ \bibnamefont {Bernevig}},\ and\ \bibinfo {author} {\bibfnamefont {C.}~\bibnamefont {Felser}},\ }\bibfield  {title} {\bibinfo {title} {Axionic charge-density wave in the weyl semimetal \ch{(TaSe4)2I}},\ }\href {https://doi.org/10.1038/s41586-019-1630-4} {\bibfield  {journal} {\bibinfo  {journal} {Nature}\ }\textbf {\bibinfo {volume} {575}},\
  \bibinfo {pages} {315} (\bibinfo {year} {2019})}\BibitemShut {NoStop}%
\bibitem [{\citenamefont {Zhang}\ \emph {et~al.}(2020)\citenamefont {Zhang}, \citenamefont {Lin}, \citenamefont {Moreo}, \citenamefont {Dong},\ and\ \citenamefont {Dagotto}}]{Zhang2020firstprinciple}%
  \BibitemOpen
  \bibfield  {author} {\bibinfo {author} {\bibfnamefont {Y.}~\bibnamefont {Zhang}}, \bibinfo {author} {\bibfnamefont {L.-F.}\ \bibnamefont {Lin}}, \bibinfo {author} {\bibfnamefont {A.}~\bibnamefont {Moreo}}, \bibinfo {author} {\bibfnamefont {S.}~\bibnamefont {Dong}},\ and\ \bibinfo {author} {\bibfnamefont {E.}~\bibnamefont {Dagotto}},\ }\bibfield  {title} {\bibinfo {title} {First-principles study of the low-temperature charge density wave phase in the quasi-one-dimensional weyl chiral compound \ch{(TaSe4)2I}},\ }\href {https://doi.org/10.1103/PhysRevB.101.174106} {\bibfield  {journal} {\bibinfo  {journal} {Phys. Rev. B}\ }\textbf {\bibinfo {volume} {101}},\ \bibinfo {pages} {174106} (\bibinfo {year} {2020})}\BibitemShut {NoStop}%
\bibitem [{\citenamefont {Shi}\ \emph {et~al.}(2021)\citenamefont {Shi}, \citenamefont {Wieder}, \citenamefont {Meyerheim}, \citenamefont {Sun}, \citenamefont {Zhang}, \citenamefont {Li}, \citenamefont {Shen}, \citenamefont {Qi}, \citenamefont {Yang}, \citenamefont {Jena}, \citenamefont {Werner}, \citenamefont {Koepernik}, \citenamefont {Parkin}, \citenamefont {Chen}, \citenamefont {Felser}, \citenamefont {Bernevig},\ and\ \citenamefont {Wang}}]{Shi2021}%
  \BibitemOpen
  \bibfield  {author} {\bibinfo {author} {\bibfnamefont {W.}~\bibnamefont {Shi}}, \bibinfo {author} {\bibfnamefont {B.~J.}\ \bibnamefont {Wieder}}, \bibinfo {author} {\bibfnamefont {H.~L.}\ \bibnamefont {Meyerheim}}, \bibinfo {author} {\bibfnamefont {Y.}~\bibnamefont {Sun}}, \bibinfo {author} {\bibfnamefont {Y.}~\bibnamefont {Zhang}}, \bibinfo {author} {\bibfnamefont {Y.}~\bibnamefont {Li}}, \bibinfo {author} {\bibfnamefont {L.}~\bibnamefont {Shen}}, \bibinfo {author} {\bibfnamefont {Y.}~\bibnamefont {Qi}}, \bibinfo {author} {\bibfnamefont {L.}~\bibnamefont {Yang}}, \bibinfo {author} {\bibfnamefont {J.}~\bibnamefont {Jena}}, \bibinfo {author} {\bibfnamefont {P.}~\bibnamefont {Werner}}, \bibinfo {author} {\bibfnamefont {K.}~\bibnamefont {Koepernik}}, \bibinfo {author} {\bibfnamefont {S.}~\bibnamefont {Parkin}}, \bibinfo {author} {\bibfnamefont {Y.}~\bibnamefont {Chen}}, \bibinfo {author} {\bibfnamefont {C.}~\bibnamefont {Felser}}, \bibinfo {author} {\bibfnamefont {B.~A.}\ \bibnamefont {Bernevig}},\ and\
  \bibinfo {author} {\bibfnamefont {Z.}~\bibnamefont {Wang}},\ }\bibfield  {title} {\bibinfo {title} {A charge-density-wave topological semimetal},\ }\href {https://doi.org/10.1038/s41567-020-01104-z} {\bibfield  {journal} {\bibinfo  {journal} {Nature Physics}\ }\textbf {\bibinfo {volume} {17}},\ \bibinfo {pages} {381} (\bibinfo {year} {2021})}\BibitemShut {NoStop}%
\bibitem [{\citenamefont {Hohenberg}(1967)}]{Hohenberg1965}%
  \BibitemOpen
  \bibfield  {author} {\bibinfo {author} {\bibfnamefont {P.~C.}\ \bibnamefont {Hohenberg}},\ }\bibfield  {title} {\bibinfo {title} {Existence of long-range order in one and two dimensions},\ }\href {https://doi.org/10.1103/PhysRev.158.383} {\bibfield  {journal} {\bibinfo  {journal} {Phys. Rev.}\ }\textbf {\bibinfo {volume} {158}},\ \bibinfo {pages} {383} (\bibinfo {year} {1967})}\BibitemShut {NoStop}%
\bibitem [{\citenamefont {McKenzie}(1995)}]{McKenzie_PRB1995}%
  \BibitemOpen
  \bibfield  {author} {\bibinfo {author} {\bibfnamefont {R.~H.}\ \bibnamefont {McKenzie}},\ }\bibfield  {title} {\bibinfo {title} {Ginzburg-landau theory of phase transitions in quasi-one-dimensional systems},\ }\href {https://doi.org/10.1103/PhysRevB.51.6249} {\bibfield  {journal} {\bibinfo  {journal} {Phys. Rev. B}\ }\textbf {\bibinfo {volume} {51}},\ \bibinfo {pages} {6249} (\bibinfo {year} {1995})}\BibitemShut {NoStop}%
\bibitem [{\citenamefont {Monien}(2001)}]{monien2001exact}%
  \BibitemOpen
  \bibfield  {author} {\bibinfo {author} {\bibfnamefont {H.}~\bibnamefont {Monien}},\ }\bibfield  {title} {\bibinfo {title} {Exact results for the crossover from gaussian to non-gaussian order parameter fluctuations in quasi-one-dimensional electronic systems},\ }\href {https://journals.aps.org/prl/abstract/10.1103/PhysRevLett.87.126402} {\bibfield  {journal} {\bibinfo  {journal} {Physical Review Letters}\ }\textbf {\bibinfo {volume} {87}},\ \bibinfo {pages} {126402} (\bibinfo {year} {2001})}\BibitemShut {NoStop}%
\bibitem [{\citenamefont {Vescoli}\ \emph {et~al.}(2000)\citenamefont {Vescoli}, \citenamefont {Zwick}, \citenamefont {Voit}, \citenamefont {Berger}, \citenamefont {Zacchigna}, \citenamefont {Degiorgi}, \citenamefont {Grioni},\ and\ \citenamefont {Gr\"uner}}]{1DbandinsulatorPhysRevLett.84.1272}%
  \BibitemOpen
  \bibfield  {author} {\bibinfo {author} {\bibfnamefont {V.}~\bibnamefont {Vescoli}}, \bibinfo {author} {\bibfnamefont {F.}~\bibnamefont {Zwick}}, \bibinfo {author} {\bibfnamefont {J.}~\bibnamefont {Voit}}, \bibinfo {author} {\bibfnamefont {H.}~\bibnamefont {Berger}}, \bibinfo {author} {\bibfnamefont {M.}~\bibnamefont {Zacchigna}}, \bibinfo {author} {\bibfnamefont {L.}~\bibnamefont {Degiorgi}}, \bibinfo {author} {\bibfnamefont {M.}~\bibnamefont {Grioni}},\ and\ \bibinfo {author} {\bibfnamefont {G.}~\bibnamefont {Gr\"uner}},\ }\bibfield  {title} {\bibinfo {title} {Dynamical properties of the one-dimensional band insulator \ch{(NbSe4)3I}},\ }\href {https://doi.org/10.1103/PhysRevLett.84.1272} {\bibfield  {journal} {\bibinfo  {journal} {Phys. Rev. Lett.}\ }\textbf {\bibinfo {volume} {84}},\ \bibinfo {pages} {1272} (\bibinfo {year} {2000})}\BibitemShut {NoStop}%
\bibitem [{\citenamefont {Stare\ifmmode \check{s}\else \v{s}\fi{}ini\ifmmode~\acute{c}\else \'{c}\fi{}}\ \emph {et~al.}(2006)\citenamefont {Stare\ifmmode \check{s}\else \v{s}\fi{}ini\ifmmode~\acute{c}\else \'{c}\fi{}}, \citenamefont {Lunkenheimer}, \citenamefont {Hemberger}, \citenamefont {Biljakovi\ifmmode~\acute{c}\else \'{c}\fi{}},\ and\ \citenamefont {Loidl}}]{DielectricPhysRevLett.96.046402}%
  \BibitemOpen
  \bibfield  {author} {\bibinfo {author} {\bibfnamefont {D.}~\bibnamefont {Stare\ifmmode \check{s}\else \v{s}\fi{}ini\ifmmode~\acute{c}\else \'{c}\fi{}}}, \bibinfo {author} {\bibfnamefont {P.}~\bibnamefont {Lunkenheimer}}, \bibinfo {author} {\bibfnamefont {J.}~\bibnamefont {Hemberger}}, \bibinfo {author} {\bibfnamefont {K.}~\bibnamefont {Biljakovi\ifmmode~\acute{c}\else \'{c}\fi{}}},\ and\ \bibinfo {author} {\bibfnamefont {A.}~\bibnamefont {Loidl}},\ }\bibfield  {title} {\bibinfo {title} {Giant dielectric response in the one-dimensional charge-ordered semiconductor \ch{(NbSe4)3I}},\ }\href {https://doi.org/10.1103/PhysRevLett.96.046402} {\bibfield  {journal} {\bibinfo  {journal} {Phys. Rev. Lett.}\ }\textbf {\bibinfo {volume} {96}},\ \bibinfo {pages} {046402} (\bibinfo {year} {2006})}\BibitemShut {NoStop}%
\bibitem [{\citenamefont {Cheng}\ \emph {et~al.}(2024)\citenamefont {Cheng}, \citenamefont {Cheng}, \citenamefont {Jiang}, \citenamefont {Xia}, \citenamefont {Song}, \citenamefont {Mootz}, \citenamefont {Luo}, \citenamefont {Perakis}, \citenamefont {Yao}, \citenamefont {Guo} \emph {et~al.}}]{cheng2024chirality}%
  \BibitemOpen
  \bibfield  {author} {\bibinfo {author} {\bibfnamefont {B.}~\bibnamefont {Cheng}}, \bibinfo {author} {\bibfnamefont {D.}~\bibnamefont {Cheng}}, \bibinfo {author} {\bibfnamefont {T.}~\bibnamefont {Jiang}}, \bibinfo {author} {\bibfnamefont {W.}~\bibnamefont {Xia}}, \bibinfo {author} {\bibfnamefont {B.}~\bibnamefont {Song}}, \bibinfo {author} {\bibfnamefont {M.}~\bibnamefont {Mootz}}, \bibinfo {author} {\bibfnamefont {L.}~\bibnamefont {Luo}}, \bibinfo {author} {\bibfnamefont {I.~E.}\ \bibnamefont {Perakis}}, \bibinfo {author} {\bibfnamefont {Y.}~\bibnamefont {Yao}}, \bibinfo {author} {\bibfnamefont {Y.}~\bibnamefont {Guo}}, \emph {et~al.},\ }\bibfield  {title} {\bibinfo {title} {Chirality manipulation of ultrafast phase switches in a correlated cdw-weyl semimetal},\ }\href {https://www.nature.com/articles/s41467-024-45036-1} {\bibfield  {journal} {\bibinfo  {journal} {Nature Communications}\ }\textbf {\bibinfo {volume} {15}},\ \bibinfo {pages} {785} (\bibinfo {year} {2024})}\BibitemShut {NoStop}%
\bibitem [{\citenamefont {Bera}\ \emph {et~al.}(2021)\citenamefont {Bera}, \citenamefont {Gayen}, \citenamefont {Mondal}, \citenamefont {Pal}, \citenamefont {Pal}, \citenamefont {Vasdev}, \citenamefont {Howlader}, \citenamefont {Jana}, \citenamefont {Maity}, \citenamefont {Ali~Saha}, \citenamefont {Das}, \citenamefont {Satpati}, \citenamefont {Nath~Pal}, \citenamefont {Mandal}, \citenamefont {Sheet},\ and\ \citenamefont {Mondal}}]{MMondalnTSI}%
  \BibitemOpen
  \bibfield  {author} {\bibinfo {author} {\bibfnamefont {A.}~\bibnamefont {Bera}}, \bibinfo {author} {\bibfnamefont {S.}~\bibnamefont {Gayen}}, \bibinfo {author} {\bibfnamefont {S.}~\bibnamefont {Mondal}}, \bibinfo {author} {\bibfnamefont {R.}~\bibnamefont {Pal}}, \bibinfo {author} {\bibfnamefont {B.}~\bibnamefont {Pal}}, \bibinfo {author} {\bibfnamefont {A.}~\bibnamefont {Vasdev}}, \bibinfo {author} {\bibfnamefont {S.}~\bibnamefont {Howlader}}, \bibinfo {author} {\bibfnamefont {M.}~\bibnamefont {Jana}}, \bibinfo {author} {\bibfnamefont {T.}~\bibnamefont {Maity}}, \bibinfo {author} {\bibfnamefont {R.}~\bibnamefont {Ali~Saha}}, \bibinfo {author} {\bibfnamefont {B.}~\bibnamefont {Das}}, \bibinfo {author} {\bibfnamefont {B.}~\bibnamefont {Satpati}}, \bibinfo {author} {\bibfnamefont {A.}~\bibnamefont {Nath~Pal}}, \bibinfo {author} {\bibfnamefont {P.}~\bibnamefont {Mandal}}, \bibinfo {author} {\bibfnamefont {G.}~\bibnamefont {Sheet}},\ and\ \bibinfo {author} {\bibfnamefont {M.}~\bibnamefont {Mondal}},\ }\bibfield
  {title} {\bibinfo {title} {Superconductivity coexisting with ferromagnetism in a quasi-one dimensional non-centrosymmetric \ch{(TaSe4)3I}},\ }\href {https://arxiv.org/abs/2111.14525} {\bibfield  {journal} {\bibinfo  {journal} {arXiv:2111.14525}\ } (\bibinfo {year} {2021})}\BibitemShut {NoStop}%
\bibitem [{\citenamefont {Gressier}\ \emph {et~al.}(1984)\citenamefont {Gressier}, \citenamefont {Meerschaut}, \citenamefont {Guemas}, \citenamefont {Rouxel},\ and\ \citenamefont {Monceau}}]{gressier1984characterization}%
  \BibitemOpen
  \bibfield  {author} {\bibinfo {author} {\bibfnamefont {P.}~\bibnamefont {Gressier}}, \bibinfo {author} {\bibfnamefont {A.}~\bibnamefont {Meerschaut}}, \bibinfo {author} {\bibfnamefont {L.}~\bibnamefont {Guemas}}, \bibinfo {author} {\bibfnamefont {J.}~\bibnamefont {Rouxel}},\ and\ \bibinfo {author} {\bibfnamefont {P.}~\bibnamefont {Monceau}},\ }\bibfield  {title} {\bibinfo {title} {Characterization of the new series of quasi one-dimensional compounds \ch{(MX4)}$_n$\ch{Y} (\ch{M= Nb, Ta; X= S, Se; Y= Br, I})},\ }\href {https://www.sciencedirect.com/science/article/pii/002245968490327X} {\bibfield  {journal} {\bibinfo  {journal} {Journal of Solid State Chemistry}\ }\textbf {\bibinfo {volume} {51}},\ \bibinfo {pages} {141} (\bibinfo {year} {1984})}\BibitemShut {NoStop}%
\bibitem [{\citenamefont {Hartmann}\ \emph {et~al.}(2015)\citenamefont {Hartmann}, \citenamefont {Zielke}, \citenamefont {Polzin}, \citenamefont {Sasaki},\ and\ \citenamefont {M\"uller}}]{mullerPhysRevLett.114.216403}%
  \BibitemOpen
  \bibfield  {author} {\bibinfo {author} {\bibfnamefont {B.}~\bibnamefont {Hartmann}}, \bibinfo {author} {\bibfnamefont {D.}~\bibnamefont {Zielke}}, \bibinfo {author} {\bibfnamefont {J.}~\bibnamefont {Polzin}}, \bibinfo {author} {\bibfnamefont {T.}~\bibnamefont {Sasaki}},\ and\ \bibinfo {author} {\bibfnamefont {J.}~\bibnamefont {M\"uller}},\ }\bibfield  {title} {\bibinfo {title} {Critical slowing down of the charge carrier dynamics at the mott metal-insulator transition},\ }\href {https://doi.org/10.1103/PhysRevLett.114.216403} {\bibfield  {journal} {\bibinfo  {journal} {Phys. Rev. Lett.}\ }\textbf {\bibinfo {volume} {114}},\ \bibinfo {pages} {216403} (\bibinfo {year} {2015})}\BibitemShut {NoStop}%
\bibitem [{\citenamefont {Kundu}\ \emph {et~al.}(2017)\citenamefont {Kundu}, \citenamefont {Ray}, \citenamefont {Dolui}, \citenamefont {Bagwe}, \citenamefont {Choudhury}, \citenamefont {Krupanidhi}, \citenamefont {Das}, \citenamefont {Raychaudhuri},\ and\ \citenamefont {Bid}}]{hemantaPhysRevLett.119.226802}%
  \BibitemOpen
  \bibfield  {author} {\bibinfo {author} {\bibfnamefont {H.~K.}\ \bibnamefont {Kundu}}, \bibinfo {author} {\bibfnamefont {S.}~\bibnamefont {Ray}}, \bibinfo {author} {\bibfnamefont {K.}~\bibnamefont {Dolui}}, \bibinfo {author} {\bibfnamefont {V.}~\bibnamefont {Bagwe}}, \bibinfo {author} {\bibfnamefont {P.~R.}\ \bibnamefont {Choudhury}}, \bibinfo {author} {\bibfnamefont {S.~B.}\ \bibnamefont {Krupanidhi}}, \bibinfo {author} {\bibfnamefont {T.}~\bibnamefont {Das}}, \bibinfo {author} {\bibfnamefont {P.}~\bibnamefont {Raychaudhuri}},\ and\ \bibinfo {author} {\bibfnamefont {A.}~\bibnamefont {Bid}},\ }\bibfield  {title} {\bibinfo {title} {Quantum phase transition in few-layer ${\mathrm{nbse}}_{2}$ probed through quantized conductance fluctuations},\ }\href {https://doi.org/10.1103/PhysRevLett.119.226802} {\bibfield  {journal} {\bibinfo  {journal} {Phys. Rev. Lett.}\ }\textbf {\bibinfo {volume} {119}},\ \bibinfo {pages} {226802} (\bibinfo {year} {2017})}\BibitemShut {NoStop}%
\bibitem [{\citenamefont {Kundu}\ \emph {et~al.}(2020)\citenamefont {Kundu}, \citenamefont {Bar}, \citenamefont {Nayak},\ and\ \citenamefont {Bansal}}]{satyakiPhysRevLett.124.095703}%
  \BibitemOpen
  \bibfield  {author} {\bibinfo {author} {\bibfnamefont {S.}~\bibnamefont {Kundu}}, \bibinfo {author} {\bibfnamefont {T.}~\bibnamefont {Bar}}, \bibinfo {author} {\bibfnamefont {R.~K.}\ \bibnamefont {Nayak}},\ and\ \bibinfo {author} {\bibfnamefont {B.}~\bibnamefont {Bansal}},\ }\bibfield  {title} {\bibinfo {title} {Critical slowing down at the abrupt mott transition: When the first-order phase transition becomes zeroth order and looks like second order},\ }\href {https://doi.org/10.1103/PhysRevLett.124.095703} {\bibfield  {journal} {\bibinfo  {journal} {Phys. Rev. Lett.}\ }\textbf {\bibinfo {volume} {124}},\ \bibinfo {pages} {095703} (\bibinfo {year} {2020})}\BibitemShut {NoStop}%
\bibitem [{\citenamefont {Machado}\ \emph {et~al.}(2023)\citenamefont {Machado}, \citenamefont {Demler}, \citenamefont {Yao},\ and\ \citenamefont {Chatterjee}}]{machado2022quantum}%
  \BibitemOpen
  \bibfield  {author} {\bibinfo {author} {\bibfnamefont {F.}~\bibnamefont {Machado}}, \bibinfo {author} {\bibfnamefont {E.~A.}\ \bibnamefont {Demler}}, \bibinfo {author} {\bibfnamefont {N.~Y.}\ \bibnamefont {Yao}},\ and\ \bibinfo {author} {\bibfnamefont {S.}~\bibnamefont {Chatterjee}},\ }\bibfield  {title} {\bibinfo {title} {Quantum noise spectroscopy of dynamical critical phenomena},\ }\href {https://doi.org/10.1103/PhysRevLett.131.070801} {\bibfield  {journal} {\bibinfo  {journal} {Phys. Rev. Lett.}\ }\textbf {\bibinfo {volume} {131}},\ \bibinfo {pages} {070801} (\bibinfo {year} {2023})}\BibitemShut {NoStop}%
\bibitem [{\citenamefont {Ghosh}\ \emph {et~al.}(2004)\citenamefont {Ghosh}, \citenamefont {Kar}, \citenamefont {Bid},\ and\ \citenamefont {Raychaudhuri}}]{ghosh2004set}%
  \BibitemOpen
  \bibfield  {author} {\bibinfo {author} {\bibfnamefont {A.}~\bibnamefont {Ghosh}}, \bibinfo {author} {\bibfnamefont {S.}~\bibnamefont {Kar}}, \bibinfo {author} {\bibfnamefont {A.}~\bibnamefont {Bid}},\ and\ \bibinfo {author} {\bibfnamefont {A.}~\bibnamefont {Raychaudhuri}},\ }\bibfield  {title} {\bibinfo {title} {A set-up for measurement of low frequency conductance fluctuation (noise) using digital signal processing techniques},\ }\href {https://arxiv.org/abs/cond-mat/0402130} {\bibfield  {journal} {\bibinfo  {journal} {arXiv preprint cond-mat/0402130}\ } (\bibinfo {year} {2004})}\BibitemShut {NoStop}%
\bibitem [{\citenamefont {Stare{\v{s}}ini{\'c}}\ \emph {et~al.}(2002)\citenamefont {Stare{\v{s}}ini{\'c}}, \citenamefont {Ki{\v{s}}}, \citenamefont {Biljakovi{\'c}}, \citenamefont {Emerling}, \citenamefont {W.~Brill}, \citenamefont {Souletie}, \citenamefont {Berger},\ and\ \citenamefont {L{\'e}vy}}]{starevsinic2002specific}%
  \BibitemOpen
  \bibfield  {author} {\bibinfo {author} {\bibfnamefont {D.}~\bibnamefont {Stare{\v{s}}ini{\'c}}}, \bibinfo {author} {\bibfnamefont {A.}~\bibnamefont {Ki{\v{s}}}}, \bibinfo {author} {\bibfnamefont {K.}~\bibnamefont {Biljakovi{\'c}}}, \bibinfo {author} {\bibfnamefont {B.}~\bibnamefont {Emerling}}, \bibinfo {author} {\bibfnamefont {J.}~\bibnamefont {W.~Brill}}, \bibinfo {author} {\bibfnamefont {J.}~\bibnamefont {Souletie}}, \bibinfo {author} {\bibfnamefont {H.}~\bibnamefont {Berger}},\ and\ \bibinfo {author} {\bibfnamefont {F.}~\bibnamefont {L{\'e}vy}},\ }\bibfield  {title} {\bibinfo {title} {Specific heats of the charge density wave compounds o-\ch{TaS3} and \ch{(TaSe)2I}},\ }\href {https://link.springer.com/article/10.1140/epjb/e2002-00263-1} {\bibfield  {journal} {\bibinfo  {journal} {The European Physical Journal B-Condensed Matter and Complex Systems}\ }\textbf {\bibinfo {volume} {29}},\ \bibinfo {pages} {71} (\bibinfo {year} {2002})}\BibitemShut {NoStop}%
\bibitem [{\citenamefont {Saint-Paul}\ \emph {et~al.}(2019)\citenamefont {Saint-Paul}, \citenamefont {Monceau} \emph {et~al.}}]{saint2019survey}%
  \BibitemOpen
  \bibfield  {author} {\bibinfo {author} {\bibfnamefont {M.}~\bibnamefont {Saint-Paul}}, \bibinfo {author} {\bibfnamefont {P.}~\bibnamefont {Monceau}}, \emph {et~al.},\ }\bibfield  {title} {\bibinfo {title} {Survey of the thermodynamic properties of the charge density wave systems},\ }\href {https://www.hindawi.com/journals/acmp/2019/2138264/} {\bibfield  {journal} {\bibinfo  {journal} {Advances in Condensed Matter Physics}\ }\textbf {\bibinfo {volume} {2019}} (\bibinfo {year} {2019})}\BibitemShut {NoStop}%
\bibitem [{\citenamefont {Scofield}(1987)}]{scofield1987ac}%
  \BibitemOpen
  \bibfield  {author} {\bibinfo {author} {\bibfnamefont {J.~H.}\ \bibnamefont {Scofield}},\ }\bibfield  {title} {\bibinfo {title} {ac method for measuring low-frequency resistance fluctuation spectra},\ }\href {https://pubs.aip.org/aip/rsi/article/58/6/985/315239/ac-method-for-measuring-low-frequency-resistance} {\bibfield  {journal} {\bibinfo  {journal} {Review of scientific instruments}\ }\textbf {\bibinfo {volume} {58}},\ \bibinfo {pages} {985} (\bibinfo {year} {1987})}\BibitemShut {NoStop}%
\bibitem [{\citenamefont {Kogan}(1996)}]{kogan_noise}%
  \BibitemOpen
  \bibfield  {author} {\bibinfo {author} {\bibfnamefont {S.}~\bibnamefont {Kogan}},\ }\href@noop {} {\emph {\bibinfo {title} {Electronic noise and fluctuations in solids}}}\ (\bibinfo  {publisher} {Cambridge University Press, England},\ \bibinfo {year} {1996})\BibitemShut {NoStop}%
\bibitem [{\citenamefont {Weissman}(1988)}]{Weissman1988Noise}%
  \BibitemOpen
  \bibfield  {author} {\bibinfo {author} {\bibfnamefont {M.~B.}\ \bibnamefont {Weissman}},\ }\bibfield  {title} {\bibinfo {title} {$\frac{1}{f}$ noise and other slow, nonexponential kinetics in condensed matter},\ }\href {https://doi.org/10.1103/RevModPhys.60.537} {\bibfield  {journal} {\bibinfo  {journal} {Rev. Mod. Phys.}\ }\textbf {\bibinfo {volume} {60}},\ \bibinfo {pages} {537} (\bibinfo {year} {1988})}\BibitemShut {NoStop}%
\bibitem [{\citenamefont {An}\ \emph {et~al.}(2021)\citenamefont {An}, \citenamefont {Ba\ifmmode~\mbox{\c{s}}\else \c{s}\fi{}ar}, \citenamefont {Stephanov},\ and\ \citenamefont {Yee}}]{Hydrodynamic.PhysRevLett.127.072301}%
  \BibitemOpen
  \bibfield  {author} {\bibinfo {author} {\bibfnamefont {X.}~\bibnamefont {An}}, \bibinfo {author} {\bibfnamefont {G.~m.~c.}\ \bibnamefont {Ba\ifmmode~\mbox{\c{s}}\else \c{s}\fi{}ar}}, \bibinfo {author} {\bibfnamefont {M.}~\bibnamefont {Stephanov}},\ and\ \bibinfo {author} {\bibfnamefont {H.-U.}\ \bibnamefont {Yee}},\ }\bibfield  {title} {\bibinfo {title} {Evolution of non-gaussian hydrodynamic fluctuations},\ }\href {https://doi.org/10.1103/PhysRevLett.127.072301} {\bibfield  {journal} {\bibinfo  {journal} {Phys. Rev. Lett.}\ }\textbf {\bibinfo {volume} {127}},\ \bibinfo {pages} {072301} (\bibinfo {year} {2021})}\BibitemShut {NoStop}%
\bibitem [{\citenamefont {Stephanov}(2009)}]{QCDPhysRevLett.102.032301}%
  \BibitemOpen
  \bibfield  {author} {\bibinfo {author} {\bibfnamefont {M.~A.}\ \bibnamefont {Stephanov}},\ }\bibfield  {title} {\bibinfo {title} {Non-gaussian fluctuations near the qcd critical point},\ }\href {https://doi.org/10.1103/PhysRevLett.102.032301} {\bibfield  {journal} {\bibinfo  {journal} {Phys. Rev. Lett.}\ }\textbf {\bibinfo {volume} {102}},\ \bibinfo {pages} {032301} (\bibinfo {year} {2009})}\BibitemShut {NoStop}%
\bibitem [{\citenamefont {Antoniou}\ \emph {et~al.}(2006)\citenamefont {Antoniou}, \citenamefont {Diakonos}, \citenamefont {Kapoyannis},\ and\ \citenamefont {Kousouris}}]{QCDPhysRevLett.97.032002}%
  \BibitemOpen
  \bibfield  {author} {\bibinfo {author} {\bibfnamefont {N.~G.}\ \bibnamefont {Antoniou}}, \bibinfo {author} {\bibfnamefont {F.~K.}\ \bibnamefont {Diakonos}}, \bibinfo {author} {\bibfnamefont {A.~S.}\ \bibnamefont {Kapoyannis}},\ and\ \bibinfo {author} {\bibfnamefont {K.~S.}\ \bibnamefont {Kousouris}},\ }\bibfield  {title} {\bibinfo {title} {Critical opalescence in baryonic qcd matter},\ }\href {https://doi.org/10.1103/PhysRevLett.97.032002} {\bibfield  {journal} {\bibinfo  {journal} {Phys. Rev. Lett.}\ }\textbf {\bibinfo {volume} {97}},\ \bibinfo {pages} {032002} (\bibinfo {year} {2006})}\BibitemShut {NoStop}%
\bibitem [{\citenamefont {Saitoh}\ \emph {et~al.}(2020)\citenamefont {Saitoh}, \citenamefont {Hatano}, \citenamefont {Ikeda},\ and\ \citenamefont {Tighe}}]{PhysRevLett.124.118001}%
  \BibitemOpen
  \bibfield  {author} {\bibinfo {author} {\bibfnamefont {K.}~\bibnamefont {Saitoh}}, \bibinfo {author} {\bibfnamefont {T.}~\bibnamefont {Hatano}}, \bibinfo {author} {\bibfnamefont {A.}~\bibnamefont {Ikeda}},\ and\ \bibinfo {author} {\bibfnamefont {B.~P.}\ \bibnamefont {Tighe}},\ }\bibfield  {title} {\bibinfo {title} {Stress relaxation above and below the jamming transition},\ }\href {https://doi.org/10.1103/PhysRevLett.124.118001} {\bibfield  {journal} {\bibinfo  {journal} {Phys. Rev. Lett.}\ }\textbf {\bibinfo {volume} {124}},\ \bibinfo {pages} {118001} (\bibinfo {year} {2020})}\BibitemShut {NoStop}%
\bibitem [{\citenamefont {Shivers}\ \emph {et~al.}(2023)\citenamefont {Shivers}, \citenamefont {Sharma},\ and\ \citenamefont {MacKintosh}}]{PhysRevLett.131.178201}%
  \BibitemOpen
  \bibfield  {author} {\bibinfo {author} {\bibfnamefont {J.~L.}\ \bibnamefont {Shivers}}, \bibinfo {author} {\bibfnamefont {A.}~\bibnamefont {Sharma}},\ and\ \bibinfo {author} {\bibfnamefont {F.~C.}\ \bibnamefont {MacKintosh}},\ }\bibfield  {title} {\bibinfo {title} {Strain-controlled critical slowing down in the rheology of disordered networks},\ }\href {https://doi.org/10.1103/PhysRevLett.131.178201} {\bibfield  {journal} {\bibinfo  {journal} {Phys. Rev. Lett.}\ }\textbf {\bibinfo {volume} {131}},\ \bibinfo {pages} {178201} (\bibinfo {year} {2023})}\BibitemShut {NoStop}%
\bibitem [{\citenamefont {Gomez}\ \emph {et~al.}(2017)\citenamefont {Gomez}, \citenamefont {Moulton},\ and\ \citenamefont {Vella}}]{gomez2017critical}%
  \BibitemOpen
  \bibfield  {author} {\bibinfo {author} {\bibfnamefont {M.}~\bibnamefont {Gomez}}, \bibinfo {author} {\bibfnamefont {D.~E.}\ \bibnamefont {Moulton}},\ and\ \bibinfo {author} {\bibfnamefont {D.}~\bibnamefont {Vella}},\ }\bibfield  {title} {\bibinfo {title} {Critical slowing down in purely elastic ‘snap-through’instabilities},\ }\href {https://www.nature.com/articles/nphys3915} {\bibfield  {journal} {\bibinfo  {journal} {Nature Physics}\ }\textbf {\bibinfo {volume} {13}},\ \bibinfo {pages} {142} (\bibinfo {year} {2017})}\BibitemShut {NoStop}%
\bibitem [{\citenamefont {Marconi}\ \emph {et~al.}(2020)\citenamefont {Marconi}, \citenamefont {M\'etayer}, \citenamefont {Acquaviva}, \citenamefont {Boyer}, \citenamefont {Gomel}, \citenamefont {Quiniou}, \citenamefont {Masoller}, \citenamefont {Giudici},\ and\ \citenamefont {Tredicce}}]{PhysRevLett.125.134102}%
  \BibitemOpen
  \bibfield  {author} {\bibinfo {author} {\bibfnamefont {M.}~\bibnamefont {Marconi}}, \bibinfo {author} {\bibfnamefont {C.}~\bibnamefont {M\'etayer}}, \bibinfo {author} {\bibfnamefont {A.}~\bibnamefont {Acquaviva}}, \bibinfo {author} {\bibfnamefont {J.~M.}\ \bibnamefont {Boyer}}, \bibinfo {author} {\bibfnamefont {A.}~\bibnamefont {Gomel}}, \bibinfo {author} {\bibfnamefont {T.}~\bibnamefont {Quiniou}}, \bibinfo {author} {\bibfnamefont {C.}~\bibnamefont {Masoller}}, \bibinfo {author} {\bibfnamefont {M.}~\bibnamefont {Giudici}},\ and\ \bibinfo {author} {\bibfnamefont {J.~R.}\ \bibnamefont {Tredicce}},\ }\bibfield  {title} {\bibinfo {title} {Testing critical slowing down as a bifurcation indicator in a low-dissipation dynamical system},\ }\href {https://doi.org/10.1103/PhysRevLett.125.134102} {\bibfield  {journal} {\bibinfo  {journal} {Phys. Rev. Lett.}\ }\textbf {\bibinfo {volume} {125}},\ \bibinfo {pages} {134102} (\bibinfo {year} {2020})}\BibitemShut {NoStop}%
\bibitem [{\citenamefont {Scheffer}\ \emph {et~al.}(2009)\citenamefont {Scheffer}, \citenamefont {Bascompte}, \citenamefont {Brock}, \citenamefont {Brovkin}, \citenamefont {Carpenter}, \citenamefont {Dakos}, \citenamefont {Held}, \citenamefont {Van~Nes}, \citenamefont {Rietkerk},\ and\ \citenamefont {Sugihara}}]{scheffer2009critical}%
  \BibitemOpen
  \bibfield  {author} {\bibinfo {author} {\bibfnamefont {M.}~\bibnamefont {Scheffer}}, \bibinfo {author} {\bibfnamefont {J.}~\bibnamefont {Bascompte}}, \bibinfo {author} {\bibfnamefont {W.~A.}\ \bibnamefont {Brock}}, \bibinfo {author} {\bibfnamefont {V.}~\bibnamefont {Brovkin}}, \bibinfo {author} {\bibfnamefont {S.~R.}\ \bibnamefont {Carpenter}}, \bibinfo {author} {\bibfnamefont {V.}~\bibnamefont {Dakos}}, \bibinfo {author} {\bibfnamefont {H.}~\bibnamefont {Held}}, \bibinfo {author} {\bibfnamefont {E.~H.}\ \bibnamefont {Van~Nes}}, \bibinfo {author} {\bibfnamefont {M.}~\bibnamefont {Rietkerk}},\ and\ \bibinfo {author} {\bibfnamefont {G.}~\bibnamefont {Sugihara}},\ }\bibfield  {title} {\bibinfo {title} {Early-warning signals for critical transitions},\ }\href {https://www.nature.com/articles/nature08227} {\bibfield  {journal} {\bibinfo  {journal} {Nature}\ }\textbf {\bibinfo {volume} {461}},\ \bibinfo {pages} {53} (\bibinfo {year} {2009})}\BibitemShut {NoStop}%
\bibitem [{\citenamefont {Scheffer}\ \emph {et~al.}(2012)\citenamefont {Scheffer}, \citenamefont {Carpenter}, \citenamefont {Lenton}, \citenamefont {Bascompte}, \citenamefont {Brock}, \citenamefont {Dakos}, \citenamefont {Van~de Koppel}, \citenamefont {Van~de Leemput}, \citenamefont {Levin}, \citenamefont {Van~Nes} \emph {et~al.}}]{scheffer2012anticipating}%
  \BibitemOpen
  \bibfield  {author} {\bibinfo {author} {\bibfnamefont {M.}~\bibnamefont {Scheffer}}, \bibinfo {author} {\bibfnamefont {S.~R.}\ \bibnamefont {Carpenter}}, \bibinfo {author} {\bibfnamefont {T.~M.}\ \bibnamefont {Lenton}}, \bibinfo {author} {\bibfnamefont {J.}~\bibnamefont {Bascompte}}, \bibinfo {author} {\bibfnamefont {W.}~\bibnamefont {Brock}}, \bibinfo {author} {\bibfnamefont {V.}~\bibnamefont {Dakos}}, \bibinfo {author} {\bibfnamefont {J.}~\bibnamefont {Van~de Koppel}}, \bibinfo {author} {\bibfnamefont {I.~A.}\ \bibnamefont {Van~de Leemput}}, \bibinfo {author} {\bibfnamefont {S.~A.}\ \bibnamefont {Levin}}, \bibinfo {author} {\bibfnamefont {E.~H.}\ \bibnamefont {Van~Nes}}, \emph {et~al.},\ }\bibfield  {title} {\bibinfo {title} {Anticipating critical transitions},\ }\href {https://www.science.org/doi/10.1126/science.1225244} {\bibfield  {journal} {\bibinfo  {journal} {science}\ }\textbf {\bibinfo {volume} {338}},\ \bibinfo {pages} {344} (\bibinfo {year} {2012})}\BibitemShut {NoStop}%
\bibitem [{\citenamefont {Niermann}\ \emph {et~al.}(2015)\citenamefont {Niermann}, \citenamefont {Grams}, \citenamefont {Becker}, \citenamefont {Bohat\'y}, \citenamefont {Schenck},\ and\ \citenamefont {Hemberger}}]{NiermannPRL2015}%
  \BibitemOpen
  \bibfield  {author} {\bibinfo {author} {\bibfnamefont {D.}~\bibnamefont {Niermann}}, \bibinfo {author} {\bibfnamefont {C.~P.}\ \bibnamefont {Grams}}, \bibinfo {author} {\bibfnamefont {P.}~\bibnamefont {Becker}}, \bibinfo {author} {\bibfnamefont {L.}~\bibnamefont {Bohat\'y}}, \bibinfo {author} {\bibfnamefont {H.}~\bibnamefont {Schenck}},\ and\ \bibinfo {author} {\bibfnamefont {J.}~\bibnamefont {Hemberger}},\ }\bibfield  {title} {\bibinfo {title} {Critical slowing down near the multiferroic phase transition in ${\mathrm{mnwo}}_{4}$},\ }\href {https://doi.org/10.1103/PhysRevLett.114.037204} {\bibfield  {journal} {\bibinfo  {journal} {Phys. Rev. Lett.}\ }\textbf {\bibinfo {volume} {114}},\ \bibinfo {pages} {037204} (\bibinfo {year} {2015})}\BibitemShut {NoStop}%
\bibitem [{\citenamefont {Hooge}(1994)}]{Hooge19941byfnoise}%
  \BibitemOpen
  \bibfield  {author} {\bibinfo {author} {\bibfnamefont {F.~N.}\ \bibnamefont {Hooge}},\ }\bibfield  {title} {\bibinfo {title} {1/f noise sources},\ }\href {https://doi.org/10.1109/16.333808} {\bibfield  {journal} {\bibinfo  {journal} {IEEE Transactions on Electron Devices}\ }\textbf {\bibinfo {volume} {41}},\ \bibinfo {pages} {1926} (\bibinfo {year} {1994})}\BibitemShut {NoStop}%
\bibitem [{\citenamefont {Kriza}\ \emph {et~al.}(1989)\citenamefont {Kriza}, \citenamefont {Mih\'aly},\ and\ \citenamefont {Gr\"uner}}]{two_fluidPhysRevLett.62.2032}%
  \BibitemOpen
  \bibfield  {author} {\bibinfo {author} {\bibfnamefont {G.}~\bibnamefont {Kriza}}, \bibinfo {author} {\bibfnamefont {G.}~\bibnamefont {Mih\'aly}},\ and\ \bibinfo {author} {\bibfnamefont {G.}~\bibnamefont {Gr\"uner}},\ }\bibfield  {title} {\bibinfo {title} {Frequency-dependent thermoelectric power in \ch{K}$_{0.3}$\ch{MoO3}},\ }\href {https://doi.org/10.1103/PhysRevLett.62.2032} {\bibfield  {journal} {\bibinfo  {journal} {Phys. Rev. Lett.}\ }\textbf {\bibinfo {volume} {62}},\ \bibinfo {pages} {2032} (\bibinfo {year} {1989})}\BibitemShut {NoStop}%
\bibitem [{\citenamefont {Tinkham}(2004)}]{tinkham2004introduction}%
  \BibitemOpen
  \bibfield  {author} {\bibinfo {author} {\bibfnamefont {M.}~\bibnamefont {Tinkham}},\ }\href@noop {} {\emph {\bibinfo {title} {Introduction to superconductivity}}}\ (\bibinfo  {publisher} {Courier Corporation},\ \bibinfo {year} {2004})\BibitemShut {NoStop}%
\bibitem [{\citenamefont {Goldenfeld}(2018)}]{goldenfeld2018lectures}%
  \BibitemOpen
  \bibfield  {author} {\bibinfo {author} {\bibfnamefont {N.}~\bibnamefont {Goldenfeld}},\ }\href@noop {} {\emph {\bibinfo {title} {Lectures on phase transitions and the renormalization group}}}\ (\bibinfo  {publisher} {CRC Press},\ \bibinfo {year} {2018})\BibitemShut {NoStop}%
\bibitem [{\citenamefont {Chaikin}\ and\ \citenamefont {Lubensky}(1995)}]{chaikin1995principles}%
  \BibitemOpen
  \bibfield  {author} {\bibinfo {author} {\bibfnamefont {P.~M.}\ \bibnamefont {Chaikin}}\ and\ \bibinfo {author} {\bibfnamefont {T.~C.}\ \bibnamefont {Lubensky}},\ }\href@noop {} {\emph {\bibinfo {title} {Principles of condensed matter physics}}},\ Vol.~\bibinfo {volume} {10}\ (\bibinfo  {publisher} {Cambridge university press Cambridge},\ \bibinfo {year} {1995})\BibitemShut {NoStop}%
\bibitem [{\citenamefont {Gilmore}(1993)}]{gilmore1993catastrophe}%
  \BibitemOpen
  \bibfield  {author} {\bibinfo {author} {\bibfnamefont {R.}~\bibnamefont {Gilmore}},\ }\href@noop {} {\emph {\bibinfo {title} {Catastrophe theory for scientists and engineers}}}\ (\bibinfo  {publisher} {Courier Corporation},\ \bibinfo {year} {1993})\BibitemShut {NoStop}%
\bibitem [{\citenamefont {J.A.~Wilson}\ and\ \citenamefont {Mahajan}(1975)}]{Wilson1975}%
  \BibitemOpen
  \bibfield  {author} {\bibinfo {author} {\bibfnamefont {F.~D.~S.}\ \bibnamefont {J.A.~Wilson}}\ and\ \bibinfo {author} {\bibfnamefont {S.}~\bibnamefont {Mahajan}},\ }\bibfield  {title} {\bibinfo {title} {Charge-density waves and superlattices in the metallic layered transition metal dichalcogenides},\ }\href {https://doi.org/https://doi.org/10.1080/00018737500101391} {\bibfield  {journal} {\bibinfo  {journal} {Advances in Physics}\ }\textbf {\bibinfo {volume} {24}},\ \bibinfo {pages} {117} (\bibinfo {year} {1975})}\BibitemShut {NoStop}%
\bibitem [{\citenamefont {Lee}\ \emph {et~al.}(2019)\citenamefont {Lee}, \citenamefont {Goh},\ and\ \citenamefont {Cho}}]{Lee2019_TaS2}%
  \BibitemOpen
  \bibfield  {author} {\bibinfo {author} {\bibfnamefont {S.-H.}\ \bibnamefont {Lee}}, \bibinfo {author} {\bibfnamefont {J.~S.}\ \bibnamefont {Goh}},\ and\ \bibinfo {author} {\bibfnamefont {D.}~\bibnamefont {Cho}},\ }\bibfield  {title} {\bibinfo {title} {Origin of the insulating phase and first-order metal-insulator transition in 1\ch{T-TaS2}},\ }\href {https://doi.org/10.1103/PhysRevLett.122.106404} {\bibfield  {journal} {\bibinfo  {journal} {Phys. Rev. Lett.}\ }\textbf {\bibinfo {volume} {122}},\ \bibinfo {pages} {106404} (\bibinfo {year} {2019})}\BibitemShut {NoStop}%
\bibitem [{\citenamefont {Pelissetto}\ and\ \citenamefont {Vicari}(2002)}]{pelissetto2002critical}%
  \BibitemOpen
  \bibfield  {author} {\bibinfo {author} {\bibfnamefont {A.}~\bibnamefont {Pelissetto}}\ and\ \bibinfo {author} {\bibfnamefont {E.}~\bibnamefont {Vicari}},\ }\bibfield  {title} {\bibinfo {title} {Critical phenomena and renormalization-group theory},\ }\href {https://www.sciencedirect.com/science/article/pii/S0370157302002193} {\bibfield  {journal} {\bibinfo  {journal} {Physics Reports}\ }\textbf {\bibinfo {volume} {368}},\ \bibinfo {pages} {549} (\bibinfo {year} {2002})}\BibitemShut {NoStop}%
\bibitem [{\citenamefont {van Smaalen}\ \emph {et~al.}(2001)\citenamefont {van Smaalen}, \citenamefont {Lam},\ and\ \citenamefont {L{\"u}decke}}]{van2001structure}%
  \BibitemOpen
  \bibfield  {author} {\bibinfo {author} {\bibfnamefont {S.}~\bibnamefont {van Smaalen}}, \bibinfo {author} {\bibfnamefont {E.~J.}\ \bibnamefont {Lam}},\ and\ \bibinfo {author} {\bibfnamefont {J.}~\bibnamefont {L{\"u}decke}},\ }\bibfield  {title} {\bibinfo {title} {Structure of the charge-density wave in \ch{(TaSe4)2I}},\ }\href {https://iopscience.iop.org/article/10.1088/0953-8984/13/44/308/meta} {\bibfield  {journal} {\bibinfo  {journal} {Journal of Physics: Condensed Matter}\ }\textbf {\bibinfo {volume} {13}},\ \bibinfo {pages} {9923} (\bibinfo {year} {2001})}\BibitemShut {NoStop}%
\bibitem [{\citenamefont {Halperin}\ \emph {et~al.}(1974)\citenamefont {Halperin}, \citenamefont {Lubensky},\ and\ \citenamefont {Ma}}]{Halperin1974}%
  \BibitemOpen
  \bibfield  {author} {\bibinfo {author} {\bibfnamefont {B.~I.}\ \bibnamefont {Halperin}}, \bibinfo {author} {\bibfnamefont {T.~C.}\ \bibnamefont {Lubensky}},\ and\ \bibinfo {author} {\bibfnamefont {S.-k.}\ \bibnamefont {Ma}},\ }\bibfield  {title} {\bibinfo {title} {First-order phase transitions in superconductors and smectic-$a$ liquid crystals},\ }\href {https://doi.org/10.1103/PhysRevLett.32.292} {\bibfield  {journal} {\bibinfo  {journal} {Phys. Rev. Lett.}\ }\textbf {\bibinfo {volume} {32}},\ \bibinfo {pages} {292} (\bibinfo {year} {1974})}\BibitemShut {NoStop}%
\bibitem [{\citenamefont {Janoschek}\ \emph {et~al.}(2013)\citenamefont {Janoschek}, \citenamefont {Garst}, \citenamefont {Bauer}, \citenamefont {Krautscheid}, \citenamefont {Georgii}, \citenamefont {B\"oni},\ and\ \citenamefont {Pfleiderer}}]{fluctuationinduced2013}%
  \BibitemOpen
  \bibfield  {author} {\bibinfo {author} {\bibfnamefont {M.}~\bibnamefont {Janoschek}}, \bibinfo {author} {\bibfnamefont {M.}~\bibnamefont {Garst}}, \bibinfo {author} {\bibfnamefont {A.}~\bibnamefont {Bauer}}, \bibinfo {author} {\bibfnamefont {P.}~\bibnamefont {Krautscheid}}, \bibinfo {author} {\bibfnamefont {R.}~\bibnamefont {Georgii}}, \bibinfo {author} {\bibfnamefont {P.}~\bibnamefont {B\"oni}},\ and\ \bibinfo {author} {\bibfnamefont {C.}~\bibnamefont {Pfleiderer}},\ }\bibfield  {title} {\bibinfo {title} {Fluctuation-induced first-order phase transition in dzyaloshinskii-moriya helimagnets},\ }\href {https://journals.aps.org/prb/abstract/10.1103/PhysRevB.87.134407} {\bibfield  {journal} {\bibinfo  {journal} {Physical Review B}\ }\textbf {\bibinfo {volume} {87}},\ \bibinfo {pages} {134407} (\bibinfo {year} {2013})}\BibitemShut {NoStop}%
\bibitem [{\citenamefont {Yu}\ and\ \citenamefont {Liu}(2021)}]{yu2021pseudo}%
  \BibitemOpen
  \bibfield  {author} {\bibinfo {author} {\bibfnamefont {J.}~\bibnamefont {Yu}}\ and\ \bibinfo {author} {\bibfnamefont {C.-X.}\ \bibnamefont {Liu}},\ }\bibfield  {title} {\bibinfo {title} {Chapter five - pseudo-gauge fields in dirac and weyl materials},\ }in\ \href {https://doi.org/https://doi.org/10.1016/bs.semsem.2021.06.003} {\emph {\bibinfo {booktitle} {Topological Insulator and Related Topics}}},\ \bibinfo {series} {Semiconductors and Semimetals}, Vol.\ \bibinfo {volume} {108},\ \bibinfo {editor} {edited by\ \bibinfo {editor} {\bibfnamefont {L.}~\bibnamefont {Li}}\ and\ \bibinfo {editor} {\bibfnamefont {K.}~\bibnamefont {Sun}}}\ (\bibinfo  {publisher} {Elsevier},\ \bibinfo {year} {2021})\ pp.\ \bibinfo {pages} {195--224}\BibitemShut {NoStop}%
\bibitem [{\citenamefont {Seidler}\ and\ \citenamefont {Solin}(1996)}]{seidler1996non}%
  \BibitemOpen
  \bibfield  {author} {\bibinfo {author} {\bibfnamefont {G.~T.}\ \bibnamefont {Seidler}}\ and\ \bibinfo {author} {\bibfnamefont {S.~A.}\ \bibnamefont {Solin}},\ }\bibfield  {title} {\bibinfo {title} {Non-gaussian 1/f noise: Experimental optimization and separation of high-order amplitude and phase correlations},\ }\href {https://journals.aps.org/prb/abstract/10.1103/PhysRevB.53.9753} {\bibfield  {journal} {\bibinfo  {journal} {Physical Review B}\ }\textbf {\bibinfo {volume} {53}},\ \bibinfo {pages} {9753} (\bibinfo {year} {1996})}\BibitemShut {NoStop}%
\bibitem [{\citenamefont {Kar}\ \emph {et~al.}(2003)\citenamefont {Kar}, \citenamefont {Raychaudhuri}, \citenamefont {Ghosh}, \citenamefont {L\"ohneysen},\ and\ \citenamefont {Weiss}}]{AKRPhysRevLett.91.216603}%
  \BibitemOpen
  \bibfield  {author} {\bibinfo {author} {\bibfnamefont {S.}~\bibnamefont {Kar}}, \bibinfo {author} {\bibfnamefont {A.~K.}\ \bibnamefont {Raychaudhuri}}, \bibinfo {author} {\bibfnamefont {A.}~\bibnamefont {Ghosh}}, \bibinfo {author} {\bibfnamefont {H.~v.}\ \bibnamefont {L\"ohneysen}},\ and\ \bibinfo {author} {\bibfnamefont {G.}~\bibnamefont {Weiss}},\ }\bibfield  {title} {\bibinfo {title} {Observation of non-gaussian conductance fluctuations at low temperatures in \ch{Si:P(B)} at the metal-insulator transition},\ }\href {https://doi.org/10.1103/PhysRevLett.91.216603} {\bibfield  {journal} {\bibinfo  {journal} {Phys. Rev. Lett.}\ }\textbf {\bibinfo {volume} {91}},\ \bibinfo {pages} {216603} (\bibinfo {year} {2003})}\BibitemShut {NoStop}%
\bibitem [{\citenamefont {Chandni}\ \emph {et~al.}(2009)\citenamefont {Chandni}, \citenamefont {Ghosh}, \citenamefont {Vijaya},\ and\ \citenamefont {Mohan}}]{chadniPhysRevLett.102.025701}%
  \BibitemOpen
  \bibfield  {author} {\bibinfo {author} {\bibfnamefont {U.}~\bibnamefont {Chandni}}, \bibinfo {author} {\bibfnamefont {A.}~\bibnamefont {Ghosh}}, \bibinfo {author} {\bibfnamefont {H.~S.}\ \bibnamefont {Vijaya}},\ and\ \bibinfo {author} {\bibfnamefont {S.}~\bibnamefont {Mohan}},\ }\bibfield  {title} {\bibinfo {title} {Criticality of tuning in athermal phase transitions},\ }\href {https://doi.org/10.1103/PhysRevLett.102.025701} {\bibfield  {journal} {\bibinfo  {journal} {Phys. Rev. Lett.}\ }\textbf {\bibinfo {volume} {102}},\ \bibinfo {pages} {025701} (\bibinfo {year} {2009})}\BibitemShut {NoStop}%
\bibitem [{\citenamefont {Koushik}\ \emph {et~al.}(2013)\citenamefont {Koushik}, \citenamefont {Kumar}, \citenamefont {Amin}, \citenamefont {Mondal}, \citenamefont {Jesudasan}, \citenamefont {Bid}, \citenamefont {Raychaudhuri},\ and\ \citenamefont {Ghosh}}]{NbNPhysRevLett.111.197001}%
  \BibitemOpen
  \bibfield  {author} {\bibinfo {author} {\bibfnamefont {R.}~\bibnamefont {Koushik}}, \bibinfo {author} {\bibfnamefont {S.}~\bibnamefont {Kumar}}, \bibinfo {author} {\bibfnamefont {K.~R.}\ \bibnamefont {Amin}}, \bibinfo {author} {\bibfnamefont {M.}~\bibnamefont {Mondal}}, \bibinfo {author} {\bibfnamefont {J.}~\bibnamefont {Jesudasan}}, \bibinfo {author} {\bibfnamefont {A.}~\bibnamefont {Bid}}, \bibinfo {author} {\bibfnamefont {P.}~\bibnamefont {Raychaudhuri}},\ and\ \bibinfo {author} {\bibfnamefont {A.}~\bibnamefont {Ghosh}},\ }\bibfield  {title} {\bibinfo {title} {Correlated conductance fluctuations close to the berezinskii-kosterlitz-thouless transition in ultrathin nbn films},\ }\href {https://doi.org/10.1103/PhysRevLett.111.197001} {\bibfield  {journal} {\bibinfo  {journal} {Phys. Rev. Lett.}\ }\textbf {\bibinfo {volume} {111}},\ \bibinfo {pages} {197001} (\bibinfo {year} {2013})}\BibitemShut {NoStop}%
\bibitem [{\citenamefont {Chatterjee}\ \emph {et~al.}(2021)\citenamefont {Chatterjee}, \citenamefont {Bisht}, \citenamefont {Reddy},\ and\ \citenamefont {Raychaudhuri}}]{sudiptaPhysRevB.104.155101}%
  \BibitemOpen
  \bibfield  {author} {\bibinfo {author} {\bibfnamefont {S.}~\bibnamefont {Chatterjee}}, \bibinfo {author} {\bibfnamefont {R.~S.}\ \bibnamefont {Bisht}}, \bibinfo {author} {\bibfnamefont {V.~R.}\ \bibnamefont {Reddy}},\ and\ \bibinfo {author} {\bibfnamefont {A.~K.}\ \bibnamefont {Raychaudhuri}},\ }\bibfield  {title} {\bibinfo {title} {Emergence of large thermal noise close to a temperature-driven metal-insulator transition},\ }\href {https://doi.org/10.1103/PhysRevB.104.155101} {\bibfield  {journal} {\bibinfo  {journal} {Phys. Rev. B}\ }\textbf {\bibinfo {volume} {104}},\ \bibinfo {pages} {155101} (\bibinfo {year} {2021})}\BibitemShut {NoStop}%
\bibitem [{\citenamefont {Antal}\ \emph {et~al.}(2001)\citenamefont {Antal}, \citenamefont {Droz}, \citenamefont {Gy\"orgyi},\ and\ \citenamefont {R\'acz}}]{Antal2001PRLExtreme}%
  \BibitemOpen
  \bibfield  {author} {\bibinfo {author} {\bibfnamefont {T.}~\bibnamefont {Antal}}, \bibinfo {author} {\bibfnamefont {M.}~\bibnamefont {Droz}}, \bibinfo {author} {\bibfnamefont {G.}~\bibnamefont {Gy\"orgyi}},\ and\ \bibinfo {author} {\bibfnamefont {Z.}~\bibnamefont {R\'acz}},\ }\bibfield  {title} {\bibinfo {title} {$1/\mathit{f}$ noise and extreme value statistics},\ }\href {https://doi.org/10.1103/PhysRevLett.87.240601} {\bibfield  {journal} {\bibinfo  {journal} {Phys. Rev. Lett.}\ }\textbf {\bibinfo {volume} {87}},\ \bibinfo {pages} {240601} (\bibinfo {year} {2001})}\BibitemShut {NoStop}%
\bibitem [{\citenamefont {Joubaud}\ \emph {et~al.}(2008)\citenamefont {Joubaud}, \citenamefont {Petrosyan}, \citenamefont {Ciliberto},\ and\ \citenamefont {Garnier}}]{GarnierPhysRevLett.100.180601}%
  \BibitemOpen
  \bibfield  {author} {\bibinfo {author} {\bibfnamefont {S.}~\bibnamefont {Joubaud}}, \bibinfo {author} {\bibfnamefont {A.}~\bibnamefont {Petrosyan}}, \bibinfo {author} {\bibfnamefont {S.}~\bibnamefont {Ciliberto}},\ and\ \bibinfo {author} {\bibfnamefont {N.~B.}\ \bibnamefont {Garnier}},\ }\bibfield  {title} {\bibinfo {title} {Experimental evidence of non-gaussian fluctuations near a critical point},\ }\href {https://doi.org/10.1103/PhysRevLett.100.180601} {\bibfield  {journal} {\bibinfo  {journal} {Phys. Rev. Lett.}\ }\textbf {\bibinfo {volume} {100}},\ \bibinfo {pages} {180601} (\bibinfo {year} {2008})}\BibitemShut {NoStop}%
\bibitem [{\citenamefont {Bogdanovich}\ and\ \citenamefont {Popovi{\'c}}(2002)}]{bogdanovich2002onset}%
  \BibitemOpen
  \bibfield  {author} {\bibinfo {author} {\bibfnamefont {S.}~\bibnamefont {Bogdanovich}}\ and\ \bibinfo {author} {\bibfnamefont {D.}~\bibnamefont {Popovi{\'c}}},\ }\bibfield  {title} {\bibinfo {title} {Onset of glassy dynamics in a two-dimensional electron system in silicon},\ }\href {https://journals.aps.org/prl/abstract/10.1103/PhysRevLett.88.236401} {\bibfield  {journal} {\bibinfo  {journal} {Physical review letters}\ }\textbf {\bibinfo {volume} {88}},\ \bibinfo {pages} {236401} (\bibinfo {year} {2002})}\BibitemShut {NoStop}%
\bibitem [{\citenamefont {Jaroszy\ifmmode~\acute{n}\else \'{n}\fi{}ski}\ \emph {et~al.}(2002)\citenamefont {Jaroszy\ifmmode~\acute{n}\else \'{n}\fi{}ski}, \citenamefont {Popovi\ifmmode~\acute{c}\else \'{c}\fi{}},\ and\ \citenamefont {Klapwijk}}]{jaroszynski2002universal}%
  \BibitemOpen
  \bibfield  {author} {\bibinfo {author} {\bibfnamefont {J.}~\bibnamefont {Jaroszy\ifmmode~\acute{n}\else \'{n}\fi{}ski}}, \bibinfo {author} {\bibfnamefont {D.}~\bibnamefont {Popovi\ifmmode~\acute{c}\else \'{c}\fi{}}},\ and\ \bibinfo {author} {\bibfnamefont {T.~M.}\ \bibnamefont {Klapwijk}},\ }\bibfield  {title} {\bibinfo {title} {Universal behavior of the resistance noise across the metal-insulator transition in silicon inversion layers},\ }\href {https://journals.aps.org/prl/abstract/10.1103/PhysRevLett.89.276401} {\bibfield  {journal} {\bibinfo  {journal} {Physical Review Letters}\ }\textbf {\bibinfo {volume} {89}},\ \bibinfo {pages} {276401} (\bibinfo {year} {2002})}\BibitemShut {NoStop}%
\bibitem [{\citenamefont {Sheldrick}(2008)}]{Sheldrick2008}%
  \BibitemOpen
  \bibfield  {author} {\bibinfo {author} {\bibfnamefont {G.~M.}\ \bibnamefont {Sheldrick}},\ }\bibfield  {title} {\bibinfo {title} {A short history of shelx},\ }\href@noop {} {\bibfield  {journal} {\bibinfo  {journal} {Acta Crystallographica Section A: Foundations of Crystallography}\ }\textbf {\bibinfo {volume} {64}},\ \bibinfo {pages} {112} (\bibinfo {year} {2008})}\BibitemShut {NoStop}%
\bibitem [{\citenamefont {Dolomanov}\ \emph {et~al.}(2009)\citenamefont {Dolomanov}, \citenamefont {Bourhis}, \citenamefont {Gildea}, \citenamefont {Howard},\ and\ \citenamefont {Puschmann}}]{Dolomanov2009}%
  \BibitemOpen
  \bibfield  {author} {\bibinfo {author} {\bibfnamefont {O.~V.}\ \bibnamefont {Dolomanov}}, \bibinfo {author} {\bibfnamefont {L.~J.}\ \bibnamefont {Bourhis}}, \bibinfo {author} {\bibfnamefont {R.~J.}\ \bibnamefont {Gildea}}, \bibinfo {author} {\bibfnamefont {J.~A.~K.}\ \bibnamefont {Howard}},\ and\ \bibinfo {author} {\bibfnamefont {H.}~\bibnamefont {Puschmann}},\ }\bibfield  {title} {\bibinfo {title} {Olex2: a complete structure solution, refinement and analysis program},\ }\href@noop {} {\bibfield  {journal} {\bibinfo  {journal} {Journal of applied crystallography}\ }\textbf {\bibinfo {volume} {42}},\ \bibinfo {pages} {339} (\bibinfo {year} {2009})}\BibitemShut {NoStop}%
\bibitem [{\citenamefont {Yang}\ \emph {et~al.}(2020)\citenamefont {Yang}, \citenamefont {Rohde}, \citenamefont {Hanff}, \citenamefont {Stange}, \citenamefont {Xiong}, \citenamefont {Shi}, \citenamefont {Bauer},\ and\ \citenamefont {Rossnagel}}]{RbMoO3.PhysRevLett.125.266402}%
  \BibitemOpen
  \bibfield  {author} {\bibinfo {author} {\bibfnamefont {L.~X.}\ \bibnamefont {Yang}}, \bibinfo {author} {\bibfnamefont {G.}~\bibnamefont {Rohde}}, \bibinfo {author} {\bibfnamefont {K.}~\bibnamefont {Hanff}}, \bibinfo {author} {\bibfnamefont {A.}~\bibnamefont {Stange}}, \bibinfo {author} {\bibfnamefont {R.}~\bibnamefont {Xiong}}, \bibinfo {author} {\bibfnamefont {J.}~\bibnamefont {Shi}}, \bibinfo {author} {\bibfnamefont {M.}~\bibnamefont {Bauer}},\ and\ \bibinfo {author} {\bibfnamefont {K.}~\bibnamefont {Rossnagel}},\ }\bibfield  {title} {\bibinfo {title} {Bypassing the structural bottleneck in the ultrafast melting of electronic order},\ }\href {https://doi.org/10.1103/PhysRevLett.125.266402} {\bibfield  {journal} {\bibinfo  {journal} {Phys. Rev. Lett.}\ }\textbf {\bibinfo {volume} {125}},\ \bibinfo {pages} {266402} (\bibinfo {year} {2020})}\BibitemShut {NoStop}%
\bibitem [{\citenamefont {Tomeljak}\ \emph {et~al.}(2009)\citenamefont {Tomeljak}, \citenamefont {Sch\"afer}, \citenamefont {St\"adter}, \citenamefont {Beyer}, \citenamefont {Biljakovic},\ and\ \citenamefont {Demsar}}]{KMoO3.PhysRevLett.102.066404}%
  \BibitemOpen
  \bibfield  {author} {\bibinfo {author} {\bibfnamefont {A.}~\bibnamefont {Tomeljak}}, \bibinfo {author} {\bibfnamefont {H.}~\bibnamefont {Sch\"afer}}, \bibinfo {author} {\bibfnamefont {D.}~\bibnamefont {St\"adter}}, \bibinfo {author} {\bibfnamefont {M.}~\bibnamefont {Beyer}}, \bibinfo {author} {\bibfnamefont {K.}~\bibnamefont {Biljakovic}},\ and\ \bibinfo {author} {\bibfnamefont {J.}~\bibnamefont {Demsar}},\ }\bibfield  {title} {\bibinfo {title} {Dynamics of photoinduced charge-density-wave to metal phase transition in \ch{K}$_{0.3}$\ch{MoO3}},\ }\href {https://doi.org/10.1103/PhysRevLett.102.066404} {\bibfield  {journal} {\bibinfo  {journal} {Phys. Rev. Lett.}\ }\textbf {\bibinfo {volume} {102}},\ \bibinfo {pages} {066404} (\bibinfo {year} {2009})}\BibitemShut {NoStop}%
\bibitem [{\citenamefont {Perfetti}\ \emph {et~al.}(2006)\citenamefont {Perfetti}, \citenamefont {Loukakos}, \citenamefont {Lisowski}, \citenamefont {Bovensiepen}, \citenamefont {Berger}, \citenamefont {Biermann}, \citenamefont {Cornaglia}, \citenamefont {Georges},\ and\ \citenamefont {Wolf}}]{TaS2.PhysRevLett.97.067402}%
  \BibitemOpen
  \bibfield  {author} {\bibinfo {author} {\bibfnamefont {L.}~\bibnamefont {Perfetti}}, \bibinfo {author} {\bibfnamefont {P.~A.}\ \bibnamefont {Loukakos}}, \bibinfo {author} {\bibfnamefont {M.}~\bibnamefont {Lisowski}}, \bibinfo {author} {\bibfnamefont {U.}~\bibnamefont {Bovensiepen}}, \bibinfo {author} {\bibfnamefont {H.}~\bibnamefont {Berger}}, \bibinfo {author} {\bibfnamefont {S.}~\bibnamefont {Biermann}}, \bibinfo {author} {\bibfnamefont {P.~S.}\ \bibnamefont {Cornaglia}}, \bibinfo {author} {\bibfnamefont {A.}~\bibnamefont {Georges}},\ and\ \bibinfo {author} {\bibfnamefont {M.}~\bibnamefont {Wolf}},\ }\bibfield  {title} {\bibinfo {title} {Time evolution of the electronic structure of 1\ch{T-(TaS2)} through the insulator-metal transition},\ }\href {https://doi.org/10.1103/PhysRevLett.97.067402} {\bibfield  {journal} {\bibinfo  {journal} {Phys. Rev. Lett.}\ }\textbf {\bibinfo {volume} {97}},\ \bibinfo {pages} {067402} (\bibinfo {year} {2006})}\BibitemShut {NoStop}%
\bibitem [{\citenamefont {Anikin}\ \emph {et~al.}(2020)\citenamefont {Anikin}, \citenamefont {Schaller}, \citenamefont {Wiederrecht}, \citenamefont {Margine}, \citenamefont {Mazin},\ and\ \citenamefont {Karapetrov}}]{NbSe2.PhysRevB.102.205139}%
  \BibitemOpen
  \bibfield  {author} {\bibinfo {author} {\bibfnamefont {A.}~\bibnamefont {Anikin}}, \bibinfo {author} {\bibfnamefont {R.~D.}\ \bibnamefont {Schaller}}, \bibinfo {author} {\bibfnamefont {G.~P.}\ \bibnamefont {Wiederrecht}}, \bibinfo {author} {\bibfnamefont {E.~R.}\ \bibnamefont {Margine}}, \bibinfo {author} {\bibfnamefont {I.~I.}\ \bibnamefont {Mazin}},\ and\ \bibinfo {author} {\bibfnamefont {G.}~\bibnamefont {Karapetrov}},\ }\bibfield  {title} {\bibinfo {title} {Ultrafast dynamics in the high-symmetry and in the charge density wave phase of 2\ch{H}-\ch{NbSe2}},\ }\href {https://doi.org/10.1103/PhysRevB.102.205139} {\bibfield  {journal} {\bibinfo  {journal} {Phys. Rev. B}\ }\textbf {\bibinfo {volume} {102}},\ \bibinfo {pages} {205139} (\bibinfo {year} {2020})}\BibitemShut {NoStop}%
\bibitem [{\citenamefont {Hooge}\ \emph {et~al.}(1981)\citenamefont {Hooge}, \citenamefont {Kleinpenning},\ and\ \citenamefont {Vandamme}}]{hooge1981experimental}%
  \BibitemOpen
  \bibfield  {author} {\bibinfo {author} {\bibfnamefont {F.}~\bibnamefont {Hooge}}, \bibinfo {author} {\bibfnamefont {T.}~\bibnamefont {Kleinpenning}},\ and\ \bibinfo {author} {\bibfnamefont {L.~K.}\ \bibnamefont {Vandamme}},\ }\bibfield  {title} {\bibinfo {title} {Experimental studies on 1/f noise},\ }\href@noop {} {\bibfield  {journal} {\bibinfo  {journal} {Reports on progress in Physics}\ }\textbf {\bibinfo {volume} {44}},\ \bibinfo {pages} {479} (\bibinfo {year} {1981})}\BibitemShut {NoStop}%
\end{thebibliography}

\begin{thebibliography}{21}%
\makeatletter
\providecommand \@ifxundefined [1]{%
 \@ifx{#1\undefined}
}%
\providecommand \@ifnum [1]{%
 \ifnum #1\expandafter \@firstoftwo
 \else \expandafter \@secondoftwo
 \fi
}%
\providecommand \@ifx [1]{%
 \ifx #1\expandafter \@firstoftwo
 \else \expandafter \@secondoftwo
 \fi
}%
\providecommand \natexlab [1]{#1}%
\providecommand \enquote  [1]{``#1''}%
\providecommand \bibnamefont  [1]{#1}%
\providecommand \bibfnamefont [1]{#1}%
\providecommand \citenamefont [1]{#1}%
\providecommand \href@noop [0]{\@secondoftwo}%
\providecommand \href [0]{\begingroup \@sanitize@url \@href}%
\providecommand \@href[1]{\@@startlink{#1}\@@href}%
\providecommand \@@href[1]{\endgroup#1\@@endlink}%
\providecommand \@sanitize@url [0]{\catcode `\\12\catcode `\$12\catcode `\&12\catcode `\#12\catcode `\^12\catcode `\_12\catcode `\%12\relax}%
\providecommand \@@startlink[1]{}%
\providecommand \@@endlink[0]{}%
\providecommand \url  [0]{\begingroup\@sanitize@url \@url }%
\providecommand \@url [1]{\endgroup\@href {#1}{\urlprefix }}%
\providecommand \urlprefix  [0]{URL }%
\providecommand \Eprint [0]{\href }%
\providecommand \doibase [0]{https://doi.org/}%
\providecommand \selectlanguage [0]{\@gobble}%
\providecommand \bibinfo  [0]{\@secondoftwo}%
\providecommand \bibfield  [0]{\@secondoftwo}%
\providecommand \translation [1]{[#1]}%
\providecommand \BibitemOpen [0]{}%
\providecommand \bibitemStop [0]{}%
\providecommand \bibitemNoStop [0]{.\EOS\space}%
\providecommand \EOS [0]{\spacefactor3000\relax}%
\providecommand \BibitemShut  [1]{\csname bibitem#1\endcsname}%
\let\auto@bib@innerbib\@empty
\bibitem [{\citenamefont {Gressier}\ \emph {et~al.}(1984)\citenamefont {Gressier}, \citenamefont {Meerschaut}, \citenamefont {Guemas}, \citenamefont {Rouxel},\ and\ \citenamefont {Monceau}}]{gressier1984characterization.S}%
  \BibitemOpen
  \bibfield  {author} {\bibinfo {author} {\bibfnamefont {P.}~\bibnamefont {Gressier}}, \bibinfo {author} {\bibfnamefont {A.}~\bibnamefont {Meerschaut}}, \bibinfo {author} {\bibfnamefont {L.}~\bibnamefont {Guemas}}, \bibinfo {author} {\bibfnamefont {J.}~\bibnamefont {Rouxel}},\ and\ \bibinfo {author} {\bibfnamefont {P.}~\bibnamefont {Monceau}},\ }\bibfield  {title} {\bibinfo {title} {Characterization of the new series of quasi one-dimensional compounds \ch{(MX4)}$_n$\ch{Y} (\ch{M= Nb, Ta; X= S, Se; Y= Br, I})},\ }\href {https://www.sciencedirect.com/science/article/pii/002245968490327X} {\bibfield  {journal} {\bibinfo  {journal} {Journal of Solid State Chemistry}\ }\textbf {\bibinfo {volume} {51}},\ \bibinfo {pages} {141} (\bibinfo {year} {1984})}\BibitemShut {NoStop}%
\bibitem [{\citenamefont {Sheldrick}(2008)}]{Sheldrick2008.S}%
  \BibitemOpen
  \bibfield  {author} {\bibinfo {author} {\bibfnamefont {G.~M.}\ \bibnamefont {Sheldrick}},\ }\bibfield  {title} {\bibinfo {title} {A short history of shelx},\ }\href {https://doi.org/https://doi.org/10.1107/S0108767307043930} {\bibfield  {journal} {\bibinfo  {journal} {Acta Crystallographica Section A: Foundations of Crystallography}\ }\textbf {\bibinfo {volume} {64}},\ \bibinfo {pages} {112} (\bibinfo {year} {2008})}\BibitemShut {NoStop}%
\bibitem [{\citenamefont {Dolomanov}\ \emph {et~al.}(2009)\citenamefont {Dolomanov}, \citenamefont {Bourhis}, \citenamefont {Gildea}, \citenamefont {Howard},\ and\ \citenamefont {Puschmann}}]{Dolomanov2009.S}%
  \BibitemOpen
  \bibfield  {author} {\bibinfo {author} {\bibfnamefont {O.~V.}\ \bibnamefont {Dolomanov}}, \bibinfo {author} {\bibfnamefont {L.~J.}\ \bibnamefont {Bourhis}}, \bibinfo {author} {\bibfnamefont {R.~J.}\ \bibnamefont {Gildea}}, \bibinfo {author} {\bibfnamefont {J.~A.~K.}\ \bibnamefont {Howard}},\ and\ \bibinfo {author} {\bibfnamefont {H.}~\bibnamefont {Puschmann}},\ }\bibfield  {title} {\bibinfo {title} {Olex2: a complete structure solution, refinement and analysis program},\ }\href {https://doi.org/https://doi.org/10.1107/S0021889808042726} {\bibfield  {journal} {\bibinfo  {journal} {Journal of applied crystallography}\ }\textbf {\bibinfo {volume} {42}},\ \bibinfo {pages} {339} (\bibinfo {year} {2009})}\BibitemShut {NoStop}%
\bibitem [{\citenamefont {Ghosh}\ \emph {et~al.}(2004)\citenamefont {Ghosh}, \citenamefont {Kar}, \citenamefont {Bid},\ and\ \citenamefont {Raychaudhuri}}]{ghosh2004set.S}%
  \BibitemOpen
  \bibfield  {author} {\bibinfo {author} {\bibfnamefont {A.}~\bibnamefont {Ghosh}}, \bibinfo {author} {\bibfnamefont {S.}~\bibnamefont {Kar}}, \bibinfo {author} {\bibfnamefont {A.}~\bibnamefont {Bid}},\ and\ \bibinfo {author} {\bibfnamefont {A.}~\bibnamefont {Raychaudhuri}},\ }\bibfield  {title} {\bibinfo {title} {A set-up for measurement of low frequency conductance fluctuation (noise) using digital signal processing techniques},\ }\href {https://doi.org/10.48550/arXiv.cond-mat/0402130} {\bibfield  {journal} {\bibinfo  {journal} {arXiv preprint cond-mat/0402130}\ } (\bibinfo {year} {2004})}\BibitemShut {NoStop}%
\bibitem [{\citenamefont {Kogan}(1996)}]{kogan_noise.S}%
  \BibitemOpen
  \bibfield  {author} {\bibinfo {author} {\bibfnamefont {S.}~\bibnamefont {Kogan}},\ }\href@noop {} {\emph {\bibinfo {title} {Electronic noise and fluctuations in solids}}}\ (\bibinfo  {publisher} {Cambridge University Press, England},\ \bibinfo {year} {1996})\BibitemShut {NoStop}%
\bibitem [{\citenamefont {Kar}\ \emph {et~al.}(2003)\citenamefont {Kar}, \citenamefont {Raychaudhuri}, \citenamefont {Ghosh}, \citenamefont {L\"ohneysen},\ and\ \citenamefont {Weiss}}]{AKRPhysRevLett.91.216603.S}%
  \BibitemOpen
  \bibfield  {author} {\bibinfo {author} {\bibfnamefont {S.}~\bibnamefont {Kar}}, \bibinfo {author} {\bibfnamefont {A.~K.}\ \bibnamefont {Raychaudhuri}}, \bibinfo {author} {\bibfnamefont {A.}~\bibnamefont {Ghosh}}, \bibinfo {author} {\bibfnamefont {H.~v.}\ \bibnamefont {L\"ohneysen}},\ and\ \bibinfo {author} {\bibfnamefont {G.}~\bibnamefont {Weiss}},\ }\bibfield  {title} {\bibinfo {title} {Observation of non-gaussian conductance fluctuations at low temperatures in si:p(b) at the metal-insulator transition},\ }\href {https://doi.org/10.1103/PhysRevLett.91.216603} {\bibfield  {journal} {\bibinfo  {journal} {Phys. Rev. Lett.}\ }\textbf {\bibinfo {volume} {91}},\ \bibinfo {pages} {216603} (\bibinfo {year} {2003})}\BibitemShut {NoStop}%
\bibitem [{\citenamefont {Koushik}\ \emph {et~al.}(2013)\citenamefont {Koushik}, \citenamefont {Kumar}, \citenamefont {Amin}, \citenamefont {Mondal}, \citenamefont {Jesudasan}, \citenamefont {Bid}, \citenamefont {Raychaudhuri},\ and\ \citenamefont {Ghosh}}]{Koushik_R.2013}%
  \BibitemOpen
  \bibfield  {author} {\bibinfo {author} {\bibfnamefont {R.}~\bibnamefont {Koushik}}, \bibinfo {author} {\bibfnamefont {S.}~\bibnamefont {Kumar}}, \bibinfo {author} {\bibfnamefont {K.~R.}\ \bibnamefont {Amin}}, \bibinfo {author} {\bibfnamefont {M.}~\bibnamefont {Mondal}}, \bibinfo {author} {\bibfnamefont {J.}~\bibnamefont {Jesudasan}}, \bibinfo {author} {\bibfnamefont {A.}~\bibnamefont {Bid}}, \bibinfo {author} {\bibfnamefont {P.}~\bibnamefont {Raychaudhuri}},\ and\ \bibinfo {author} {\bibfnamefont {A.}~\bibnamefont {Ghosh}},\ }\bibfield  {title} {\bibinfo {title} {Correlated conductance fluctuations close to the berezinskii-kosterlitz-thouless transition in ultrathin nbn films},\ }\href {https://doi.org/10.1103/PhysRevLett.111.197001} {\bibfield  {journal} {\bibinfo  {journal} {Phys. Rev. Lett.}\ }\textbf {\bibinfo {volume} {111}},\ \bibinfo {pages} {197001} (\bibinfo {year} {2013})}\BibitemShut {NoStop}%
\bibitem [{\citenamefont {Chandni}\ \emph {et~al.}(2009)\citenamefont {Chandni}, \citenamefont {Ghosh}, \citenamefont {Vijaya},\ and\ \citenamefont {Mohan}}]{chadniPhysRevLett.102.025701.S}%
  \BibitemOpen
  \bibfield  {author} {\bibinfo {author} {\bibfnamefont {U.}~\bibnamefont {Chandni}}, \bibinfo {author} {\bibfnamefont {A.}~\bibnamefont {Ghosh}}, \bibinfo {author} {\bibfnamefont {H.~S.}\ \bibnamefont {Vijaya}},\ and\ \bibinfo {author} {\bibfnamefont {S.}~\bibnamefont {Mohan}},\ }\bibfield  {title} {\bibinfo {title} {Criticality of tuning in athermal phase transitions},\ }\href {https://doi.org/10.1103/PhysRevLett.102.025701} {\bibfield  {journal} {\bibinfo  {journal} {Phys. Rev. Lett.}\ }\textbf {\bibinfo {volume} {102}},\ \bibinfo {pages} {025701} (\bibinfo {year} {2009})}\BibitemShut {NoStop}%
\bibitem [{\citenamefont {Kundu}\ \emph {et~al.}(2020)\citenamefont {Kundu}, \citenamefont {Bar}, \citenamefont {Nayak},\ and\ \citenamefont {Bansal}}]{satyakiPhysRevLett.124.095703.S}%
  \BibitemOpen
  \bibfield  {author} {\bibinfo {author} {\bibfnamefont {S.}~\bibnamefont {Kundu}}, \bibinfo {author} {\bibfnamefont {T.}~\bibnamefont {Bar}}, \bibinfo {author} {\bibfnamefont {R.~K.}\ \bibnamefont {Nayak}},\ and\ \bibinfo {author} {\bibfnamefont {B.}~\bibnamefont {Bansal}},\ }\bibfield  {title} {\bibinfo {title} {Critical slowing down at the abrupt mott transition: When the first-order phase transition becomes zeroth order and looks like second order},\ }\href {https://doi.org/10.1103/PhysRevLett.124.095703} {\bibfield  {journal} {\bibinfo  {journal} {Phys. Rev. Lett.}\ }\textbf {\bibinfo {volume} {124}},\ \bibinfo {pages} {095703} (\bibinfo {year} {2020})}\BibitemShut {NoStop}%
\bibitem [{\citenamefont {Chatterjee}\ \emph {et~al.}(2021)\citenamefont {Chatterjee}, \citenamefont {Bisht}, \citenamefont {Reddy},\ and\ \citenamefont {Raychaudhuri}}]{sudiptaPhysRevB.104.155101.S}%
  \BibitemOpen
  \bibfield  {author} {\bibinfo {author} {\bibfnamefont {S.}~\bibnamefont {Chatterjee}}, \bibinfo {author} {\bibfnamefont {R.~S.}\ \bibnamefont {Bisht}}, \bibinfo {author} {\bibfnamefont {V.~R.}\ \bibnamefont {Reddy}},\ and\ \bibinfo {author} {\bibfnamefont {A.~K.}\ \bibnamefont {Raychaudhuri}},\ }\bibfield  {title} {\bibinfo {title} {Emergence of large thermal noise close to a temperature-driven metal-insulator transition},\ }\href {https://doi.org/10.1103/PhysRevB.104.155101} {\bibfield  {journal} {\bibinfo  {journal} {Phys. Rev. B}\ }\textbf {\bibinfo {volume} {104}},\ \bibinfo {pages} {155101} (\bibinfo {year} {2021})}\BibitemShut {NoStop}%
\bibitem [{\citenamefont {Yang}\ \emph {et~al.}(2020)\citenamefont {Yang}, \citenamefont {Rohde}, \citenamefont {Hanff}, \citenamefont {Stange}, \citenamefont {Xiong}, \citenamefont {Shi}, \citenamefont {Bauer},\ and\ \citenamefont {Rossnagel}}]{RbMoO3.PhysRevLett.125.266402.S}%
  \BibitemOpen
  \bibfield  {author} {\bibinfo {author} {\bibfnamefont {L.~X.}\ \bibnamefont {Yang}}, \bibinfo {author} {\bibfnamefont {G.}~\bibnamefont {Rohde}}, \bibinfo {author} {\bibfnamefont {K.}~\bibnamefont {Hanff}}, \bibinfo {author} {\bibfnamefont {A.}~\bibnamefont {Stange}}, \bibinfo {author} {\bibfnamefont {R.}~\bibnamefont {Xiong}}, \bibinfo {author} {\bibfnamefont {J.}~\bibnamefont {Shi}}, \bibinfo {author} {\bibfnamefont {M.}~\bibnamefont {Bauer}},\ and\ \bibinfo {author} {\bibfnamefont {K.}~\bibnamefont {Rossnagel}},\ }\bibfield  {title} {\bibinfo {title} {Bypassing the structural bottleneck in the ultrafast melting of electronic order},\ }\href {https://doi.org/10.1103/PhysRevLett.125.266402} {\bibfield  {journal} {\bibinfo  {journal} {Phys. Rev. Lett.}\ }\textbf {\bibinfo {volume} {125}},\ \bibinfo {pages} {266402} (\bibinfo {year} {2020})}\BibitemShut {NoStop}%
\bibitem [{\citenamefont {Tomeljak}\ \emph {et~al.}(2009)\citenamefont {Tomeljak}, \citenamefont {Sch\"afer}, \citenamefont {St\"adter}, \citenamefont {Beyer}, \citenamefont {Biljakovic},\ and\ \citenamefont {Demsar}}]{KMoO3.PhysRevLett.102.066404.S}%
  \BibitemOpen
  \bibfield  {author} {\bibinfo {author} {\bibfnamefont {A.}~\bibnamefont {Tomeljak}}, \bibinfo {author} {\bibfnamefont {H.}~\bibnamefont {Sch\"afer}}, \bibinfo {author} {\bibfnamefont {D.}~\bibnamefont {St\"adter}}, \bibinfo {author} {\bibfnamefont {M.}~\bibnamefont {Beyer}}, \bibinfo {author} {\bibfnamefont {K.}~\bibnamefont {Biljakovic}},\ and\ \bibinfo {author} {\bibfnamefont {J.}~\bibnamefont {Demsar}},\ }\bibfield  {title} {\bibinfo {title} {Dynamics of photoinduced charge-density-wave to metal phase transition in \ch{K}$_{0.3}$\ch{MoO3}},\ }\href {https://doi.org/10.1103/PhysRevLett.102.066404} {\bibfield  {journal} {\bibinfo  {journal} {Phys. Rev. Lett.}\ }\textbf {\bibinfo {volume} {102}},\ \bibinfo {pages} {066404} (\bibinfo {year} {2009})}\BibitemShut {NoStop}%
\bibitem [{\citenamefont {Perfetti}\ \emph {et~al.}(2006)\citenamefont {Perfetti}, \citenamefont {Loukakos}, \citenamefont {Lisowski}, \citenamefont {Bovensiepen}, \citenamefont {Berger}, \citenamefont {Biermann}, \citenamefont {Cornaglia}, \citenamefont {Georges},\ and\ \citenamefont {Wolf}}]{TaS2.PhysRevLett.97.067402.S}%
  \BibitemOpen
  \bibfield  {author} {\bibinfo {author} {\bibfnamefont {L.}~\bibnamefont {Perfetti}}, \bibinfo {author} {\bibfnamefont {P.~A.}\ \bibnamefont {Loukakos}}, \bibinfo {author} {\bibfnamefont {M.}~\bibnamefont {Lisowski}}, \bibinfo {author} {\bibfnamefont {U.}~\bibnamefont {Bovensiepen}}, \bibinfo {author} {\bibfnamefont {H.}~\bibnamefont {Berger}}, \bibinfo {author} {\bibfnamefont {S.}~\bibnamefont {Biermann}}, \bibinfo {author} {\bibfnamefont {P.~S.}\ \bibnamefont {Cornaglia}}, \bibinfo {author} {\bibfnamefont {A.}~\bibnamefont {Georges}},\ and\ \bibinfo {author} {\bibfnamefont {M.}~\bibnamefont {Wolf}},\ }\bibfield  {title} {\bibinfo {title} {Time evolution of the electronic structure of 1\ch{T-(TaS2)} through the insulator-metal transition},\ }\href {https://doi.org/10.1103/PhysRevLett.97.067402} {\bibfield  {journal} {\bibinfo  {journal} {Phys. Rev. Lett.}\ }\textbf {\bibinfo {volume} {97}},\ \bibinfo {pages} {067402} (\bibinfo {year} {2006})}\BibitemShut {NoStop}%
\bibitem [{\citenamefont {Zong}\ \emph {et~al.}(2019)\citenamefont {Zong}, \citenamefont {Dolgirev}, \citenamefont {Kogar}, \citenamefont {Erge\ifmmode~\mbox{\c{c}}\else \c{c}\fi{}en}, \citenamefont {Yilmaz}, \citenamefont {Bie}, \citenamefont {Rohwer}, \citenamefont {Tung}, \citenamefont {Straquadine}, \citenamefont {Wang}, \citenamefont {Yang}, \citenamefont {Shen}, \citenamefont {Li}, \citenamefont {Yang}, \citenamefont {Park}, \citenamefont {Hoffmann}, \citenamefont {Ofori-Okai}, \citenamefont {Kozina}, \citenamefont {Wen}, \citenamefont {Wang}, \citenamefont {Fisher}, \citenamefont {Jarillo-Herrero},\ and\ \citenamefont {Gedik}}]{N.Gedik.PhysRevLett.123.097601}%
  \BibitemOpen
  \bibfield  {author} {\bibinfo {author} {\bibfnamefont {A.}~\bibnamefont {Zong}}, \bibinfo {author} {\bibfnamefont {P.~E.}\ \bibnamefont {Dolgirev}}, \bibinfo {author} {\bibfnamefont {A.}~\bibnamefont {Kogar}}, \bibinfo {author} {\bibfnamefont {E.}~\bibnamefont {Erge\ifmmode~\mbox{\c{c}}\else \c{c}\fi{}en}}, \bibinfo {author} {\bibfnamefont {M.~B.}\ \bibnamefont {Yilmaz}}, \bibinfo {author} {\bibfnamefont {Y.-Q.}\ \bibnamefont {Bie}}, \bibinfo {author} {\bibfnamefont {T.}~\bibnamefont {Rohwer}}, \bibinfo {author} {\bibfnamefont {I.-C.}\ \bibnamefont {Tung}}, \bibinfo {author} {\bibfnamefont {J.}~\bibnamefont {Straquadine}}, \bibinfo {author} {\bibfnamefont {X.}~\bibnamefont {Wang}}, \bibinfo {author} {\bibfnamefont {Y.}~\bibnamefont {Yang}}, \bibinfo {author} {\bibfnamefont {X.}~\bibnamefont {Shen}}, \bibinfo {author} {\bibfnamefont {R.}~\bibnamefont {Li}}, \bibinfo {author} {\bibfnamefont {J.}~\bibnamefont {Yang}}, \bibinfo {author} {\bibfnamefont {S.}~\bibnamefont {Park}}, \bibinfo {author} {\bibfnamefont
  {M.~C.}\ \bibnamefont {Hoffmann}}, \bibinfo {author} {\bibfnamefont {B.~K.}\ \bibnamefont {Ofori-Okai}}, \bibinfo {author} {\bibfnamefont {M.~E.}\ \bibnamefont {Kozina}}, \bibinfo {author} {\bibfnamefont {H.}~\bibnamefont {Wen}}, \bibinfo {author} {\bibfnamefont {X.}~\bibnamefont {Wang}}, \bibinfo {author} {\bibfnamefont {I.~R.}\ \bibnamefont {Fisher}}, \bibinfo {author} {\bibfnamefont {P.}~\bibnamefont {Jarillo-Herrero}},\ and\ \bibinfo {author} {\bibfnamefont {N.}~\bibnamefont {Gedik}},\ }\bibfield  {title} {\bibinfo {title} {Dynamical slowing-down in an ultrafast photoinduced phase transition},\ }\href {https://doi.org/10.1103/PhysRevLett.123.097601} {\bibfield  {journal} {\bibinfo  {journal} {Phys. Rev. Lett.}\ }\textbf {\bibinfo {volume} {123}},\ \bibinfo {pages} {097601} (\bibinfo {year} {2019})}\BibitemShut {NoStop}%
\bibitem [{\citenamefont {Schmitt}\ \emph {et~al.}(2008)\citenamefont {Schmitt}, \citenamefont {Kirchmann}, \citenamefont {Bovensiepen}, \citenamefont {Moore}, \citenamefont {Rettig}, \citenamefont {Krenz}, \citenamefont {Chu}, \citenamefont {Ru}, \citenamefont {Perfetti}, \citenamefont {Lu} \emph {et~al.}}]{schmitt.science.1160778}%
  \BibitemOpen
  \bibfield  {author} {\bibinfo {author} {\bibfnamefont {F.}~\bibnamefont {Schmitt}}, \bibinfo {author} {\bibfnamefont {P.~S.}\ \bibnamefont {Kirchmann}}, \bibinfo {author} {\bibfnamefont {U.}~\bibnamefont {Bovensiepen}}, \bibinfo {author} {\bibfnamefont {R.~G.}\ \bibnamefont {Moore}}, \bibinfo {author} {\bibfnamefont {L.}~\bibnamefont {Rettig}}, \bibinfo {author} {\bibfnamefont {M.}~\bibnamefont {Krenz}}, \bibinfo {author} {\bibfnamefont {J.-H.}\ \bibnamefont {Chu}}, \bibinfo {author} {\bibfnamefont {N.}~\bibnamefont {Ru}}, \bibinfo {author} {\bibfnamefont {L.}~\bibnamefont {Perfetti}}, \bibinfo {author} {\bibfnamefont {D.}~\bibnamefont {Lu}}, \emph {et~al.},\ }\bibfield  {title} {\bibinfo {title} {Transient electronic structure and melting of a charge density wave in \ch{TbTe3}},\ }\href {https://www.science.org/doi/10.1126/science.1160778} {\bibfield  {journal} {\bibinfo  {journal} {Science}\ }\textbf {\bibinfo {volume} {321}},\ \bibinfo {pages} {1649} (\bibinfo {year} {2008})}\BibitemShut {NoStop}%
\bibitem [{\citenamefont {Anikin}\ \emph {et~al.}(2020)\citenamefont {Anikin}, \citenamefont {Schaller}, \citenamefont {Wiederrecht}, \citenamefont {Margine}, \citenamefont {Mazin},\ and\ \citenamefont {Karapetrov}}]{NbSe2.PhysRevB.102.205139.S}%
  \BibitemOpen
  \bibfield  {author} {\bibinfo {author} {\bibfnamefont {A.}~\bibnamefont {Anikin}}, \bibinfo {author} {\bibfnamefont {R.~D.}\ \bibnamefont {Schaller}}, \bibinfo {author} {\bibfnamefont {G.~P.}\ \bibnamefont {Wiederrecht}}, \bibinfo {author} {\bibfnamefont {E.~R.}\ \bibnamefont {Margine}}, \bibinfo {author} {\bibfnamefont {I.~I.}\ \bibnamefont {Mazin}},\ and\ \bibinfo {author} {\bibfnamefont {G.}~\bibnamefont {Karapetrov}},\ }\bibfield  {title} {\bibinfo {title} {Ultrafast dynamics in the high-symmetry and in the charge density wave phase of 2\ch{H}-\ch{NbSe2}},\ }\href {https://doi.org/10.1103/PhysRevB.102.205139} {\bibfield  {journal} {\bibinfo  {journal} {Phys. Rev. B}\ }\textbf {\bibinfo {volume} {102}},\ \bibinfo {pages} {205139} (\bibinfo {year} {2020})}\BibitemShut {NoStop}%
\bibitem [{\citenamefont {Hooge}(1994)}]{333808}%
  \BibitemOpen
  \bibfield  {author} {\bibinfo {author} {\bibfnamefont {F.}~\bibnamefont {Hooge}},\ }\bibfield  {title} {\bibinfo {title} {1/f noise sources},\ }\href {https://doi.org/10.1109/16.333808} {\bibfield  {journal} {\bibinfo  {journal} {IEEE Transactions on Electron Devices}\ }\textbf {\bibinfo {volume} {41}},\ \bibinfo {pages} {1926} (\bibinfo {year} {1994})}\BibitemShut {NoStop}%
\bibitem [{\citenamefont {Hooge}\ \emph {et~al.}(1981)\citenamefont {Hooge}, \citenamefont {Kleinpenning},\ and\ \citenamefont {Vandamme}}]{Hooge1981}%
  \BibitemOpen
  \bibfield  {author} {\bibinfo {author} {\bibfnamefont {F.~N.}\ \bibnamefont {Hooge}}, \bibinfo {author} {\bibfnamefont {T.~G.~M.}\ \bibnamefont {Kleinpenning}},\ and\ \bibinfo {author} {\bibfnamefont {L.~K.~J.}\ \bibnamefont {Vandamme}},\ }\bibfield  {title} {\bibinfo {title} {Experimental studies on 1/f noise},\ }\href {https://doi.org/10.1088/0034-4885/44/5/001} {\bibfield  {journal} {\bibinfo  {journal} {Reports on Progress in Physics}\ }\textbf {\bibinfo {volume} {44}},\ \bibinfo {pages} {479} (\bibinfo {year} {1981})}\BibitemShut {NoStop}%
\bibitem [{\citenamefont {Weissman}(1988)}]{Weissman1988Noise.S}%
  \BibitemOpen
  \bibfield  {author} {\bibinfo {author} {\bibfnamefont {M.~B.}\ \bibnamefont {Weissman}},\ }\bibfield  {title} {\bibinfo {title} {$\frac{1}{f}$ noise and other slow, nonexponential kinetics in condensed matter},\ }\href {https://doi.org/10.1103/RevModPhys.60.537} {\bibfield  {journal} {\bibinfo  {journal} {Rev. Mod. Phys.}\ }\textbf {\bibinfo {volume} {60}},\ \bibinfo {pages} {537} (\bibinfo {year} {1988})}\BibitemShut {NoStop}%
\bibitem [{\citenamefont {Gruner}(2018)}]{gruner2018density.S}%
  \BibitemOpen
  \bibfield  {author} {\bibinfo {author} {\bibfnamefont {G.}~\bibnamefont {Gruner}},\ }\href@noop {} {\emph {\bibinfo {title} {Density waves in solids}}}\ (\bibinfo  {publisher} {CRC press},\ \bibinfo {year} {2018})\BibitemShut {NoStop}%
\bibitem [{\citenamefont {Seidler}\ and\ \citenamefont {Solin}(1996)}]{SeidlerPhysRevB.53.9753}%
  \BibitemOpen
  \bibfield  {author} {\bibinfo {author} {\bibfnamefont {G.~T.}\ \bibnamefont {Seidler}}\ and\ \bibinfo {author} {\bibfnamefont {S.~A.}\ \bibnamefont {Solin}},\ }\bibfield  {title} {\bibinfo {title} {Non-gaussian 1/f noise: Experimental optimization and separation of high-order amplitude and phase correlations},\ }\href {https://doi.org/10.1103/PhysRevB.53.9753} {\bibfield  {journal} {\bibinfo  {journal} {Phys. Rev. B}\ }\textbf {\bibinfo {volume} {53}},\ \bibinfo {pages} {9753} (\bibinfo {year} {1996})}\BibitemShut {NoStop}%
\end{thebibliography}

%

\end{document}